\documentclass[11pt,a4paper,preprint,amsmath,amssymb,tightenlines]{article}
\pdfoutput=1
\usepackage{jheppub}
\usepackage[english]{babel}
\usepackage{graphicx}
\usepackage{bm}
\usepackage{subcaption}
\usepackage{slashed}
\usepackage{makecell}

\usepackage{tikz}
\usetikzlibrary{calc,positioning,arrows.meta,decorations.pathmorphing,decorations.markings}
\usepackage[compat=1.1.0]{tikz-feynman}
\tikzset{
    position/.style args={#1:#2 from #3}{
        at=(#3.#1), anchor=#1+180, shift=(#1:#2)
    }
}

\makeatletter
\pgfdeclaredecoration{gluon}{coil}
{
  \state{coil}[switch if less than=%
    0.5\pgfdecorationsegmentlength+%
    \pgfdecorationsegmentaspect\pgfdecorationsegmentamplitude+%
    \pgfdecorationsegmentaspect\pgfdecorationsegmentamplitude to last,
               width=+\pgfdecorationsegmentlength]
  {
    \pgfpathcurveto{\pgfpoint@oncoil{0}{0.555}{1}}{\pgfpoint@oncoil{0.545}{1}{2}}{\pgfpoint@oncoil{1}{1}{3}}
    \pgfpathcurveto{\pgfpoint@oncoil{1.555}{1}{4}}{\pgfpoint@oncoil{2}{0.555}{5}}{\pgfpoint@oncoil{2}{0}{6}}
    \pgfpathcurveto{\pgfpoint@oncoil{2}{-0.555}{7}}{\pgfpoint@oncoil{1.555}{-1}{8}}{\pgfpoint@oncoil{1}{-1}{9}}
    \pgfpathcurveto{\pgfpoint@oncoil{0.545}{-1}{10}}{\pgfpoint@oncoil{0}{-0.555}{11}}{\pgfpoint@oncoil{0}{0}{12}}
  }
  \state{last}[next state=final]
  {
    \pgfpathcurveto{\pgfpoint@oncoil{0}{0.555}{1}}{\pgfpoint@oncoil{0.545}{1}{2}}{\pgfpoint@oncoil{1}{1}{3}}
    \pgfpathcurveto{\pgfpoint@oncoil{1.555}{1}{4}}{\pgfpoint@oncoil{2}{0.555}{5}}{\pgfpoint@oncoil{2}{0}{6}}
  }
  \state{final}{}
}
\def\pgfpoint@oncoil#1#2#3{%
  \pgf@x=#1\pgfdecorationsegmentamplitude%
  \pgf@x=\pgfdecorationsegmentaspect\pgf@x%
  \pgf@y=#2\pgfdecorationsegmentamplitude%
  \pgf@xa=0.083333333333\pgfdecorationsegmentlength%
  \advance\pgf@x by#3\pgf@xa%
}
\makeatother
\tikzset{->-/.style={decoration={markings,mark=at position #1 with {\arrow{stealth}}},postaction={decorate}}}

\usepackage{xspace}
\usepackage{hyperref}
\graphicspath{{figs/}{./}}

\allowdisplaybreaks[4]

\usepackage{listings}
\definecolor{codegreen}{rgb}{0,0.6,0}
\definecolor{codegray}{rgb}{0.5,0.5,0.5}
\definecolor{codepurple}{rgb}{0.58,0,0.82}
\definecolor{backcolour}{rgb}{0.95,0.95,0.95}
\lstdefinestyle{mystyle}{
    backgroundcolor=\color{backcolour},   
    commentstyle=\color{codegreen},
    keywordstyle=\color{magenta},
    numberstyle=\tiny\color{codegray},
    stringstyle=\color{codepurple},
    basicstyle=\ttfamily\footnotesize,
    breakatwhitespace=false,         
    breaklines=true,                 
    captionpos=b,                    
    keepspaces=true,                 
    numbers=left,                    
    numbersep=5pt,                  
    showspaces=false,                
    showstringspaces=false,
    showtabs=false,                  
    tabsize=2
}
\allowdisplaybreaks[4]
\newcommand{\nn}{\nonumber}

\newcommand{\sjt}{ \widetilde{\phantom{\,}sj} }

\newcommand\evenoverset[2]{\overset{\scriptstyle #1\mathstrut}{\scriptstyle #2}}

\definecolor{darkyellow}{rgb}{0.5, 0.5, 0.0}
\definecolor{darkpurple}{rgb}{0.5, 0.2, 0.8}
\definecolor{darkblue}{rgb}{0.0, 0.0, 0.8}
\definecolor{darkgreen}{rgb}{0.0, 0.4, 0.0}
\definecolor{darkred}{rgb}{0.5, 0.0, 0.0}

\hypersetup{
    linktocpage,
     colorlinks,
     citecolor=darkred,
     linkcolor= darkred,
     urlcolor=darkred
}

\title{Heavy Jet Mass in Hadronic Higgs Decays}

\author[a]{Luen Clingerman}
\emailAdd{lmalshi@g.harvard.edu}
\affiliation[a]{Department of Physics, Harvard University, Cambridge, MA 02138, USA}

\author[b]{and Xiaoyuan Zhang}
\emailAdd{xyz2@mit.edu}
\affiliation[b]{Center for Theoretical Physics - a Leinweber Institute, Massachusetts Institute of Technology, Cambridge, MA 02139, USA}

\preprint{\vbox{
		\hbox{MIT-CTP/6065}
}}

\abstract{
    The heavy jet mass distributions in hadronic Higgs decays are computed to next-to-next-to-next-to-leading logarithmic order (N${}^3$LL${}^\prime$) in the dijet limit and NNLL in the trijet limit, matched to the next-to-next-to-leading order (NNLO). 
    Both resummation results are obtained from the factorization theorems in Soft-Collinear Effective Theory. 
    In particular, we study the Sudakov shoulders in the trijet region, originating from the incomplete cancellation of infrared singularities between final states with different parton multiplicities, and resum the induced large logarithms to all orders.
    The shoulder resummation yields sizable corrections and improves the perturbative stability of the distribution.
    Our results provide state-of-the-art predictions for heavy jet mass in $H \to gg$ and $H\to q\bar{q}$ and can be applied to precision Higgs measurements at future $e^+e^-$ colliders.
}

\begin{document}
\maketitle
\flushbottom
\newpage

\section{Introduction}

The discovery of the Higgs boson at the Large Hadron Collider (LHC)~\cite{ATLAS:2012yve,CMS:2012qbp} completed the particle content of the Standard Model (SM) and established the experimental study of the Higgs sector as the central program of high-energy physics. Precise measurements of the Higgs boson, including its couplings to gauge bosons and fermions as well as its self-coupling, will provide us valuable information on understanding the Standard Model and searching for new physics.
The LHC and its high-luminosity upgrade will pin down the dominant Higgs couplings only at the few-percent level~\cite{deBlas:2019rxi}, and a qualitative leap in precision will require the clean environment of a lepton-collider Higgs factory. Among the proposed machines, FCC-ee~\cite{TLEPDesignStudyWorkingGroup:2013myl}, the CEPC~\cite{CEPCStudyGroup:2018rmc,CEPCStudyGroup:2018ghi,CEPCStudyGroup:2023quu}, and a future high-energy muon collider~\cite{MuonCollider:2022xlm,Forslund:2022xjq} all envisage running as dedicated Higgs factories, collecting samples of order $10^{6}$ Higgs bosons in a clean $e^+e^-$ or $\mu^+\mu^-$ environment. 
In these programs, the hadronic decays $H\to gg$ and $H\to q\bar q$ play a uniquely important role: together they account for roughly two-thirds of the total Higgs width and therefore drive the precision with which the Yukawa couplings and the effective coupling between Higgs and gluons can be extracted. Meeting the precision for these measurements places stringent demands on the theoretical description of hadronic Higgs decays.

Theoretical progress on Higgs physics over the last decade has been remarkable. On the production side, the dominant gluon-fusion channel has been pushed to
N$^3$LO in the heavy-top limit~\cite{Anastasiou:2014vaa,Anastasiou:2015vya,%
Anastasiou:2016cez,Mistlberger:2018etf} and partially known at N$^4$LO~\cite{Mistlberger:2025ksa}. The fully differential predictions are also available to the same order, see Refs.~\cite{Cieri:2018oms,Dulat:2018bfe,Chen:2021isd} for examples. 
On the decay side, the inclusive partial widths for $H\to gg$ and $H\to q\bar q$ have been computed to N$^4$LO in QCD~\cite{Baikov:2005rw,Davies:2017xsp,Herzog:2017dtz}, and fully differential predictions are available  through N$^3$LO~\cite{Anastasiou:2011qx,DelDuca:2015zqa,%
Mondini:2019gid,Chen:2023fba,Fox:2025cuz}.
These advances are mirrored by the applications of event shapes observables in hadronic Higgs decays, see Refs.~\cite{Coloretti:2022jcl,Gehrmann-DeRidder:2023uld,Fox:2025qmp,Fox:2025txz,Fox:2025vxj,Gao:2019mlt,Ju:2023dfa,Luo:2019nig,Gao:2020vyx,Yang:2024gcn} for some examples.

In this paper, we study one of the event shape observables called heavy jet mass (HJM). HJM is introduced based on another observable called thrust. Given an $e^+e^-$ collision, we find a unit three-vector $\hat{n}$ that maximizes the following quantity 
\begin{equation}
T = \max_{\hat n}\frac{\sum_i |\vec p_i\cdot \hat n|}{\sum_i |\vec p_i|}\,,
\label{eq:thrustdef}
\end{equation}
with $\vec p_i$ the three-momentum of $i$-th final-state particles,
and we often define $\tau=1-T$ as thrust and the vector $\vec{n}$ as thrust axis. The plane orthogonal to the thrust axis divides the final-state particles into two hemispheres, $\mathcal{H}_{L,R}$ and one can calculate their invariant masses. HJM is defined to be the larger invariant mass, normalized to the center-of-mass energy $Q$,
\begin{equation}
\rho= \max_{i\in\{L,R\}}\frac{M_i^2}{Q^2}\,, \quad M_i^2\,=\, \Big(\sum_{k\in \mathcal{H}_i} p_k\Big)^{2}\,.
\end{equation}
Both thrust and HJM are well studied for $e^+e^-\to \gamma^\star/Z\to q\bar{q}$, the dominant process at the Large Electron Positron (LEP) collider. The fixed-order predictions for these two observables have been pushed to next-to-leading order (NLO) with Catani-Seymour dipole subtraction~\cite{Catani:1996jh,Catani:1996vz} and to NNLO with \textsc{Antenna} subtraction~\cite{Gehrmann-DeRidder:2005btv,Gehrmann-DeRidder:2014hxk,Aveleira:2025svg} and \textsc{Colorful} subtraction~\cite{Somogyi:2006da,Somogyi:2006db,Aglietti:2008fe}. 
Similar to many QCD observables, when $\tau,\rho\to 0$ there are large logarithmic enhancements at each order of the perturbation theory. These logarithms originate from soft emissions and collinear splittings of a dijet configuration (two back-to-back jets in the final state) and take the form $\sum_{n=1}^{\infty}\sum_{m=0}^{2n-1}c_{nm}\alpha_s^n \left[\ln^m x/x\right]_{+}$ with $x=\tau,\rho$ and $c_{nm}$ constant coefficients. To ensure the perturbation theory is not spoiled by these logs in the dijet limit, a resummation to all orders is required. In Refs.~\cite{Schwartz:2007ib,Becher:2008cf,Chien:2010kc,Abbate:2010xh,Benitez:2024nav}, this limit has been resummed to N${}^3$LL${}^\prime$ order within Soft-Collinear Effective Theory (SCET) framework~\cite{Bauer:2000yr,Bauer:2000ew,Bauer:2001yt,Bauer:2001ct,Beneke:2002ph}. Other methods have also been applied to thrust resummation~\cite{Catani:1992ua,Banfi:2001bz,Banfi:2014sua}. The effects of different resummation methods are shown to be negligible in Ref.~\cite{Almeida:2014uva} and recently revisited in Refs.~\cite{Aglietti:2025jdj,Buonocore:2026kgu}.

Among the Higgs decay channels, the hadronic ones, with the final-state hadrons initiated by gluons or quarks, are particularly interesting in QCD. (for a review, see Ref.~\cite{Spira:2016ztx})
We will refer to them as $Hgg$ and $Hq\bar q$ decays in this paper. To study them, we use the Higgs effective field theory (HEFT)~\cite{PhysRevLett.39.1304,Shifman:1978zn,Inami:1982xt} with top quarks integrated out. The HEFT Lagrangian is written as,
\begin{equation}
    \mathcal{L}_{\text{eff}}=-\frac{1}{4}\lambda(\mu) \mathcal{O}_g+\sum_{q} \frac{y_q(\mu)}{\sqrt{2}} \mathcal{O}_q\,,\qquad \mathcal{O}_g=H \text{Tr}\left(G^{\mu\nu} G_{\mu\nu} \right),\quad \mathcal{O}_q=H \bar \psi_q \psi_q\,.
\end{equation}
Here $ G^{\mu\nu}$ is the gluon field strength tensor, and $H$ and $\psi_q$ stand for Higgs and quark fields respectively. $\lambda(\mu)$ is the Wilson coefficient with respect to the effective $\mathcal{O}_g$ operator and $y_q(\mu)$ is simply the Yukawa coupling. We take the massless quark limit (except for the top quark) while keeping the Yukawa coupling non-zero. Following Ref.~\cite{Gao:2019mlt}, in the massless quark limit, the chiral symmetry guarantees that there is no interference between $\mathcal{O}_g$ and $\mathcal{O}_q$ operators. This allows one to calculate the matrix elements and observables separately in $Hgg$ and $Hq\bar q$ decays. 
The thrust and heavy jet mass in hadronic Higgs decays have been computed to NNLO with \textsc{Antenna} subtraction~\cite{Coloretti:2022jcl,Gehrmann-DeRidder:2023uld,Fox:2025qmp}. In the dijet limit, thrust has been resummed to N${}^3$LL${}^\prime$ or approximate N${}^4$LL using SCET~\cite{Ju:2023dfa}; both thrust and HJM are resummed to NNLL using \textsc{Ares} method and matched to NNLO~\cite{Fox:2025txz,Fox:2025vxj}. 
In this work, one of the main results is to obtain the HJM at N${}^3$LL${}^\prime$+NNLO accuracy.

Beyond the dijet limit, the HJM distribution exhibits a second, qualitatively different source of large logarithms. At leading order in $e^+e^-\to q\bar q$ (or $H\to g g$, $H\to q\bar{q}$), the HJM value is restricted to $\rho\le 1/3$ by three-parton final states,
with the boundary $\rho=1/3$ corresponding to a symmetric three-jet configuration. Real emissions at higher orders progressively populate the region $\rho>1/3$, so that the upper endpoint of the support of $d\sigma/d\rho$ grows order by order in perturbation theory. 
As a result, the differential distribution develops logarithms of the distance to the boundary on both sides of $\rho=1/3$, schematically
\begin{align}
\label{eq:sh_log_form}
\frac{d\sigma}{d\rho}\Bigg|_{\rho\to 1/3}&=\theta\left(\frac{1}{3}-\rho\right)\left(\frac{1}{3}-\rho\right)\sum_{n=1}^{\infty}\sum_{m=0}^{2n-2}\alpha_s^{n}\,d_{nm}^{-}\,
\ln^{m}\left(\frac{1}{3}-\rho\right)\nn\\
&+\theta\left(\rho-\frac{1}{3}\right)\left(\rho-\frac{1}{3}\right)\sum_{n=1}^{\infty}\sum_{m=0}^{2n-2}\alpha_s^{n}\,d_{nm}^{+}\,
\ln^{m}\left(\rho-\frac{1}{3}\right)\,,
\end{align}
with constant coefficients $d_{nm}^{\pm}$. Because the logarithms are multiplied by a linear factor $\pm(1/3-\rho)$, arising from the phase space, the distribution itself remains finite. Instead its derivatives diverge and if one takes the second derivative of the distribution, the logarithmic structure is similar to the dijet limit.
This is the Sudakov shoulder first identified by Catani and Webber in 1997~\cite{Catani:1997xc}. 
Physically, the logarithms arise from soft and collinear radiation off the symmetric three-jet configuration sitting precisely at the kinematic edge, in close analogy with the soft-collinear origin of the dijet logarithms. Consequently, the fixed-order prediction is unreliable in a neighborhood of $\rho=1/3$, and a trustworthy description of the shoulder region requires resumming these logarithms to all orders, which can again be organized within SCET. For $e^+e^-\to q\bar{q}$, the shoulder logs are resummed to NNLL accuracy~\cite{Bhattacharya:2023qet}. In this paper, we generalize the framework to hadronic Higgs decays $Hgg$ and $Hq\bar q$.

The HJM distribution in hadronic Higgs decays opens up several
phenomenological opportunities at a future Higgs factory. First, it offers a theoretically clean and largely independent handle on the strong coupling constant. Event shapes have long served as a classic testing ground for $\alpha_s$ extractions at $e^+e^-$ colliders. These determinations are, however, the subject of a long-standing tension: the values obtained from thrust, $C$-parameter and HJM are lower than the world average~\cite{Salam:2001bd,MovillaFernandez:2001ed,Gardi:2002bg,Dinsdale:2004zw,Becher:2008cf,Dissertori:2009ik,Abbate:2010xh,Chien:2010kc,Alioli:2012fc,Hoang:2014wka,Hoang:2015hka,Banfi:2023mes,Benitez:2024nav,Benitez:2025vsp,dEnterria:2022hzv,Huston:2023ofk}. The origin of this discrepancy remains an active question. Several recent studies have
revisited the problem from complementary angles, reassessing trijet power corrections~\cite{Luisoni:2020efy,Caola:2021kzt,Caola:2022vea,Nason:2023asn,Nason:2025qbx}, hadron-mass and recoil effects~\cite{Salam:2001bd,Nason:2023asn}, and the impact of the resummation scheme~\cite{Aglietti:2025jdj,Buonocore:2026kgu}. In this context, an independent determination from the gluon-initiated $Hgg$ channel is particularly attractive. It allows probing QCD dynamics in a genuinely complementary regime, with a different sensitivity to non-perturbative corrections, and the high-accuracy predictions presented in this work are a prerequisite for exploiting it.
Secondly, the HJM distribution also provides a way to measure $Hgg$ effective coupling $\lambda(\mu)$ and Yukawa coupling $y_q$ from future collider data. In a realistic analysis, the full HJM distribution will be the sum of both $Hgg$ and $Hq\bar{q}$ channels.  
Since $\Gamma(H\to gg)$ scales with the squared effective coupling $\lambda^2(\mu)$ and $\Gamma(H\to q\bar{q})$ with the squared bottom Yukawa $y_q^2$, a simultaneous fit measures the relative size of these widths and therefore constrains the ratio $\lambda(\mu)/y_q$ directly from the distribution~\cite{Gao:2016jcm}.

This paper is organized as follows. In Sec.~\ref{sec:dijet}, we revisit the dijet factorization theorem and present the dijet resummed distributions, matched to the fixed-order data with profile scales. 
In Sec.~\ref{sec:nlo}, we calculate NLO thrust and HJM in the symmetric trijet limit, $\tau,\rho\sim 1/3$, and obtain the analytic expressions at leading power.
In Sec.~\ref{sec:shoulder}, we review the shoulder factorization theorem and derive the shoulder resummed distributions with shoulder profile scales. 
In Sec.~\ref{sec:numerical_result}, we report the final results, joint dijet and shoulder resummation for HJM in both $Hgg$ and $Hq\bar{q}$ processes.
We conclude in Sec.~\ref{sec:conclusion}.

\section{Dijet resummation}
\label{sec:dijet}

In this section, we review the kinematics for HJM in the dijet limit ($\rho\to 0$) and the factorization theorem in SCET. Using the fixed-order ingredients computed in the literature, we present the N${}^3$LL${}^\prime_\text{dij}+$NNLO resummation for both $Hgg$ and $Hq\bar{q}$ processes.

\subsection{Dijet kinematics and factorization}

We start by reviewing the dijet kinematics for thrust and HJM. In the dijet limit $\tau,\rho\to 0$, the final state collapses onto two narrow, almost back-to-back jets of small invariant mass, accompanied by soft wide-angle radiation.
Recalling the definition given in the introduction, the thrust axis $\hat n$ divides the final-state particles into the two hemispheres $\mathcal{H}_{L}$ and $\mathcal{H}_{R}$, whose total four-momenta we denote by $P_{L}^\mu$ and $P_{R}^\mu$. The hemisphere masses are then $M_{L,R}^{2}=P_{L,R}^{2}$ and HJM is the larger of the two normalized to $Q^{2}$. In the same limit thrust is small as well and, up to power corrections, reduces to the sum of the two hemisphere masses, $Q^{2}\tau\simeq M_{L}^{2}+M_{R}^{2}$, so that both observables are controlled by the same back-to-back dijet configuration and share the kinematic analysis below. From the perspective of SCET, we identify the following hierarchy of scales,
\begin{equation}
    Q^{2}\;\gg\;  Q^{2}\rho \;\gg\; Q^{2}\rho^{2}\,,
    \label{eq:dijet_hierarchy}
\end{equation}
associated respectively with the hard production of the two back-to-back partons, the collinear dynamics inside each jet, and the soft radiation exchanged between them. To organize this hierarchy, we introduce two light-like reference vectors $n^\mu$ and $\bar n^\mu$ aligned with the jet directions, satisfying $n^{2}=\bar n^{2}=0$ and $n\cdot\bar n=2$, and decompose any momentum as
\begin{equation}
    p^{\mu}=(n\cdot p)\,\frac{\bar n^{\mu}}{2}+(\bar n\cdot p)\,\frac{n^{\mu}}{2}+p_{\perp}^{\mu}\,.
\end{equation}
This allows us to write the momentum as $p^\mu=(n\cdot p,\,\bar n\cdot p,\,p_{\perp})$. Introducing the power-counting parameter $\lambda\equiv\sqrt{\rho}$, the leading-power modes are the collinear, anti-collinear, and soft fields, with momenta scaling as
\begin{equation}
    p_{c}\sim Q\,(\lambda^{2},1,\lambda)\,,\qquad
    p_{\bar c}\sim Q\,(1,\lambda^{2},\lambda)\,,\qquad
    p_{s}\sim Q\,(\lambda^{2},\lambda^{2},\lambda^{2})\,.
\end{equation}
The collinear and anti-collinear modes describe the energetic partons initiating the two jets together with their collinear radiation, generating the hemisphere masses $M_{L,R}^{2}\sim Q^{2}\lambda^{2}$, while the soft mode describes the wide-angle radiation, which shifts each hemisphere mass by an amount $Q\,k_{L,R}\sim Q^{2}\lambda^{2}$.
Following Ref.~\cite{Chien:2010kc}, both thrust and HJM can be written in terms of the projection of the double differential hemisphere mass distribution, namely
\begin{align}
    \frac{d\sigma}{d\tau}&=Q^{2}\int dM_{L}^{2}\,dM_{R}^{2}\,\frac{d^{2}\sigma}{dM_{L}^{2}\,dM_{R}^{2}}\,\delta\!\left(Q^{2}\tau-M_{L}^{2}-M_{R}^{2}\right)\,,\\
    \frac{d\sigma}{d\rho}&=Q^{2}\int dM_{L}^{2}\,dM_{R}^{2}\,\frac{d^{2}\sigma}{dM_{L}^{2}\,dM_{R}^{2}}\Big[\delta\!\left(Q^{2}\rho-M_{L}^{2}\right)\theta\!\left(M_{L}^{2}-M_{R}^{2}\right)+(L\leftrightarrow R)\Big]\,.
\end{align}
Here thrust integrates over the sum of the two hemisphere masses, whereas HJM selects the larger one.

In SCET, the factorization theorem for the double differential distribution is well understood. In the dijet limit, it factorizes into hard, jet, and soft functions~\cite{Schwartz:2007ib,Becher:2008cf,Chien:2010kc},
\begin{multline}
    \frac{1}{\sigma_{0}}\frac{d^{2}\sigma_{\text{dij}}}{dM_{L}^{2}\,dM_{R}^{2}}
    =H^{\text{dij}}_{i}(Q^{2},\mu)\int dp_{L}^{2}\,dp_{R}^{2}\,dk_{L}\,dk_{R}\;
    J_{i}\!\left(p_{L}^{2},\mu\right)
    J_{i}\!\left(p_{R}^{2},\mu\right)
    S^{\text{dij}}_{i}\!\left(k_{L},k_{R},\mu\right)\,\\
    \times\delta\!\left(M_{L}^{2}-p_{L}^{2}-Q k_{L}\right)
    \delta\!\left(M_{R}^{2}-p_{R}^{2}-Q k_{R}\right)\,,
    \label{eq:hemifact}
\end{multline}
where $i=q,g$ labels the partonic channel ($q$ for $Hq\bar{q}$, $g$ for $Hgg$). 
$\sigma_{0}$ is the Born total cross-section for the given process. 
The hard function $H^{\text{dij}}_{i}$ encodes the virtual corrections at the hard scale $Q$, obtained by matching the HEFT current onto the two-jet SCET operator. $J_{i}$ is the inclusive jet function that depends on invariant mass and resums collinear radiation. 
The hemisphere soft function $S^{\text{dij}}_{i}$ describes the wide-angle soft radiation, and $k_{L,R}$ are the soft momenta projected onto the two hemispheres.
The two delta functions implement the leading-power decomposition of each hemisphere mass into collinear and soft pieces, $M_{L,R}^{2}=p_{L,R}^{2}+Q k_{L,R}$; carrying out the $p_{L,R}^{2}$ integrals against them yields a more compact convolution form with $J_{i}(M_{L,R}^{2}-Q k_{L,R},\mu)$. 
The factorization theorem Eq.~\eqref{eq:hemifact} is identical in form to the one for $\gamma^{\star}\to q\bar q$; the dependence on decay channel enters only through the three ingredients $H^{\text{dij}}_{i}$, $J_{i}$, and $S^{\text{dij}}_{i}$, which we now discuss in turn.

\subsubsection{Dijet hard functions}
The hard function arises from two successive matching steps. Integrating the top quark out of full QCD first generates the HEFT operators $\mathcal{O}_{g}$ and $\mathcal{O}_{q}$, whose Wilson coefficients are the effective couplings $\lambda(\mu)$ and $y_{q}(\mu)$ introduced above. The $Hgg$ coupling carries the top-quark Wilson coefficient $C_{t}(\mu)\equiv C(m_{t},\mu)$ produced in this step, $\lambda(\mu)=-\alpha_{s}(\mu)\,C_{t}(\mu)/(3\pi v)$ with $v$ the Higgs vacuum expectation value~\cite{Coloretti:2022jcl,Gehrmann-DeRidder:2023uld}. The HEFT operators are then matched onto the two-jet SCET operators built from gauge-invariant collinear fields along the back-to-back directions $n$ and $\bar n$,
\begin{align}
\mathcal{O}_{g}&\;\to\;C_{g}(Q^{2},\mu)\,H\,\mathcal{B}_{n\perp}^{a,\mu}\,\mathcal{B}_{\bar n\perp,\mu}^{a}\,,\\
\mathcal{O}_{q}&\;\to\;C_{q}(Q^{2},\mu)\,H\,\bar\chi_{n}\,\chi_{\bar n}\,,
\end{align}
where $\chi_n\equiv W_n^\dagger \xi_n$ is  the collinear quark and $\mathcal{B}^{\mu}_{n\perp}\equiv \frac{1}{g}\left[\frac{1}{\bar n\cdot \mathcal{P}}W_n^\dagger [i\bar n\cdot D_n, iD_{n\perp}^\mu] W_n\right]$ is the collinear gluon. They are both dressed by collinear Wilson lines $W_n$ (and analogously along $\bar n$) and form a complete set of collinear gauge-invariant building blocks \cite{Marcantonini:2008qn} in SCET.
The SCET matching coefficients $C_{g}$ and $C_{q}$ are the gluon and quark scalar form factors, and the dijet hard functions are their moduli squared,
\begin{equation}
H^{\text{dij}}_{g}(Q^{2},\mu)=\big|C_{g}(Q^{2},\mu)\big|^{2}\,,\qquad
H^{\text{dij}}_{q}(Q^{2},\mu)=\big|C_{q}(Q^{2},\mu)\big|^{2}\,,
\end{equation}
which satisfy the renormalization-group equation~\cite{Becher:2006mr}
\begin{equation}
\frac{d}{d\ln\mu}\,H^{\text{dij}}_{i}(Q^{2},\mu)
=\left[\,2\,\Gamma_{i}^{\text{cusp}}(\alpha_{s})\,\ln\frac{Q^{2}}{\mu^{2}}+2\,\gamma^{\text{dij}}_{H,i}(\alpha_{s})\,\right]H^{\text{dij}}_{i}(Q^{2},\mu)\,.
\label{eq:hard_rge}
\end{equation}
$\Gamma_{i}^{\text{cusp}}$ is the cusp anomalous dimension~\cite{Korchemsky:1987wg,Moch:2004pa} (with $i=g$ for $Hgg$ and $i=q$ for $Hq\bar q$) and $\gamma^{\text{dij}}_{H,i}$ is the corresponding hard non-cusp anomalous dimension~\cite{Moch:2005id, Moch:2005tm, Idilbi:2005ni, Idilbi:2006dg, Becher:2006mr,Becher:2007ty, Ahrens:2009cxz}.
Up to three loops, $\Gamma_{q}^{\text{cusp}}$ and $\Gamma_{g}^{\text{cusp}}$ are related by Casimir scaling $C_F\leftrightarrow C_A$.
The solution of Eq.~\eqref{eq:hard_rge} resums the logarithms of $Q^{2}/\mu^{2}$ by evolving the hard function from the hard matching scale $\mu_{h}\sim Q$, where it is free of large logarithms, down to the common factorization scale $\mu$,
\begin{equation}
H^{\text{dij}}_{i}(Q^{2},\mu)=H^{\text{dij}}_{i}(Q^{2},\mu_{h})\,
\exp\!\left[\,4S_{i}(\mu_{h},\mu)-2A^{\text{dij}}_{H,i}(\mu_{h},\mu)\,\right]
\left(\frac{Q^{2}}{\mu_{h}^{2}}\right)^{-2A_{\Gamma,i}(\mu_{h},\mu)}\,,
\label{eq:hard_rge_sol}
\end{equation}
where the Sudakov kernels $S_{i}$,  $A_{\Gamma,i}$ and $A^{\text{dij}}_{H,i}$ are built from the cusp $\Gamma_{i}^{\text{cusp}}$ and the hard non-cusp $\gamma^\text{dij}_{H,i}$ anomalous dimensions. Explicitly, they are defined as
\begin{align}
S_{i}(\nu,\mu)&=-\int_{\alpha_{s}(\nu)}^{\alpha_{s}(\mu)}d\alpha\,\frac{\Gamma_{i}^{\text{cusp}}(\alpha)}{\beta(\alpha)}\int_{\alpha_{s}(\nu)}^{\alpha}\frac{d\alpha'}{\beta(\alpha')}\,,\nn\\
A_{\Gamma,i}(\nu,\mu)&=-\int_{\alpha_{s}(\nu)}^{\alpha_{s}(\mu)}d\alpha\,\frac{\Gamma_{i}^{\text{cusp}}(\alpha)}{\beta(\alpha)}\,,\qquad
A^{\text{dij}}_{H,i}(\nu,\mu)=-\int_{\alpha_{s}(\nu)}^{\alpha_{s}(\mu)}d\alpha\,\frac{\gamma^{\text{dij}}_{H,i}(\alpha)}{\beta(\alpha)}\,,
\label{eq:SAfunctions}
\end{align}
where $\beta(\alpha)\equiv d\alpha_{s}/d\ln\mu$ is the QCD beta function~\cite{Tarasov:1980au,Larin:1993tp,vanRitbergen:1997va,Czakon:2004bu,Herzog:2017ohr}. The kernels for jet function and soft functions have the same structure, with $\gamma^{\text{dij}}_{H,i}$ replaced by the jet and soft non-cusp anomalous dimensions, respectively. We have collected the analytic expressions for these kernels in the appendix.

We emphasize that the top-quark Wilson coefficient $C_{t}(\mu)$ does not enter the dijet hard functions $H^{\text{dij}}_{i}$. It appears only inside the overall coupling $\lambda(\mu)$, as a multiplicative factor $|C_{t}(\mu)|^{2}$ common to the differential rate $d\Gamma_{Hgg}/d\rho$ and to the total width $\Gamma_{Hgg}$. 
Since we will present the HJM distribution normalized to the total decay rate, we do not need to consider $|C_{t}(\mu)|^{2}$.
The same treatment has been used in Ref.~\cite{Lee:2026zyl}.

\subsubsection{Jet functions} 
For $Hgg$ the jets are gluon jets, and they are described by the inclusive gluon jet function. For $Hq\bar q$ the two jets are quark jets. Both jet functions are known to three loops~\cite{Bauer:2003pi,Bosch:2004th,Becher:2006qw,Becher:2009th,Becher:2010pd,Bruser:2018rad,Banerjee:2018ozf}. It turns out more convenient to study the jet functions in Laplace space via
\begin{equation}
\tilde{j}_i(\nu, \mu)\equiv \int_0^\infty dp^2 e^{-\nu p^2} J_i(p^2,\mu)
\end{equation}
and the RGE equation reads~\cite{Becher:2008cf},
\begin{equation}
\frac{d}{d\ln\mu}\,\tilde{j}\left(\ln\frac{\nu}{\mu^2},\mu\right)
= \left[-2\,\Gamma_{i}^{\text{cusp}}(\alpha_s)\,\ln\frac{\nu}{\mu^2} - 2\,\gamma_{J,i}(\alpha_s)\right]
\tilde{j}\left(\ln\frac{\nu}{\mu^2},\mu\right)\,.
\label{eq:jet_rge}
\end{equation}
Here $\gamma_{J,i}$ is the jet non-cusp anomalous dimension. Solving the RGE, we obtain the solution in Laplace space and then we can perform the inverse Laplace transformation. This leads to
\begin{equation}
    J_{i}(p^{2},\mu)=\exp\!\left[-4S_{i}(\mu_{j},\mu)+2A_{J,i}(\mu_{j},\mu)\right]\widetilde{j}_{i}(\partial_{\eta_{j}},\mu_{j})\,\frac{1}{p^{2}}\left(\frac{p^{2}}{\mu_{j}^{2}}\right)^{\eta_{j}}\frac{e^{-\gamma_{E}\eta_{j}}}{\Gamma(\eta_{j})}\,,
    \label{eq:jet_rge_sol}
\end{equation}
where the jet boundary $\widetilde{j}_{i}(L,\mu)$ is the Laplace transform of the perturbative jet function. It is a polynomial in derivative operator $\partial_{\eta_j}$ at any fixed order in $\alpha_{s}$, with $\eta_{j}=2A_{\Gamma,i}(\mu_{j},\mu)$.

\subsubsection{Dijet soft functions}
The hemisphere dijet soft function is defined in terms of two soft Wilson lines along the jet directions $n$ and $\bar{n}$. In hadronic Higgs decays, we will have both quark case and gluon case. The plane perpendicular to the jet direction divides the soft emissions into two hemispheres and we use $k_L$ and $k_R$ to denote the momentum deposited in each hemisphere and projected onto the corresponding light-cone direction. Regarding the observables, thrust only depends on the sum of soft momentum $k=k_L+k_R$, while HJM will have non-trivial dependence on both.
Explicitly, the thrust soft function is defined as
\begin{equation}
S^{\text{dij}}_{T,i}(k,\mu)=\int dk_{L}\,dk_{R}\,\delta(k-k_{L}-k_{R})\,S^{\text{dij}}_{i}(k_{L},k_{R},\mu)\,.
\label{eq:soft_thrust_def}
\end{equation}
and its RGE in Laplace space is
\begin{equation}
\frac{d}{d\ln\mu}\,\widetilde{s}^{\text{dij}}_{T,i}\left(\ln\frac{\nu}{\mu},\mu\right)
= \left[4\,\Gamma_i^{\mathrm{cusp}}(\alpha_s)\,\ln\frac{\nu}{\mu} - 2\,\gamma^{\text{dij}}_{S,i}(\alpha_s)\right]
\widetilde{s}^{\text{dij}}_{T,i}\left(\ln\frac{\nu}{\mu},\mu\right).
\end{equation}
Similar to jet function, we can express the solution in terms of RG kernels,
\begin{equation}
    S^{\text{dij}}_{T,i}(k,\mu)=\exp\!\left[4S_{i}(\mu_{s},\mu)+2A^{\text{dij}}_{S,i}(\mu_{s},\mu)\right]\widetilde{s}^{\text{dij}}_{T,i}(\partial_{\eta_{s}},\mu_{s})\,\frac{1}{k}\left(\frac{k}{\mu_{s}}\right)^{\eta_{s}}\frac{e^{-\gamma_{E}\eta_{s}}}{\Gamma(\eta_{s})}\,,
    \label{eq:soft_rge_sol}
\end{equation}
with $\eta_{s}=-4A_{\Gamma,i}(\mu_{s},\mu)$, and $\widetilde{s}^{\text{dij}}_{T,i}(L,\mu)$ the Laplace transform of the perturbative thrust soft function, known up to three loops~\cite{Monni:2011gb,Kelley:2011ng,Hornig:2011iu,Baranowski:2020xlp,Baranowski:2024vxg}.

On the contrary, HJM requires the full dijet soft function. In Laplace space, we assign two Laplace variables $\nu_L,\nu_R$ and define the short-hand notation for logs $L_{L,R}=-\ln(\mu\,\nu_{L,R}\,e^{\gamma_{E}})$. Following Ref.~\cite{Chien:2010kc}, we decompose the HJM soft function into three components,
\begin{equation}
    \tilde s^\text{dij}_{i}(L_{L},L_{R},\mu)=\tilde s^\text{dij}_{\mu,i}(L_{L},\mu)\,\tilde s^\text{dij}_{\mu,i}(L_{R},\mu)\,\tilde s_{\text{NGL},i}(L_{L}-L_{R})\,,
    \label{eq:softfact}
\end{equation}
where the $\mu$-dependent part $\tilde s^\text{dij}_{\mu,i}$ satisfies the same RGE as the thrust soft function with adjusted anomalous dimensions, $\gamma^\text{dij}_{S,i}\to\frac{1}{2}\gamma^\text{dij}_{S,i}$ and adjusted arguments, $L_{L,R}\to 2 L_{L,R}$. The non-global logarithm $\tilde s_{\text{NGL},i}$ encodes the soft interference between the two hemispheres and is known fully at two loops~\cite{Kelley:2011ng,Hornig:2011iu} and partially at three loops~\cite{Baranowski:2024vxg}. Below we summarize the available perturbative results for this soft function.

First of all, it is convenient to write the NGLs in exponentiated form. Its natural argument is the coupling evaluated at the scale $\mu_{\text{NGL}}=e^{-\gamma_{E}}(\nu_{L}\nu_{R})^{-1/2}$, at which $L_{L}+L_{R}=0$. Defining $L\equiv L_{L}-L_{R}=\ln(\nu_{R}/\nu_{L})$, we have
\begin{equation}
    \tilde s_{\text{NGL},i}(L)=\exp\!\left[\left(\frac{\alpha_{s}(\mu_{\text{NGL}})}{4\pi}\right)\,c^{S}_{1,i}+\left(\frac{\alpha_{s}(\mu_{\text{NGL}})}{4\pi}\right)^{2} s^{\text{NGL}}_{2,i}(L)+\left(\frac{\alpha_{s}(\mu_{\text{NGL}})}{4\pi}\right)^{3} s^{\text{NGL}}_{3,i}(L)+\cdots\right]\,.
    \label{eq:sngl_exp}
\end{equation}
Here $c^{S}_{1,i}=-C_F \pi^2$ is the one-loop soft constant and the first non-global logarithm appears at two loops through $s^{\text{NGL}}_{2,i}(L)$. In the quark case, the analytic expression at two loops can be found in Refs.~\cite{Kelley:2011ng,Hornig:2011iu} and below we use the form from Ref.~\cite{Hornig:2011iu}.
Writing $s^{\text{NGL}}_{2,q}(L)=2\,t_{2}(b)$ with $b\equiv e^{L}=\nu_{R}/\nu_{L}$, the analytic expression reads,
\begin{align}
    t_{2}(b)=&-\frac{2\pi^{2}}{3}C_{F}C_{A}\ln^{2}b
    +2\ln\!\left(\frac{b+b^{-1}}{2}\right)\!\left[C_{F}C_{A}\,\frac{11\pi^{2}-3-18\zeta_{3}}{9}+C_{F}T_{F}n_{f}\,\frac{6-4\pi^{2}}{9}\right]\nonumber\\
    &+2C_{F}T_{F}n_{f}\big[f_{Q}(b)+f_{Q}(b^{-1})-2f_{Q}(1)\big]
    +2C_{F}C_{A}\big[f_{N}(b)+f_{N}(b^{-1})-2f_{N}(1)\big]+s_{2}\,,
    \label{eq:t2}
\end{align}
where the abelian and non-abelian functions are
\begin{align}
    f_{Q}(b)=&\;\frac{2\ln b}{3(b-1)}-\frac{b\ln^{2}b}{3(b-1)^{2}}-\frac{3-2\pi^{2}}{9}\ln\!\big(b+b^{-1}\big)+\frac{2}{3}\ln^{2}b\,\ln(1-b)+\frac{8}{3}\ln b\,\mathrm{Li}_{2}(b)-4\,\mathrm{Li}_{3}(b)\,,\nonumber\\
    f_{N}(b)=&\;-\frac{\pi^{4}}{36}-\frac{\ln b}{3(b-1)}+\frac{b\ln^{2}b}{6(b-1)^{2}}+\frac{3-11\pi^{2}+18\zeta_{3}}{18}\ln\!\big(b+b^{-1}\big)-\frac{11}{6}\ln^{2}b\,\ln(1-b)\nonumber\\
    &+\frac{\ln^{4}b}{24}-\frac{\pi^{2}}{3}\mathrm{Li}_{2}(1-b)+\mathrm{Li}_{2}^{2}(1-b)-\frac{22}{3}\ln b\,\mathrm{Li}_{2}(b)+2\ln b\,\mathrm{Li}_{3}(1-b)+11\,\mathrm{Li}_{3}(b)\,.
    \label{eq:fQfN}
\end{align}
In Eq.~\eqref{eq:t2}, $f_{Q}(1)$ and $f_{N}(1)$ stand for the $b\to1$ limits of the corresponding functions. The constant $s_2$ is 
\begin{equation}
s_2=C_F \left[\left(\frac{40}{81} + \frac{77\pi^2}{27} - \frac{52}{9}\zeta_3\right) T_F n_f + \left(\frac{7\pi^4}{15} - \frac{1070}{81} - \frac{871\pi^2}{108} + \frac{143}{9}\zeta_3\right) C_A\right]
\end{equation}
The complete form of three-loop soft function is not known, and thus we model it by the even polynomial truncated at $\mathcal{O}(L^{4})$,
\begin{equation}
    s^{\text{NGL}}_{3,i}(L)=2\big(s_{3}+s_{3,2}\,L^{2}+s_{3,4}\,L^{4}\big)\,.
    \label{eq:sngl3}
\end{equation}
$s_3$ is the same as thrust soft function,
\begin{multline}
	s_3 = C_F \left(34.47129\, C_F T_F n_f + 419.8619\, C_A T_F n_f - 376.88789\, C_A^2\right)\\
+ C_F n_f^2 T_F^2 \left(\frac{132704}{6561} - \frac{25952}{243}\zeta_3 - \frac{200\pi^2}{243} + \frac{164\pi^4}{1215}\right) \,,
\end{multline}
and we set $s_{3,2}=s_{3,4}=0$. But it is straightforward to update the calculation once they are available.
For gluon soft function with $i=g$, we can directly apply the Casimir scaling, because the wide-angle radiation only resolves the total color charge of the back-to-back dipole.
The same replacement applies to the soft anomalous dimension to the orders relevant for N${}^3$LL${}^\prime$ resummation.

\subsubsection{Normalization} 
Different from $e^+e^-\to q\bar{q}$ process, the normalization in hadronic Higgs decays, which is usually the Born total decay rate, involves an additional coupling. We need to specify the scale in this coupling. In $Hgg$ process, the Born total decay rate is
\begin{equation}
	\Gamma_{Hgg}(\mu)=\frac{\lambda(\mu)^2 m_H^3 N_A}{64 \pi}\,.
\end{equation}
Using the $\lambda-$RGE with anomalous dimensions $\gamma_\lambda(\alpha_s(\mu))$,
\begin{equation}
	\frac{d\lambda(\mu)}{d\ln\mu}=\gamma_\lambda(\alpha_s(\mu))\lambda(\mu)\,,
\end{equation}
we can express $\lambda(\mu)$ in terms of $\lambda(\mu_h)$ via
\begin{equation}
	\lambda(\mu)=\lambda(\mu_h)\, e^{A_{\gamma_\lambda}(\mu,\mu_h)}\,.
\end{equation}
Using this solution, we can pass $\mu$ to $\mu_h$, and then pass $\mu_h$ to $Q=m_H$. Since $\mu$ is the resummation scale, the evolution from $\mu$ to $\mu_h$ should be kept in the RG kernel. On the other hand, $\mu_h$ to $Q=m_H$ evolution is similar to the hard function boundary $H_i^{\text{dij}}(Q^2,\mu_h)$, and thus needs to be expanded in $\alpha_s(\mu_h)$. Explicitly, we have
\begin{equation}\label{eq:lambda_res}
	\lambda(\mu)^2=\lambda(Q)^2\,\times \left[e^{2 A_{\gamma_\lambda}(\mu_h,Q)}\right]_{\text{expanded in }\alpha_s(\mu_h)}\,\times e^{2 A_{\gamma_\lambda}(\mu,\mu_h)}\,.
\end{equation}
Therefore, in the resummation, we will normalize the distribution to $\Gamma_{Hgg}(Q)$. One can also include the $K$ factor that takes Born decay rate to full decay rate. The rest of Eq.~\eqref{eq:lambda_res} will go into the boundary and the RG kernel, respectively.

Similarly, for $Hq\bar{q}$ process, we have the Yukawa coupling in the Born total decay rate,
\begin{equation}
	\Gamma_{Hq\bar{q}}(\mu)=\frac{y_q(\mu)^2 m_H C_A}{16 \pi}\,.
\end{equation}
The Yukawa coupling RGE is
\begin{equation}
	\frac{dy_q(\mu)}{d\ln\mu}=\gamma_y(\alpha_s(\mu))y_q(\mu)\,,
\end{equation}
which leads to 
\begin{equation}\label{eq:yukawa_res}
	y_q^2(\mu)=y_q^2(Q)\,\times \left[e^{2 A_{\gamma_y}(\mu_h,Q)}\right]_{\text{expanded in }\alpha_s(\mu_h)}\,\times e^{2 A_{\gamma_y}(\mu,\mu_h)}\,.
\end{equation}
We perform similar treatment as the $Hgg$ effective coupling $\lambda(\mu)$.

\subsection{Fixed-order expansion}
\label{sec:dijet_fixed_order}

Using the factorization theorem for thrust and HJM in the dijet limit, we can derive their singular expansion up to $\mathcal{O}(\alpha_s^3)$. For $Hgg$ HJM, we find
\begin{align}
    \frac{\rho}{\sigma_{0}}&\frac{d\sigma^{\text{sing}}_{Hgg}}{d\rho}=\left(\frac{\alpha_{s}}{4\pi}\right) \left(-8 C_A \ln \rho-\frac{22 C_A}{3}+\frac{8 n_f T_F}{3}\right)+\left(\frac{\alpha_{s}}{4\pi}\right)^2 \Bigg\{32 C_A^2 \ln^3\rho\nn\\
    &+\left(132 C_A^2-48 C_A n_f T_F\right)\ln^2\rho+ \left(-\frac{16}{3} C_A n_f T_F+\left(-98-8 \pi ^2\right) C_A^2+\frac{32}{3} n_f^2 T_F^2\right)\ln \rho\nn\\
    &+\left(\frac{1072}{9}+\frac{16 \pi ^2}{9}\right) C_A n_f T_F+\left(8 \zeta_3-\frac{1666}{9}-\frac{44 \pi ^2}{9}\right) C_A^2+8 C_F n_f T_F-\frac{160}{9} n_f^2 T_F^2\Bigg\}\nn\\
    &+\left(\frac{\alpha_{s}}{4\pi}\right)^3\Bigg\{ -64 C_A^3 \ln^5\rho+\left(\frac{640}{3} C_A^2 n_f T_F-\frac{1760 C_A^3}{3}\right)\ln^4\rho\nn\\
    &+\left[\frac{18368}{27} C_A^2 n_f T_F-\frac{4736}{27} C_A n_f^2 T_F^2+\left(\frac{128 \pi ^2}{3}-\frac{10088}{27}\right) C_A^3\right]\ln^3\rho \nn\\
    &+\left[\left(-\frac{6496}{3}-64 \pi ^2\right) C_A^2 n_f T_F+192 C_A n_f^2 T_F^2-144 C_A C_F n_f T_F\right.\nn\\
    &\left.+\left(-224 \zeta_3+\frac{33580}{9}+176 \pi ^2\right) C_A^3+\frac{256}{9} n_f^3 T_F^3\right]\ln^2\rho \nn\\
    &+\left[\left(960 \zeta_3-\frac{13352}{27}-\frac{3280 \pi ^2}{27}\right) C_A^2 n_f T_F+(592-640 \zeta_3) C_A C_F n_f T_F\right.\nn\\
    &+\left(\frac{19040}{27}+\frac{256 \pi ^2}{9}\right) C_A n_f^2 T_F^2+\left(-880 \zeta_3-\frac{38624}{27}+\frac{1448 \pi ^2}{27}-\frac{316 \pi ^4}{45}\right) C_A^3\nn\\
    &\left.+\frac{224}{3} C_F n_f^2 T_F^2-\frac{2560}{27} n_f^3 T_F^3\right]\ln \rho +\left(\frac{4576 \zeta_3}{3}+\frac{121094}{27}-\frac{17500 \pi ^2}{81}+\frac{124 \pi ^4}{27}\right) C_A^2 n_f T_F\nn\\
    &+\left(-\frac{3712 \zeta_3}{9}-\frac{10772}{9}+\frac{5872 \pi ^2}{81}\right) C_A n_f^2 T_F^2+\left(-704 \zeta_3+\frac{9878}{9}+\frac{16 \pi ^2}{3}\right) C_A C_F n_f T_F\nn\\
    &+\left(-\frac{8360 \zeta_3}{9}+\frac{80 \pi ^2 \zeta_3}{3}-224 \zeta_5-\frac{398477}{81}+\frac{14798 \pi ^2}{81}-\frac{341 \pi ^4}{27}\right) C_A^3\nn\\
    &+\left(256 \zeta_3-\frac{3400}{9}\right) C_F n_f^2 T_F^2-4 C_F^2 n_f T_F+\left(\frac{6400}{81}-\frac{512 \pi ^2}{81}\right) n_f^3 T_F^3\Bigg\}\,.
\end{align}
For $Hgg$ thrust, we find
\begin{align}
    \frac{\tau}{\sigma_{0}}&\frac{d\sigma^{\text{sing}}_{Hgg}}{d\tau}=\left(\frac{\alpha_{s}}{4\pi}\right) \left(-8 C_A \ln \tau-\frac{22 C_A}{3}+\frac{8 n_f T_F}{3}\right)+\left(\frac{\alpha_{s}}{4\pi}\right)^2 \Bigg\{ 32 C_A^2 \ln^3\tau\nn\\
    &+\left(132 C_A^2-48 C_A n_f T_F\right)\ln^2\tau+\ln \tau \left(-\frac{16}{3} C_A n_f T_F+\left(-98-\frac{40 \pi ^2}{3}\right) C_A^2+\frac{32}{3} n_f^2 T_F^2\right)\nn\\
    &+\left(\frac{1072}{9}+\frac{32 \pi ^2}{9}\right) C_A n_f T_F+\left(40 \zeta_3-\frac{1666}{9}-\frac{88 \pi ^2}{9}\right) C_A^2+8 C_F n_f T_F-\frac{160}{9} n_f^2 T_F^2 \Bigg\}\nn\\
    &+\left(\frac{\alpha_{s}}{4\pi}\right)^3 \Bigg\{ -64 C_A^3 \ln^5\tau+ \left(\frac{640}{3} C_A^2 n_f T_F-\frac{1760 C_A^3}{3}\right) \ln^4\tau \nn\\
    &+\left[\frac{18368}{27} C_A^2 n_f T_F-\frac{4736}{27} C_A n_f^2 T_F^2+\left(\frac{256 \pi ^2}{3}-\frac{10088}{27}\right) C_A^3\right]\ln^3\tau\nn\\
    &+\left[\left(-\frac{6496}{3}-\frac{416 \pi ^2}{3}\right) C_A^2 n_f T_F+192 C_A n_f^2 T_F^2-144 C_A C_F n_f T_F\right.\nn\\
    &\left.+\left(-992 \zeta_3+\frac{33580}{9}+\frac{1144 \pi ^2}{3}\right) C_A^3+\frac{256}{9} n_f^3 T_F^3\right]\ln^2\tau\nn\\
    &+\left[\left(1856 \zeta_3-\frac{13352}{27}-\frac{1792 \pi ^2}{9}\right) C_A^2 n_f T_F+(592-640 \zeta_3) C_A C_F n_f T_F\right.\nn\\
    &+\left(\frac{19040}{27}+\frac{1408 \pi ^2}{27}\right) C_A n_f^2 T_F^2+\left(-3344 \zeta_3-\frac{38624}{27}+\frac{2072 \pi ^2}{27}-\frac{124 \pi ^4}{45}\right) C_A^3\nn\\
    &\left.+\frac{224}{3} C_F n_f^2 T_F^2-\frac{2560}{27} n_f^3 T_F^3\right]\ln \tau +\left(\frac{17920 \zeta_3}{9}+\frac{121094}{27}-\frac{11048 \pi ^2}{81}-\frac{52 \pi ^4}{135}\right) C_A^2 n_f T_F\nn\\
    &+\left(-\frac{1664 \zeta_3}{3}-\frac{10772}{9}+\frac{5216 \pi ^2}{81}\right) C_A n_f^2 T_F^2+\left(-704 \zeta_3+\frac{9878}{9}+\frac{32 \pi ^2}{3}\right) C_A C_F n_f T_F\nn\\
    &+\left(-\frac{9608 \zeta_3}{9}+128 \pi ^2 \zeta_3-1376 \zeta_5-\frac{398477}{81}+\frac{5110 \pi ^2}{81}+\frac{143 \pi ^4}{135}\right) C_A^3\nn\\
    &+\left(256 \zeta_3-\frac{3400}{9}\right) C_F n_f^2 T_F^2-4 C_F^2 n_f T_F+\left(\frac{6400}{81}-\frac{640 \pi ^2}{81}\right) n_f^3 T_F^3 \Bigg\}\,.
\end{align}
For $Hq\bar{q}$ HJM, we find
\begin{align}
    \frac{\rho}{\sigma_{0}}&\frac{d\sigma^{\text{sing}}_{Hq\bar{q}}}{d\rho}=\left(\frac{\alpha_{s}}{4\pi}\right)\left(-8 C_F \ln \rho-6 C_F\right)\nn\\
    &+\left(\frac{\alpha_{s}}{4\pi}\right)^2\Bigg\{ 32 C_F^2 \ln^3\rho+ \left(44 C_A C_F-16 C_F n_f T_F+72 C_F^2\right)\ln^2\rho \nn\\
    &+\left[\left(\frac{8 \pi ^2}{3}-\frac{338}{9}\right) C_A C_F+\frac{88}{9} C_F n_f T_F+\left(-44-\frac{32 \pi ^2}{3}\right) C_F^2\right]\ln \rho +(24 \zeta_3-57) C_A C_F\nn\\
    &+20 C_F n_f T_F+\left(-16 \zeta_3-63-4 \pi ^2\right) C_F^2 \Bigg\}\nn\\
    &+\left(\frac{\alpha_{s}}{4\pi}\right)^3\Bigg\{ -64 C_F^3 \ln^5\rho+ \left(-\frac{880}{3} C_A C_F^2+\frac{320}{3} C_F^2 n_f T_F-240 C_F^3\right)\ln^4\rho\nn\\
    &+ \left[\frac{4928}{27} C_A C_F n_f T_F+\left(-\frac{464}{9}-\frac{64 \pi ^2}{3}\right) C_A C_F^2-\frac{6776}{27} C_A^2 C_F+\frac{448}{9} C_F^2 n_f T_F\right.\nn\\
    &\left.-\frac{896}{27} C_F n_f^2 T_F^2+\left(32+64 \pi ^2\right) C_F^3\right]\ln^3\rho+ \left[\left(\frac{32 \pi ^2}{3}-\frac{4096}{9}\right) C_A C_F n_f T_F\right.\nn\\
    &+\left(-288 \zeta_3+1462+\frac{280 \pi ^2}{3}\right) C_A C_F^2+\left(\frac{6394}{9}-\frac{88 \pi ^2}{3}\right) C_A^2 C_F\nn\\
    &\left.+\left(-536-\frac{128 \pi ^2}{3}\right) C_F^2 n_f T_F+\frac{544}{9} C_F n_f^2 T_F^2+\left(64 \zeta_3+648+96 \pi ^2\right) C_F^3\right]\ln^2\rho\nn\\
    &+\left[\left(64 \zeta_3+\frac{5384}{81}-\frac{1024 \pi ^2}{27}\right) C_A C_F n_f T_F+\left(-\frac{4384 \zeta_3}{3}-\frac{6100}{9}+\frac{880 \pi ^2}{27}+\frac{116 \pi ^4}{45}\right) C_A C_F^2\right.\nn\\
    &\left.+\left(176 \zeta_3-\frac{22646}{81}+\frac{680 \pi ^2}{9}-\frac{88 \pi ^4}{45}\right) C_A^2 C_F+\left(\frac{896 \zeta_3}{3}+\frac{2576}{9}-\frac{560 \pi ^2}{27}\right) C_F^2 n_f T_F\right.\nn\\
    &\left.+\left(\frac{1120}{81}+\frac{128 \pi ^2}{27}\right) C_F n_f^2 T_F^2+\left(384 \zeta_3-86-\frac{176 \pi ^2}{3}-\frac{344 \pi ^4}{45}\right) C_F^3\right]\ln \rho \nn\\
    &+\left(\frac{4672 \zeta_3}{9}+\frac{17888}{27}-\frac{10952 \pi ^2}{81}+\frac{16 \pi ^4}{5}\right) C_A C_F n_f T_F\nn\\
    &+\left(-\frac{3232 \zeta_3}{9}-80 \zeta_5-\frac{23804}{27}+\frac{12854 \pi ^2}{81}-\frac{134 \pi ^4}{15}\right) C_A^2 C_F\nn\\
    &+\left(-\frac{4808 \zeta_3}{9}+16 \pi ^2 \zeta_3-240 \zeta_5-\frac{8509}{6}+\frac{152 \pi ^2}{3}+\frac{17 \pi ^4}{15}\right) C_A C_F^2\nn\\
    &+\left(\frac{2944 \zeta_3}{9}+\frac{1768}{3}-\frac{68 \pi ^2}{3}-\frac{16 \pi ^4}{15}\right) C_F^2 n_f T_F+\left(-\frac{1024 \zeta_3}{9}-\frac{3056}{27}+\frac{2240 \pi ^2}{81}\right) C_F n_f^2 T_F^2\nn\\
    &+\left(-80 \zeta_3+\frac{32 \pi ^2 \zeta_3}{3}+96 \zeta_5-\frac{841}{2}-32 \pi ^2-\frac{18 \pi ^4}{5}\right) C_F^3\Bigg\}\,.
\end{align}
For $Hq\bar{q}$ thrust, we find
\begin{align}
    \frac{\tau}{\sigma_{0}}&\frac{d\sigma^{\text{sing}}_{Hq\bar{q}}}{d\tau}=\left(\frac{\alpha_{s}}{4\pi}\right)\left(-8 C_F \ln \tau-6 C_F\right)\nn\\
    &+\left(\frac{\alpha_{s}}{4\pi}\right)^2\Bigg\{ 32 C_F^2 \ln^3\tau+ \left(44 C_A C_F-16 C_F n_f T_F+72 C_F^2\right) \ln^2\tau \nn\\
    &+\left[\left(\frac{8 \pi ^2}{3}-\frac{338}{9}\right) C_A C_F+\frac{88}{9} C_F n_f T_F+\left(-44-16 \pi ^2\right) C_F^2\right] \ln \tau +(24 \zeta_3-57) C_A C_F\nn\\
    &+20 C_F n_f T_F+\left(16 \zeta_3-63-8 \pi ^2\right) C_F^2 \Bigg\}\nn\\
    &+\left(\frac{\alpha_{s}}{4\pi}\right)^3\Bigg\{ -64 C_F^3 \ln^5\tau+  \left(-\frac{880}{3} C_A C_F^2+\frac{320}{3} C_F^2 n_f T_F-240 C_F^3\right)\ln^4\tau\nn\\
    &+\left[\frac{4928}{27} C_A C_F n_f T_F+\left(-\frac{464}{9}-\frac{64 \pi ^2}{3}\right) C_A C_F^2-\frac{6776}{27} C_A^2 C_F+\frac{448}{9} C_F^2 n_f T_F\right.\nn\\
    &\left.-\frac{896}{27} C_F n_f^2 T_F^2+\left(32+\frac{320 \pi ^2}{3}\right) C_F^3\right]\ln^3\tau + \left[\left(\frac{32 \pi ^2}{3}-\frac{4096}{9}\right) C_A C_F n_f T_F\right.\nn\\
    &\left.+\left(-288 \zeta_3+1462+\frac{544 \pi ^2}{3}\right) C_A C_F^2+\left(\frac{6394}{9}-\frac{88 \pi ^2}{3}\right) C_A^2 C_F\right.\nn\\
    &\left.+\left(-536-\frac{224 \pi ^2}{3}\right) C_F^2 n_f T_F+\frac{544}{9} C_F n_f^2 T_F^2+\left(-704 \zeta_3+648+192 \pi ^2\right) C_F^3\right]\ln^2\tau\nn\\
    &+ \left[\left(64 \zeta_3+\frac{5384}{81}-\frac{1024 \pi ^2}{27}\right) C_A C_F n_f T_F+\left(-\frac{7552 \zeta_3}{3}-\frac{6100}{9}+\frac{788 \pi ^2}{27}+\frac{356 \pi ^4}{45}\right) C_A C_F^2\right.\nn\\
    &\left.+\left(176 \zeta_3-\frac{22646}{81}+\frac{680 \pi ^2}{9}-\frac{88 \pi ^4}{45}\right) C_A^2 C_F+\left(\frac{2048 \zeta_3}{3}+\frac{2576}{9}-\frac{496 \pi ^2}{27}\right) C_F^2 n_f T_F\right.\nn\\
    &\left.+\left(\frac{1120}{81}+\frac{128 \pi ^2}{27}\right) C_F n_f^2 T_F^2+\left(-768 \zeta_3-86-52 \pi ^2-\frac{392 \pi ^4}{45}\right) C_F^3\right]\ln \tau\nn\\
    &+\left(\frac{4672 \zeta_3}{9}+\frac{17888}{27}-\frac{10304 \pi ^2}{81}+\frac{272 \pi ^4}{135}\right) C_A C_F n_f T_F\nn\\
    &+\left(-\frac{3232 \zeta_3}{9}-80 \zeta_5-\frac{23804}{27}+\frac{13250 \pi ^2}{81}-\frac{766 \pi ^4}{135}\right) C_A^2 C_F\nn\\
    &+\left(-\frac{4624 \zeta_3}{9}-240 \zeta_5-\frac{8509}{6}-\frac{37 \pi ^2}{9}+\frac{391 \pi ^4}{45}\right) C_A C_F^2\nn\\
    &+\left(\frac{2816 \zeta_3}{9}+\frac{1768}{3}-\frac{4 \pi ^2}{9}-\frac{128 \pi ^4}{45}\right) C_F^2 n_f T_F+\left(-\frac{1024 \zeta_3}{9}-\frac{3056}{27}+\frac{1952 \pi ^2}{81}\right) C_F n_f^2 T_F^2\nn\\
    &+\left(-120 \zeta_3+128 \pi ^2 \zeta_3-1056 \zeta_5-\frac{841}{2}-65 \pi ^2-\frac{26 \pi ^4}{15}\right) C_F^3 \Bigg\}\,.
\end{align}

\begin{figure}[!htbp]
    \centering
    \includegraphics[width=0.46\linewidth]{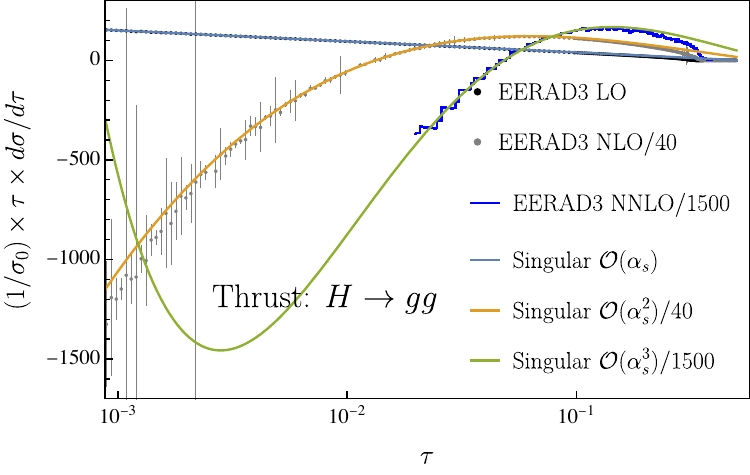}\quad
    \includegraphics[width=0.46\linewidth]{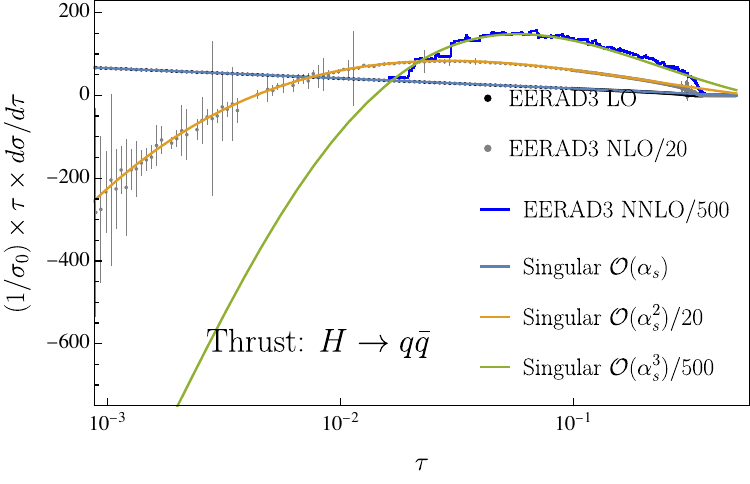}\\
    \includegraphics[width=0.46\linewidth]{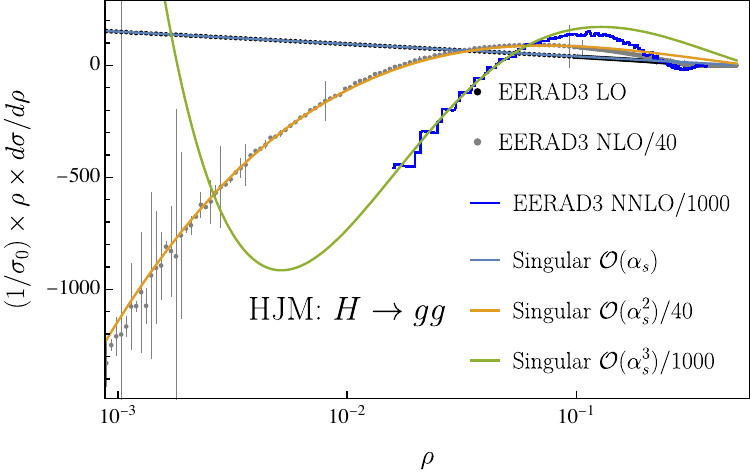}\quad
    \includegraphics[width=0.46\linewidth]{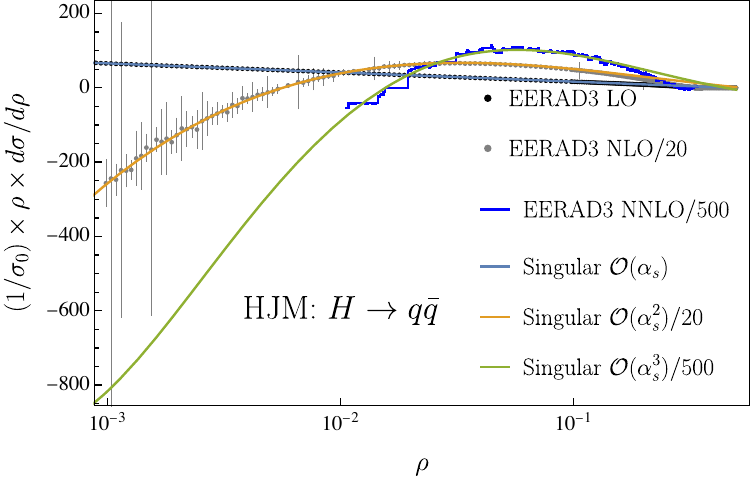}
    \caption{Comparison of singular expansions in the dijet limit with \textsc{Eerad3} data. We show both thrust (top panel) and HJM (bottom panel); both $Hgg$ decay (left panel) and $Hq\bar{q}$ decay (right panel). To show LO, NLO and NNLO on the same figure, we rescale the NLO and NNLO results by a constant factor, written in the legend.}
    \label{fig:dijet_singular_comparison}
\end{figure}

In Fig.~\ref{fig:dijet_singular_comparison}, we compare the analytic expressions with the numerical data from \textsc{Eerad3}~\cite{Gehrmann-DeRidder:2014hxk,Aveleira:2025svg}. Here LO, NLO and NNLO refer to the coefficients in the expansion of $\alpha_s/(4\pi)$. 
LO and NLO are obtained using the public code \textsc{Eerad3} with around $4\times 10^9$ samples, in both linear and log bins. The NNLO data were taken from Ref.~\cite{Fox:2025qmp}, which has limited bins in the small $\tau$ region.
We find good agreement for both thrust and HJM, and for both $Hgg$ and $Hq\bar{q}$ processes.

\subsection{Canonical scales and N$^{3}$LL${}^\prime$ resummation}
\label{sec:dijet_resummation}

In this subsection, we come back to the resummed distributions. It is usually convenient to introduce the cumulant distributions
\begin{equation}
    \Sigma^{\tau}_{i}(\tau)=\frac{1}{\sigma_{0}}\int_{0}^{\tau}d\tau'\,\frac{d\sigma_{i}}{d\tau'}\,,\qquad
    \Sigma^{\rho}_{i}(\rho)=\frac{1}{\sigma_{0}}\int_{0}^{\rho}d\rho'\,\frac{d\sigma_{i}}{d\rho'}\,,
\end{equation}
from which the differential distributions are recovered by differentiation. Putting all the resummed ingredients discussed in Sec.~\ref{sec:dijet} into the factorization theorem in Eq.~\eqref{eq:hemifact}, the resummed thrust distribution reads
\begin{multline}
    \Sigma^{\tau}_{i}(\tau)=W_{i}(\mu_{h},\mu_{j},\mu_{s})\, \mathcal{N}(Q^2,\mu_h) \,H^{\text{dij}}_{i}(Q^{2},\mu_{h})\,\left[\widetilde{j}_{i}\!\Big(\ln\frac{\mu_{s}Q}{\mu_{j}^{2}}+\partial_{\eta},\mu_{j}\Big)\right]^{2}\\
    \times\widetilde{s}^{\text{dij}}_{T,i}(\partial_{\eta},\mu_{s})\times\left(\frac{\tau Q}{\mu_{s}}\right)^{\eta}\frac{e^{-\gamma_{E}\eta}}{\Gamma(\eta+1)}\,,\qquad \eta=4A_{\Gamma,i}(\mu_{j},\mu_{s})\,.
    \label{eq:thrust_resummed}
\end{multline}
with the RG evolution 
\begin{multline}
    W_{i}(\mu_{h},\mu_{j},\mu_{s})=\exp\!\big[4S_{i}(\mu_{h},\mu_{j})+4S_{i}(\mu_{s},\mu_{j})+2A^{\text{dij}}_{S,i}(\mu_{s},\mu_{h})+4A_{J,i}(\mu_{j},\mu_{h})\big]\\
    \times \left(\frac{Q^{2}}{\mu_{h}^{2}}\right)^{-2A_{\Gamma,i}(\mu_{h},\mu_{j})}\,.
    \label{eq:evol_factor}
\end{multline}
% Here $A_{\gamma_{n}}(\mu,\mu_h)$ is the RG evolution for $Hgg$ effective coupling ($n=\lambda$) or Yukawa coupling ($n=y$). 
Here $\mathcal{N}(Q^2,\mu_h)$ is boundary of $\lambda$-RGE or Yukawa RGE, expanded in $\alpha_s(\mu_h)$ as in Eq.~\eqref{eq:lambda_res} and Eq.~\eqref{eq:yukawa_res}.
$i=q,g$ represent $Hq\bar{q}$ and $Hgg$ channels respectively. For HJM, we find
\begin{multline}
    \Sigma^{\rho}_{i}(\rho)=W_{i}(\mu_{h},\mu_{j},\mu_{s})\,\mathcal{N}(Q^2,\mu_h)\, H^{\text{dij}}_{i}(Q^{2},\mu_{h})\,\widetilde{j}_{i}\!\Big(\ln\frac{\mu_{s}Q}{\mu_{j}^{2}}+\partial_{\eta_{1}},\mu_{j}\Big)\widetilde{j}_{i}\!\Big(\ln\frac{\mu_{s}Q}{\mu_{j}^{2}}+\partial_{\eta_{2}},\mu_{j}\Big)\\
    \times\widetilde{s}^{\text{dij}}_{\mu,i}(\partial_{\eta_{1}},\mu_{s})\,\widetilde{s}^{\text{dij}}_{\mu,i}(\partial_{\eta_{2}},\mu_{s})\widetilde{s}_{\text{NGL},i}(\partial_{\eta_{1}}-\partial_{\eta_{2}})\left(\frac{\rho Q}{\mu_{s}}\right)^{\eta_{1}+\eta_{2}}\frac{e^{-\gamma_{E}\eta_{1}}}{\Gamma(\eta_{1}+1)}\frac{e^{-\gamma_{E}\eta_{2}}}{\Gamma(\eta_{2}+1)}\,,
    \label{eq:hjm_resummed}
\end{multline}
with $\eta_{1}=\eta_{2}=2A_{\Gamma,i}(\mu_{j},\mu_{s})$. Replacing $\widetilde{s}_{\text{NGL},i}(\partial_{\eta_{1}}-\partial_{\eta_{2}})\to\widetilde{s}_{\text{NGL},i}(0)$ collapses the two hemispheres into a product and recovers the thrust structure in Eq.~\eqref{eq:thrust_resummed} with $\widetilde{s}^{\text{dij}}_{T,i}=\left(\widetilde{s}^{\text{dij}}_{\mu,i}\right)^{2}\,\widetilde{s}_{\text{NGL},i}(0)$.

We verify the RG invariance of the resummed distributions as follows.
First of all, Eq.~\eqref{eq:thrust_resummed} and Eq.~\eqref{eq:hjm_resummed} are independent of the arbitrary reference scale $\mu$, where we perform the resummation. This follows from relation among different anomalous dimensions and we eliminate all explicit $\mu$ dependence in the distributions. Secondly, expanding N${}^k$LL${}^\prime$ to $\mathcal{O}(\alpha_s^k)$, the result is free of any residual $\mu_{h}$, $\mu_{j}$, $\mu_{s}$ dependence, and the differential of $\Sigma_i$ reproduces exactly the singular fixed-order expansion in Sec.~\ref{sec:dijet_fixed_order}.

The large logarithms in hard, jet and soft functions suggest a canonical choice of the resummation scales
\begin{equation}
    \mu_{h}=Q\,,\qquad \mu_{j}^{\text{dij}}=\sqrt{x}\,Q\,, \qquad \mu_{s}^{\text{dij}}=x\,Q\,,
\end{equation}
with $x=\tau$ or $\rho$. Note that in very small $\tau$ or $\rho$, $\mu_s^{\text{dij}}$ as the smallest scale, can become non-perturbative, and thus the perturbative resummation breaks down. To avoid the Landau pole problem, we freeze the soft scale to be $\mu_s^{\text{dij}} \geq \mu_s^{\text{min}}=1.1$ GeV. In addition, we introduce the scale variation parameters $e_h$, $e_j$ and $e_s$ to estimate the perturbative uncertainty. Explicitly, we adopt the following dijet canonical scale
\begin{equation}
    \mu_{h}=e_h Q\,,\qquad \mu_{s}^{\text{dij}}=\sqrt{\left(e_s \mu_h x \right)^2+\left(\mu_s^{\text{min}}\right)^2} \qquad \mu_{j}^{\text{dij}}=e_j \sqrt{\mu_h \mu_{s}^{\text{dij}}}\,,
    \label{eq:canonical_scales}
\end{equation}
We correlate the variations such that: (1) the hierarchy among hard, jet and soft scales is preserved during variations; (2) when $\tau,\rho\to 0$, $\mu_s^{\text{dij}}$ always converges to $\mu_s^{\text{min}}$ regardless of any variation. For Higgs decays, we consider $Q=m_H$, where the Higgs is at rest. 

\begin{table}[!htbp]
    \centering
    \setlength{\tabcolsep}{15pt}
    {\small
    \renewcommand{\arraystretch}{1.2}
    \begin{tabular}{|c|c|c|c|c|c|c|}
        \hline
        Order & $\Gamma^{i}_{\text{cusp}}$ & $\gamma^{H,J,S}_{i}$ & $H^{\text{dij}}_{i},\widetilde{j}_{i},\widetilde{s}^{\text{dij}}_{i}$ & $\widetilde{s}_{\text{NGL},i}$ & $\beta$ & matching\\
        \hline
        LL              & 1-loop & --     & tree   & --     & 1-loop & --   \\
        \hline
        NLL             & 2-loop & 1-loop & tree   & --     & 2-loop & --   \\
        \hline
        NNLL            & 3-loop & 2-loop & 1-loop & 1-loop     & 3-loop & LO   \\
        \hline
        N$^{3}$LL       & 4-loop & 3-loop & 2-loop & 2-loop & 4-loop & NLO   \\
        \Xhline{1pt}
        NLL$^{\prime}$  & 2-loop & 1-loop & 1-loop & 1-loop    & 2-loop & LO   \\
        \hline
        NNLL$^{\prime}$ & 3-loop & 2-loop & 2-loop & 2-loop & 3-loop & NLO  \\
        \hline
        N$^{3}$LL$^{\prime}$ & 4-loop & 3-loop & 3-loop & 3-loop & 4-loop & NNLO \\
        \hline
    \end{tabular}}
    \caption{Order counting for the dijet resummation and its matching to fixed-order. We show both the unprime orders and prime orders, where prime orders contain higher-loop boundaries.}
    \label{tab:resummation_orders}
\end{table}

\begin{figure}[!htbp]
\centering
    \includegraphics[width=0.48\linewidth]{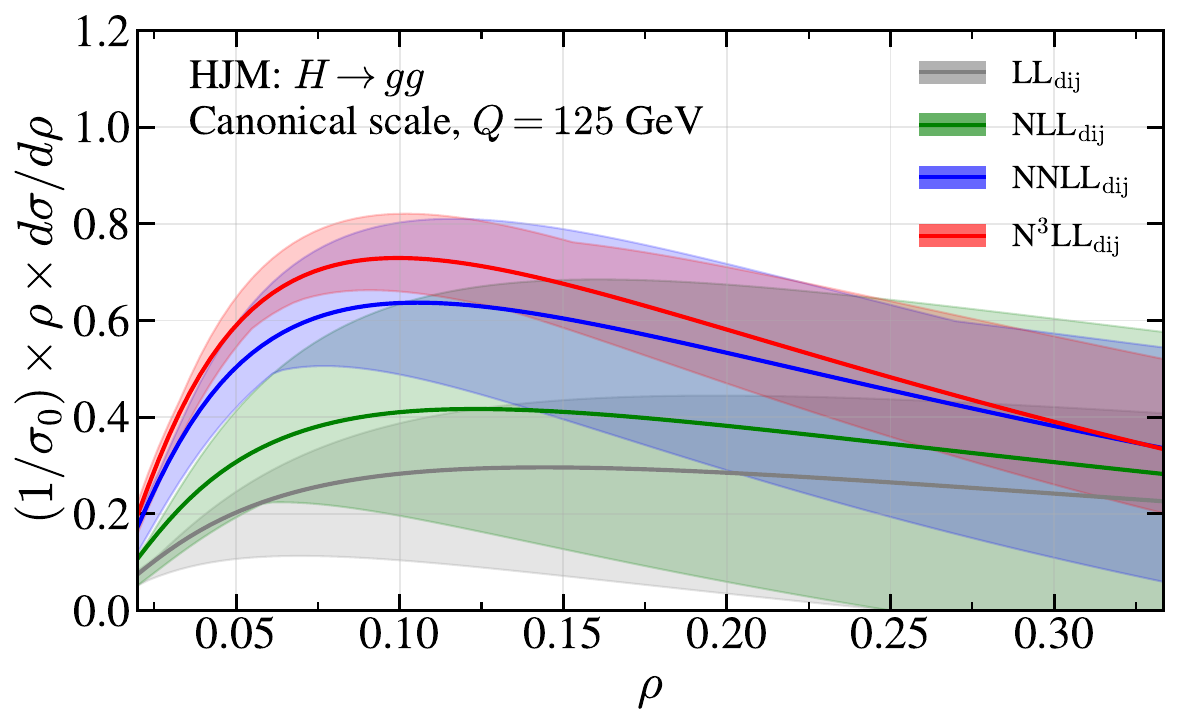}
    \includegraphics[width=0.48\linewidth]{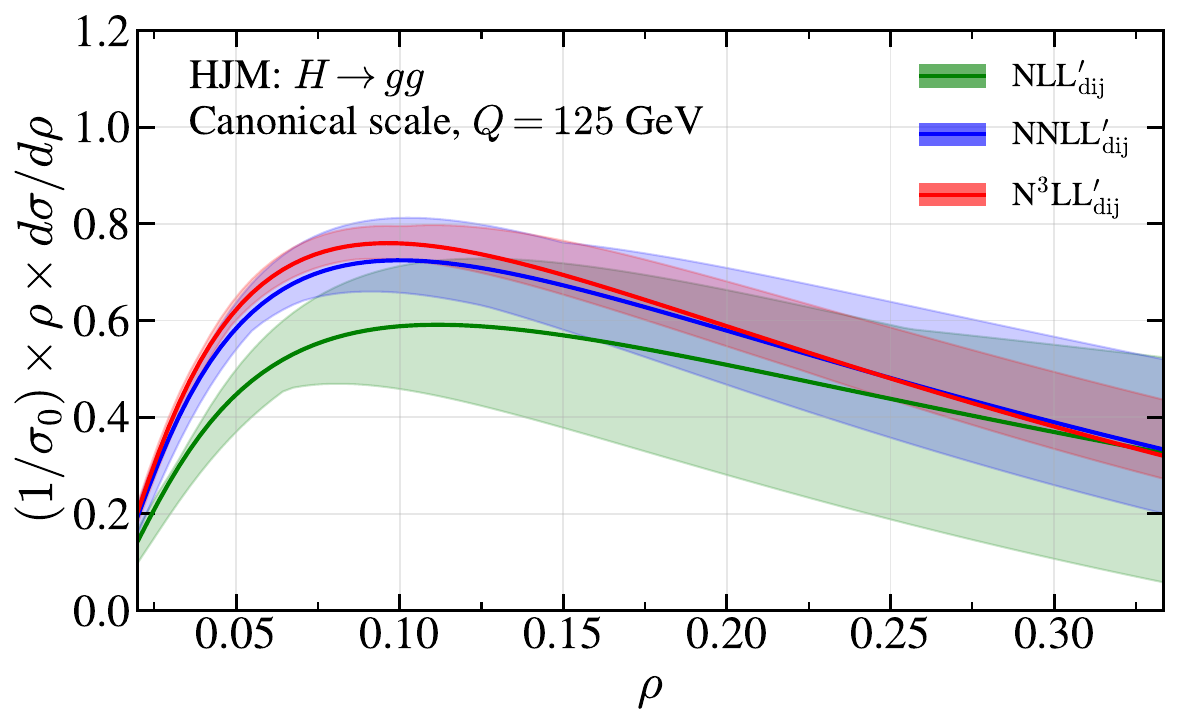}\\
   \includegraphics[width=0.48\linewidth]{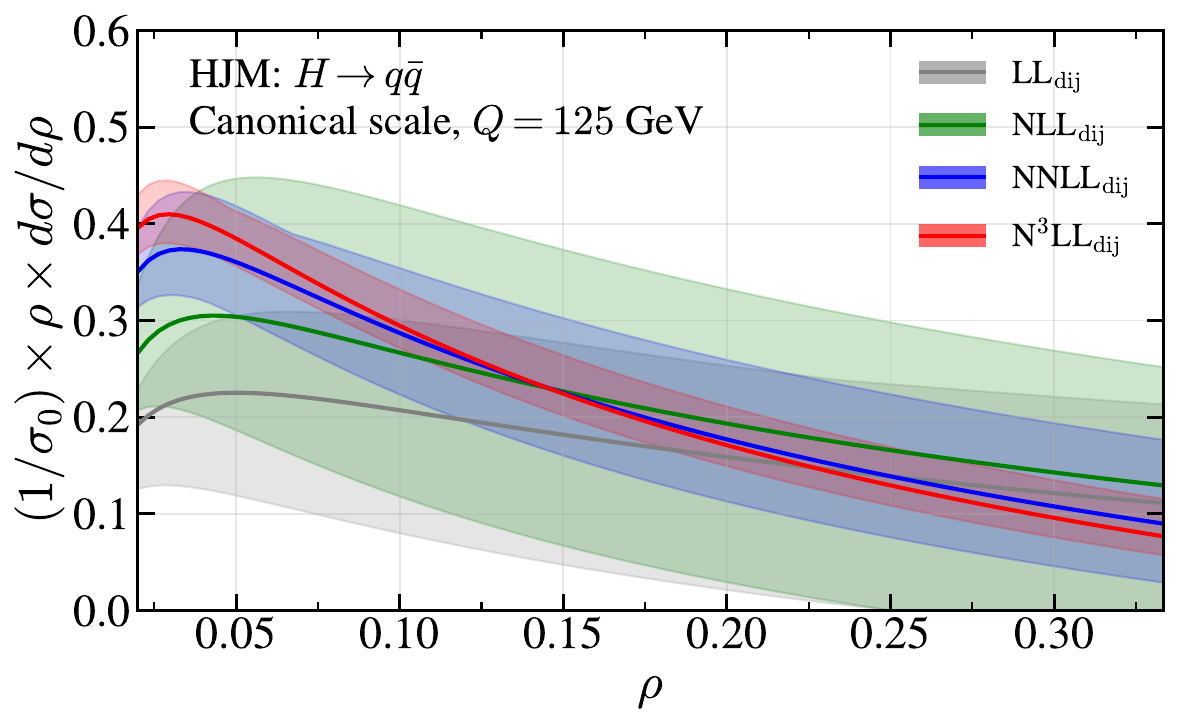}
    \includegraphics[width=0.48\linewidth]{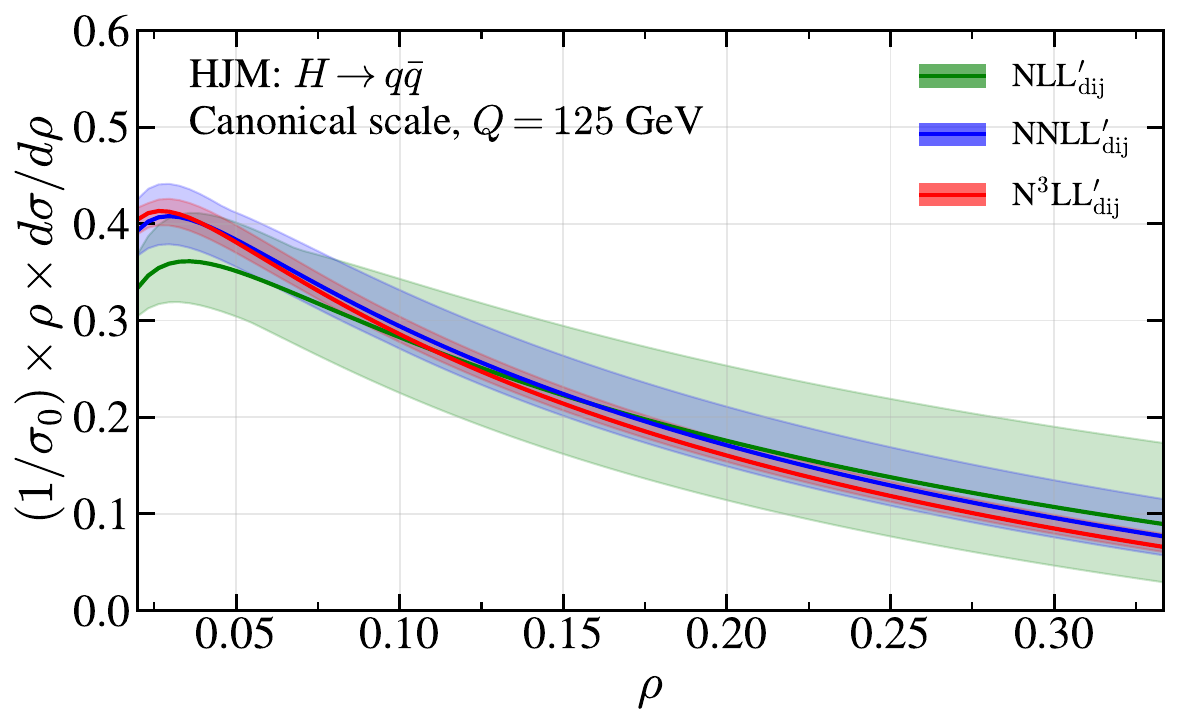}\\
    \caption{Canonical resummation of HJM in hadronic Higgs decays. The top panel shows $Hgg$ process and the bottom panel shows $Hq\bar{q}$ process. The left panel gives unprime orders and the right panel is prime orders. We use $Q=m_H=125$ GeV in all cases. The distributions are normalized to Born total decay rate, $\sigma_0=\Gamma_{Hgg}(Q)$ and $\Gamma_{Hq\bar{q}}(Q)$ respectively.}
    \label{fig:dijet_can_resum_HJM}
\end{figure}

\begin{figure}[!htbp]
\centering
    \includegraphics[width=0.48\linewidth]{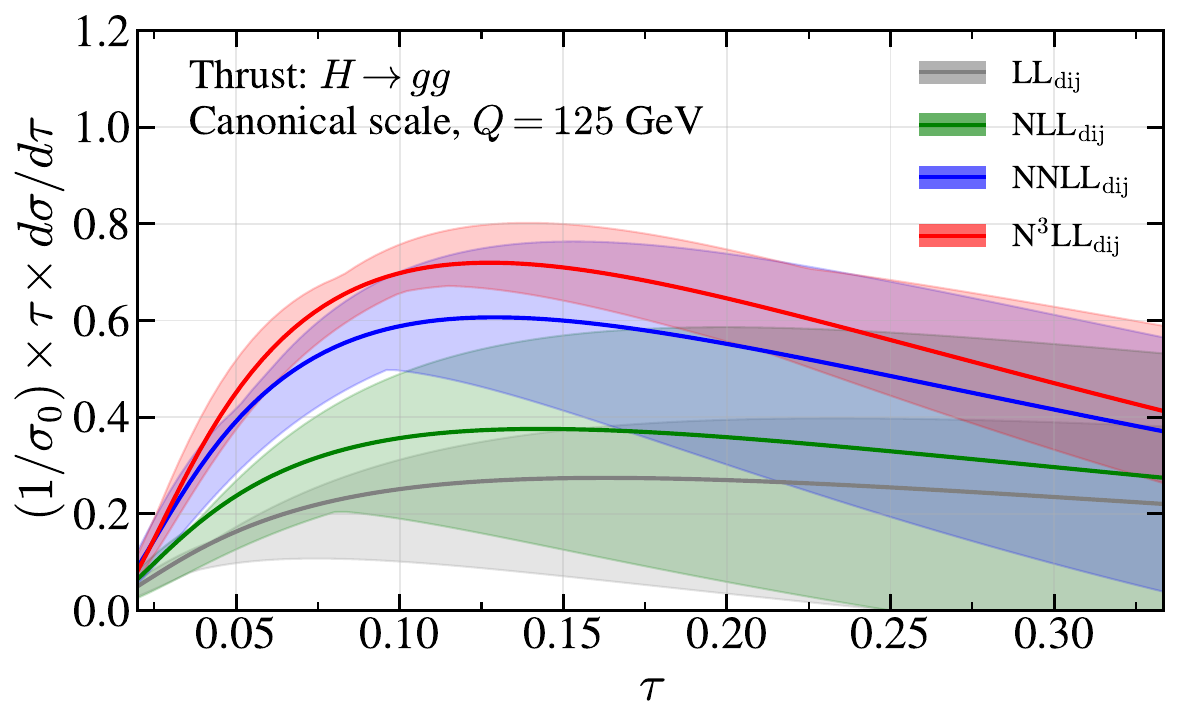}
    \includegraphics[width=0.48\linewidth]{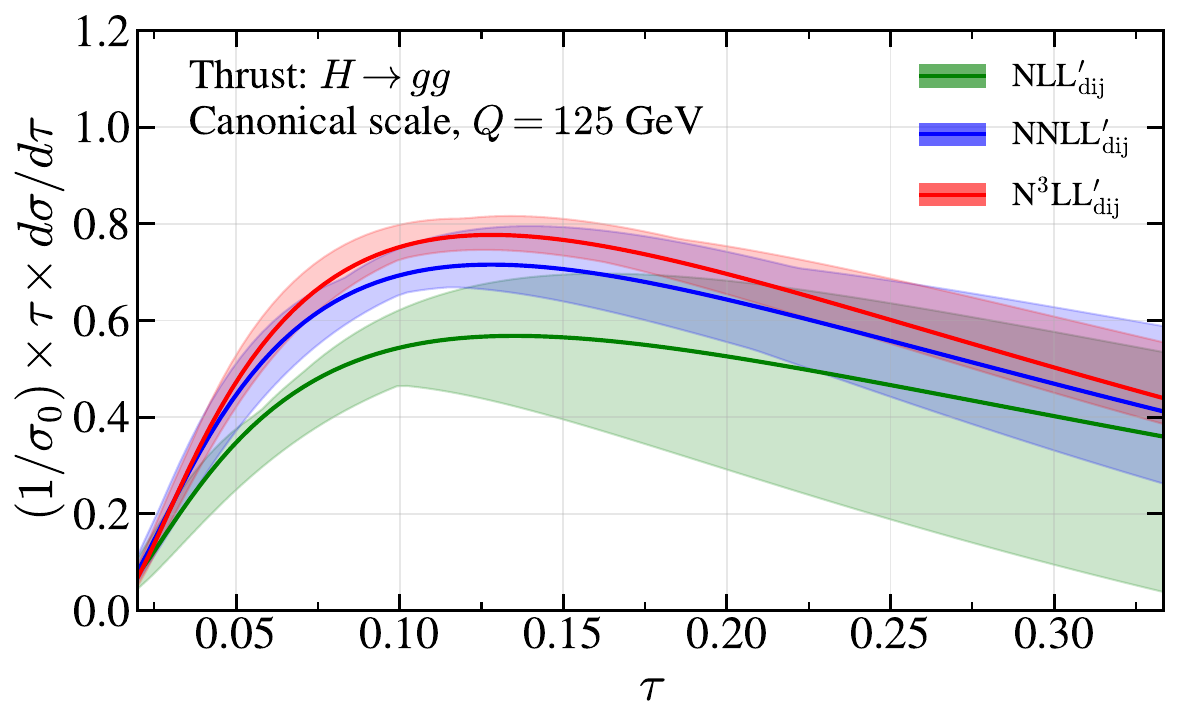}\\
   \includegraphics[width=0.48\linewidth]{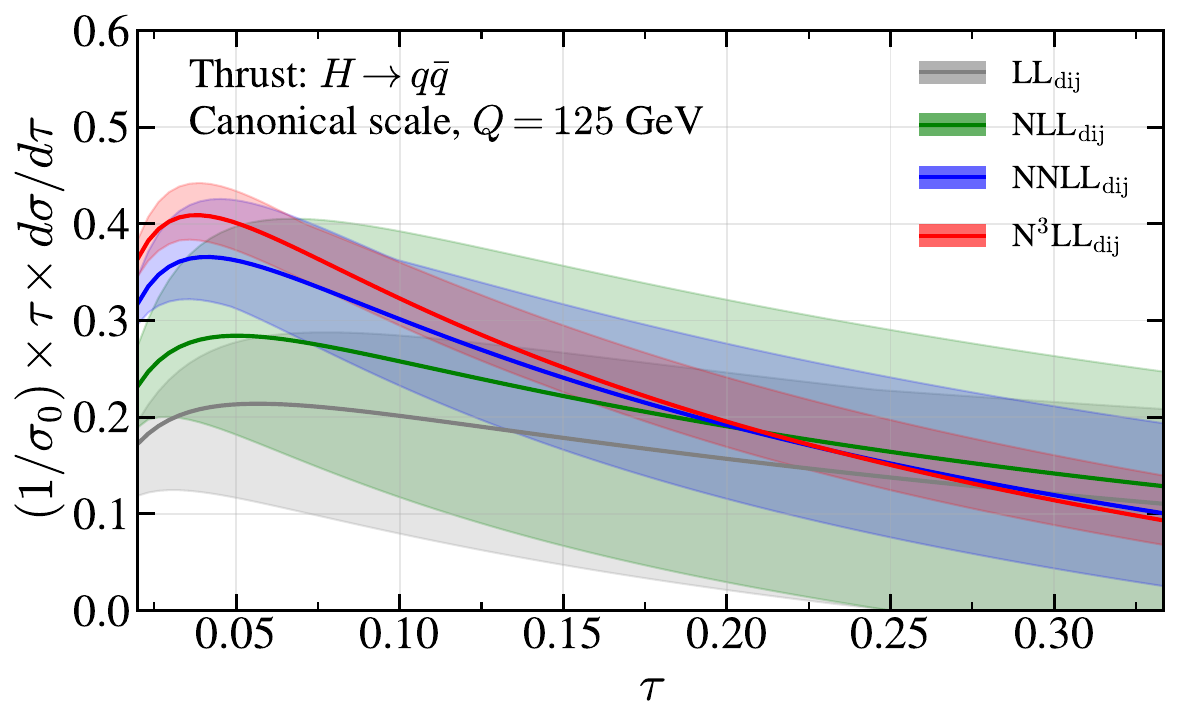}
    \includegraphics[width=0.48\linewidth]{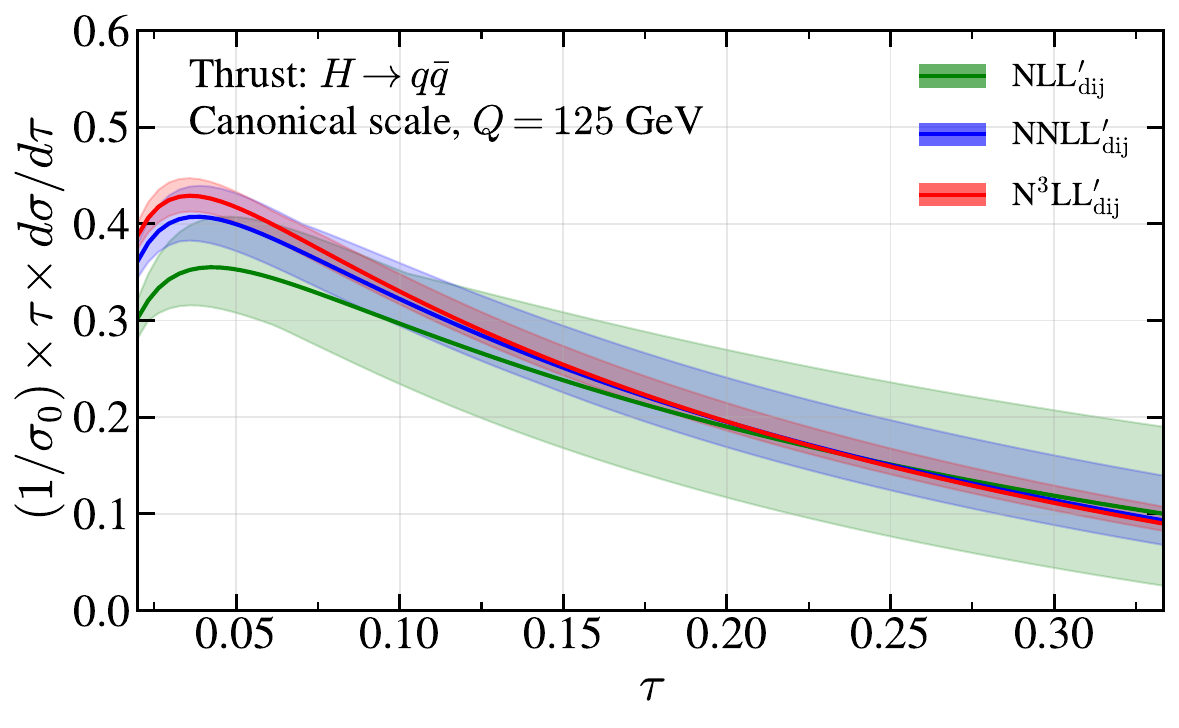}\\
    \caption{The same as Fig.~\ref{fig:dijet_can_resum_HJM}, but for thrust.}
    \label{fig:dijet_can_resum_Thrust}
\end{figure}

The ingredients required to reach a given logarithmic accuracy are summarized in Tab.~\ref{tab:resummation_orders}. Note that the difference between unprimed N$^{k}$LL and N$^{k}$LL$^{\prime}$ is the order in the boundary of $H^{\text{dij}}_{i}$, $\widetilde{j}_{i}$ and $\widetilde{s}^\text{dij}_{i}$, where the latter uses one higher order.
In Fig.~\ref{fig:dijet_can_resum_HJM}, we present the canonical resummation for HJM in both $Hgg$ and $Hq\bar{q}$ processes. We show both unprime orders and prime orders and find good convergence in both cases. Compared to $e^+e^-\to q\bar{q}$ process, we observe the peak in larger $\rho$, especially in $Hgg$ process.
The kinks appearing around $\rho=1/3$ are the Sudakov shoulders which we will discuss in Sec.~\ref{sec:shoulder}. For completeness, we also present the canonical resummation for thrust in Fig.~\ref{fig:dijet_can_resum_Thrust}.

\subsection{Dijet profile functions}
\label{sec:dijet_profile}

In this subsection, we revisit the dijet profile functions for resummation scales and describe the matching procedure. Here matching refers to including the non-singular contribution from fixed-order calculation in addition to the dijet resummation. The main formula is 
\begin{equation}
    \frac{d\sigma^\text{dij, match}_{i}}{d\rho}=\frac{d\sigma^{\text{dij, res}}_{i}}{d\rho}+\left(\frac{d\sigma^\text{FO}_{i}}{d\rho}-\frac{d\sigma^\text{dij, sing}_{i}}{d\rho}\right)\,=\frac{d\sigma^{\text{dij, res}}_{i}}{d\rho}+\frac{d\sigma^{\text{dij, ns}}_{i}}{d\rho}\,,
    \label{eq:dijet_matching}
\end{equation}
where $d\sigma^\text{dij, sing}_{i}$ is the singular expansion of resummed distribution, to the same order in $\alpha_s$ as $d\sigma^\text{FO}_{i}$. We also define the non-singular cross-section as $d\sigma^\text{dij, ns}_{i}=d\sigma^\text{FO}_{i}-d\sigma^\text{dij, sing}_{i}$. 
In this work, we compute the LO distributions analytically, and obtain the NLO correction from the program \textsc{Eerad3} and NNLO from Ref.~\cite{Fox:2025qmp}.

In order for matching to work, we need to turn off the dijet resummation in the region where dijet logs no longer dominate the distribution. In the meantime, we also need to freeze the scales when they enter the non-perturbative regime, $x\sim\Lambda_\text{QCD}/Q$. As studied in Ref.~\cite{Abbate:2010xh}, we adopt the following 5-region soft scale, 
\begin{equation}
\mu_s^{\text{dij, prof}}(x)=\begin{cases}
\mu_s^{\rm min}\,, & 0\le x\le t_0\,,\\[3pt]
\zeta\big(\mu_s^{\rm min},0;\,0,e_s\mu_h;\,t_0,t_1; \,x\big)\,, & t_0<x\le t_1\,,\\[3pt]
e_s\,\mu_h\,x\,, & t_1<x\le t_2\,,\\[3pt]
\zeta\big(0,e_s\mu_h;\,\mu_h,0;\,t_2,t_s; \,x\big)\,, & t_2<x\le t_s\,,\\[3pt]
\mu_h\,, & t_s<x\le \frac12\,,
\end{cases}
\label{eq:soft_profile}
\end{equation}
subject to the floor $\mu_s\ge\mu_s^{\rm min}$. The central region reproduces the canonical
soft scale $\mu_s^{\rm dij}=e_s\mu_hx$, while the two outer plateaus freeze $\mu_s$ at the
nonperturbative value $\mu_s^{\rm min}$ and merge it into the hard scale, respectively. The
lower transition points $t_0=n_0/Q$ and $t_1=n_1/Q$ scale as $1/Q$, so that the
nonperturbative region sets in at fixed values of the physical momentum $xQ\sim n_{0,1}$; the
upper points $t_2$ and $t_s$ are fixed in $x$ and delimit the approach to fixed order.
The interpolating function $\zeta$ connects two straight lines with a continuous value and
first derivative. Joining $a_1+b_1x$ at $t_1$ to $a_2+b_2x$ at $t_2$, we define
\begin{equation}
\zeta(a_1,b_1;a_2,b_2;t_1,t_2; x)=\begin{cases}
\hat a_1+b_1(x-t_1)+e_1(x-t_1)^2\,, & t_1\le x\le t_m\,,\\[3pt]
\hat a_2+b_2(x-t_2)+e_2(x-t_2)^2\,, & t_m<x\le t_2\,,
\end{cases}
\label{eq:zeta_def}
\end{equation}
with the notation $\hat a_i=a_i+b_it_i$, the midpoint $t_m=(t_1+t_2)/2$, and
\begin{equation}
e_1=\frac{4(\hat a_2-\hat a_1)-(3b_1+b_2)\,\Delta t}{2\,\Delta t^2}\,,\qquad
e_2=\frac{4(\hat a_1-\hat a_2)+(b_1+3b_2)\,\Delta t}{2\,\Delta t^2}\,,\qquad \Delta t\equiv t_2-t_1\,,
\end{equation}
chosen so that $\zeta$ reduces to the bounding lines at the endpoints and is smooth at $t_m$.
For jet scale, we simplify the form in Ref.~\cite{Abbate:2010xh} and take the following,
\begin{equation}
\mu_j^{\text{dij, prof}}(x)=e_j\,\sqrt{\mu_h\,\mu_s^{\text{dij, prof}}(x)}\,.
\label{eq:jet_profile}
\end{equation}
Since we have $\mu_s^{\text{dij, prof}}\ge \mu_s^{\text{min}}$, the jet scale is also away from Landau pole. Finally, the hard scale stays the same as canonical scale, $\mu_h=e_hQ$.

The default values of profile parameters in hadronic Higgs decay, collected in Tab.~\ref{tab:profile_params}, mostly follow recent HJM analysis~\cite{Benitez:2025vsp}: $\mu_s^{\rm min}=1.1\,$GeV, $n_0=2$, $n_1=10$, $t_2=0.25$ and $t_s=0.40$. 
The central profile corresponds to
$e_h,\, e_s,\,e_j=1$, for which $\mu_{h,j,s}$ reduce to the canonical
scales in Eq.~\eqref{eq:canonical_scales} throughout the resummation region; the perturbative
uncertainty is estimated by varying the parameters independently over the ranges in
Tab.~\ref{tab:profile_params} and taking the envelope. 

\begin{table}[!htbp]
\centering
\setlength{\tabcolsep}{12pt}
\begin{tabular}{|c|c|c|}
\hline
parameter & default & variation range\\
\hline
$\mu_s^{\rm min}$ & $1.1\,$GeV & -- \\
$n_0$ & $2$ & $[1.5,2.5]$ \\
$n_1$ & $10$ & $[8.5,11.5]$ \\
$t_2$ & $0.25$ & $[0.225,0.275]$ \\
$t_s$ & $0.40$ & $[0.375,0.425]$ \\
$e_h$ & $1$ & $[1/2,2]$ \\
$e_s$ & $1$ & $[1/2,2]$ \\
$e_j$ & $1$ & $[1/2,2]$ \\
\hline
\end{tabular}
\caption{Default values and variation ranges of the dijet profile parameters.
The transition parameters $n_{0,1}$, $t_2$, $t_s$ follow
Ref.~\cite{Benitez:2025vsp}; the hard, soft and jet normalizations
$e_{h,s,j}$ are varied by the conventional factor of two.}
\label{tab:profile_params}
\end{table}

\subsection{Results}

\begin{figure}[!htbp]
	\centering
	\includegraphics[width=0.48\linewidth]{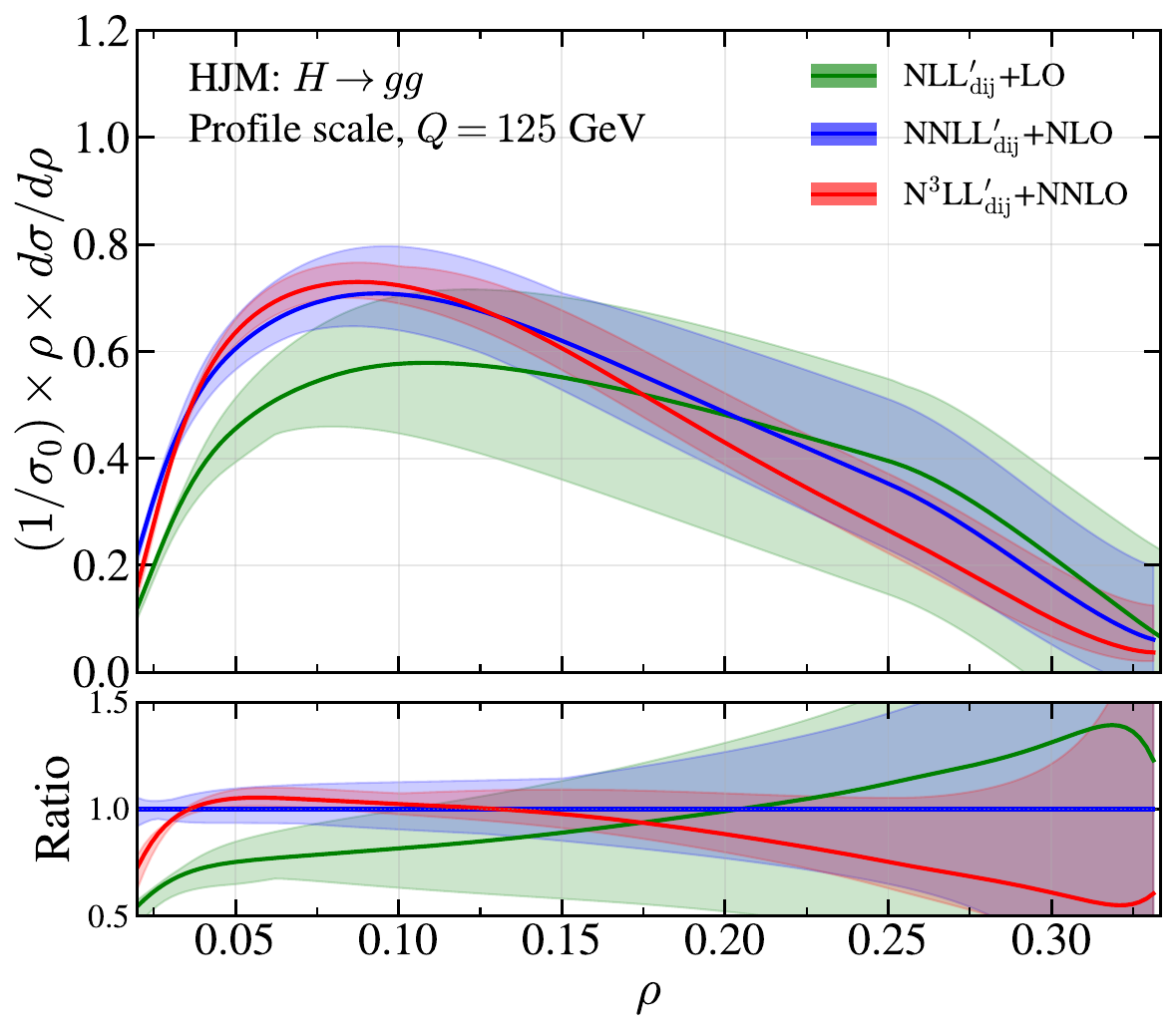}
	\includegraphics[width=0.48\linewidth]{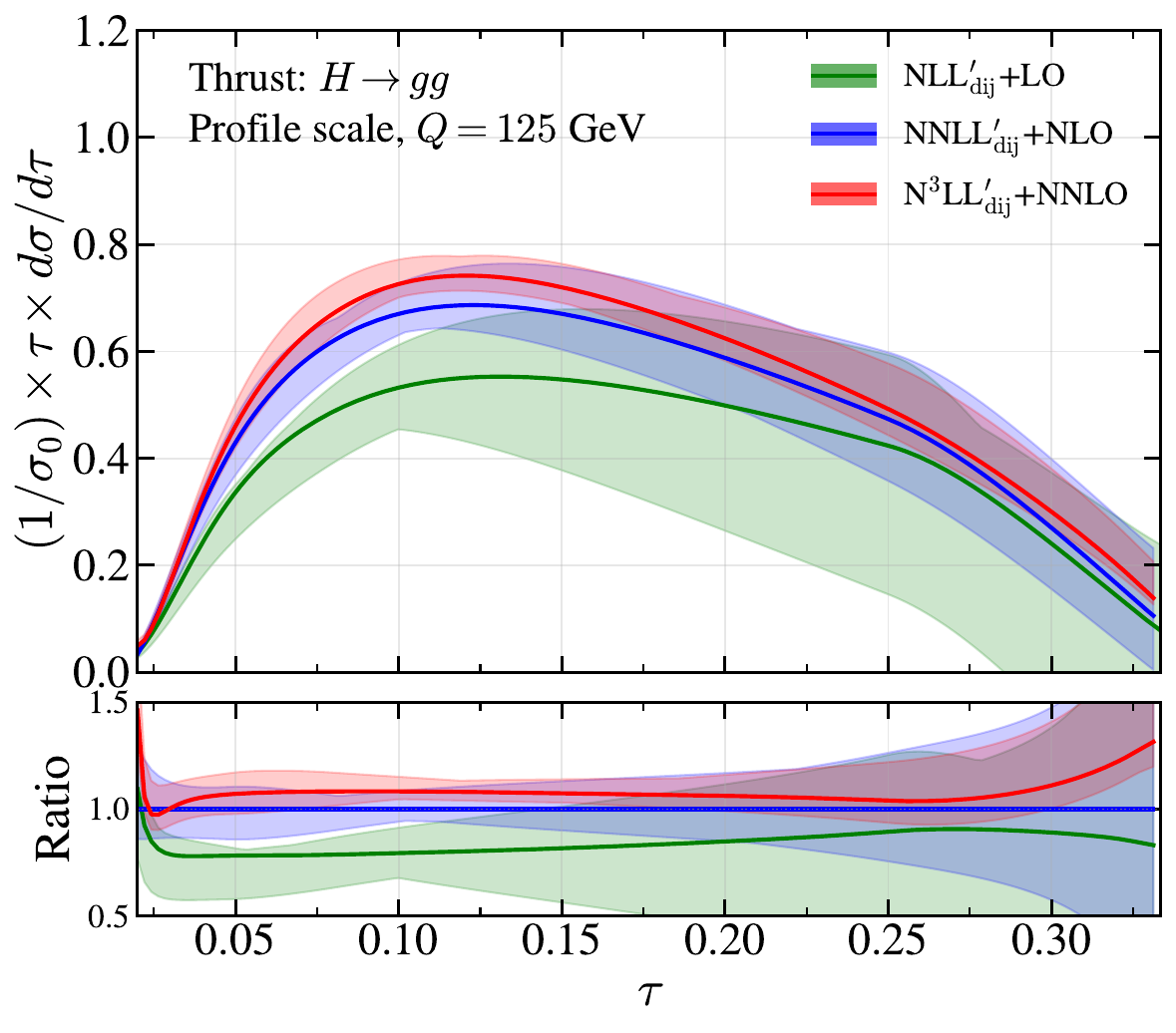}\\
	\includegraphics[width=0.48\linewidth]{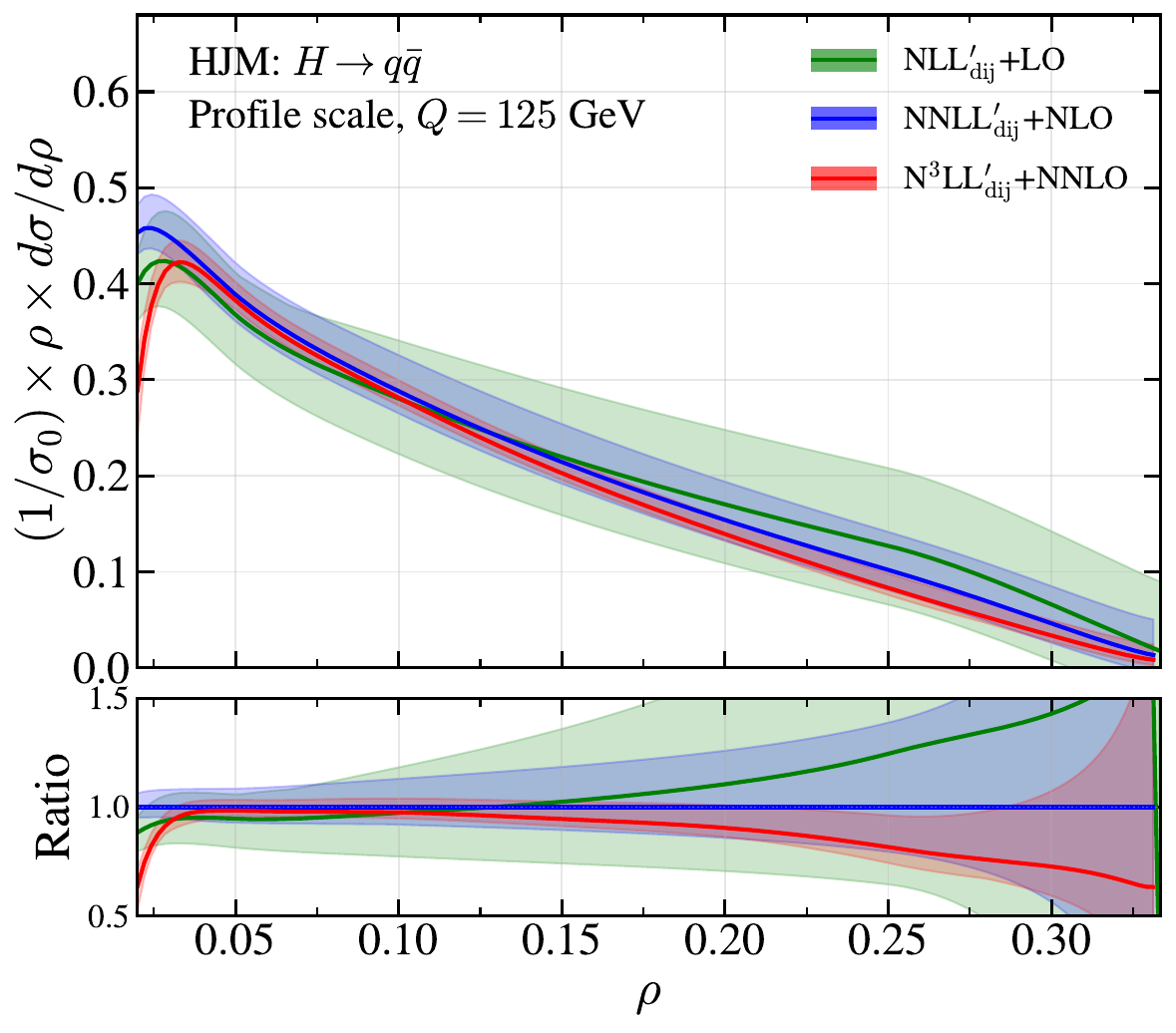}
	\includegraphics[width=0.48\linewidth]{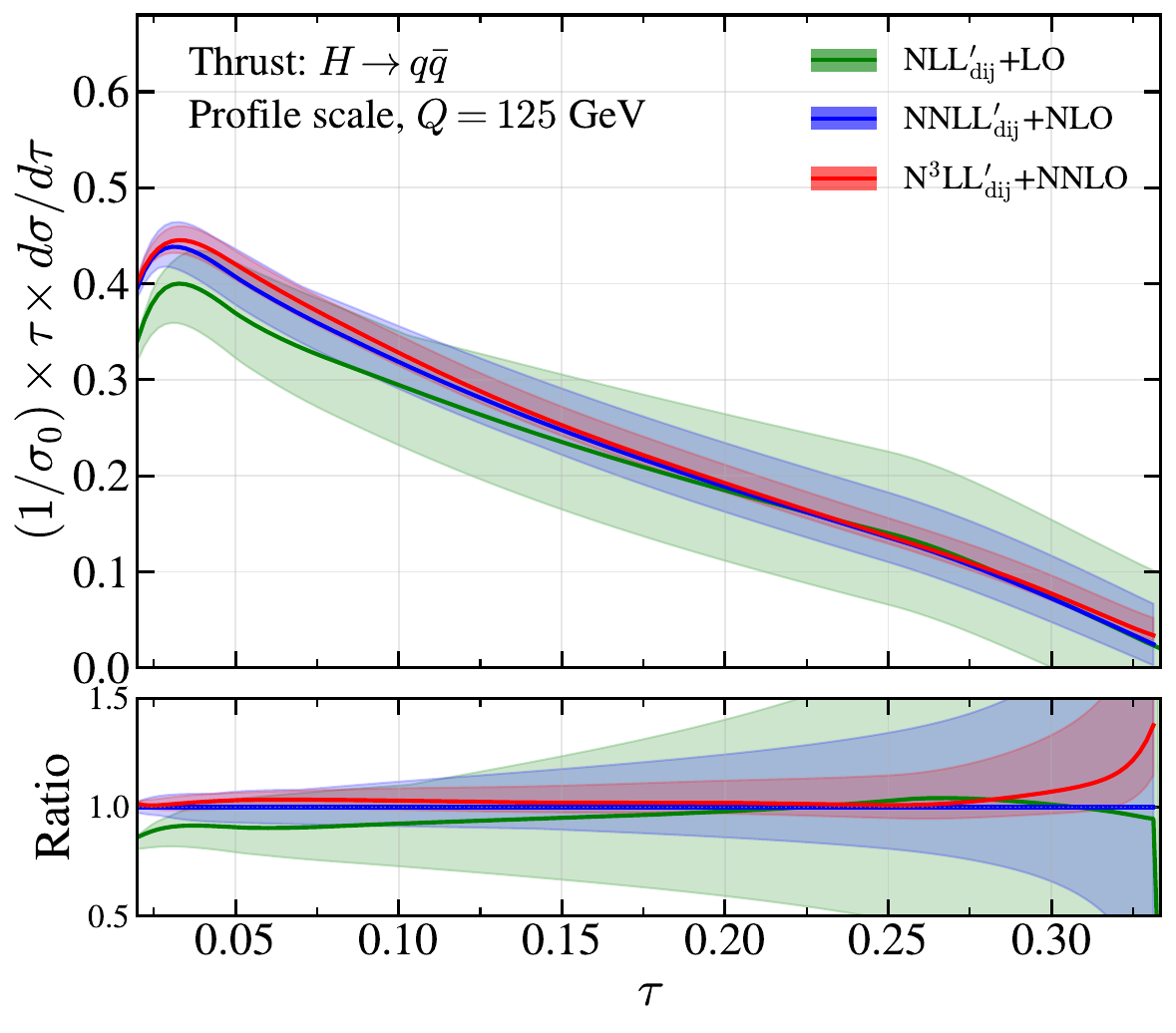}\\
	\caption{Dijet resummation matched to fixed-order for HJM and thrust in both $Hgg$ and $Hq\bar{q}$ processes. Dijet profile scales are used in the matching procedure and the uncertainty bands are the envelop of individual profile parameter variations. We find good order-by-order convergence in all cases.}
	\label{fig:dijet_prof_resum}
\end{figure}

In Fig.~\ref{fig:dijet_prof_resum}, we show the final dijet resummation result for HJM in hadronic Higgs decays. Again we include thrust distributions for completeness. For future $e^+e^-$ colliders like FCC-ee, we will mainly focus on $e^+e^-\to Z H$ process with $H$ almost on-shell, and thus we consider the center-of-mass energy $Q=125$ GeV in our calculation. We present the prime orders, from NLL${}^\prime_\text{dij}+$LO to N${}^3$LL${}^\prime_\text{dij}+$NNLO. The uncertainty bands are the envelope of individual variations listed in Tab.~\ref{tab:profile_params}.
For all four cases, we observe good order-by-order convergence in the dijet resummation region. In the small $\tau,\rho$ region, the peak location in $Hq\bar{q}$ is similar to $e^+e^-\to q\bar{q}$, around $\tau,\rho\sim 0.025$; however, $Hgg$ has the peak location around $\tau,\rho\sim 0.1$, much larger than the quark case. A similar observation is also found in the collinear resummation of projected energy correlators in Ref.~\cite{Lee:2026zyl}, which reveals that the gluonic non-perturbative power correction parameter $\Omega_{1g}$ is larger than the quark one $\Omega_{1q}$.
In the lower panels, we take the ratio of each resummation band to the central value of the middle order, NNLL${}^\prime_\text{dij}+$NLO, and illustrate the effect of the highest order. Note that the central value corrections in the trijet region are larger than in the dijet region. This indicates that an important ingredient is missing, which is the Sudakov shoulder contributions that we will study in the next sections.

\section{NLO calculations in the trijet region}
\label{sec:nlo}

In this section, we calculate the logarithmic terms of NLO HJM in the trijet limit $\rho\to 1/3$. We will expand the $1\to 4$ matrix elements around the symmetric trijet configuration and the four-particle phase space to leading power, and integrate it against the HJM measurement. In Ref.~\cite{Bhattacharya:2022dtm}, similar calculation has been performed for $\gamma^\star\to q\bar q$, and now we extend it to hadronic Higgs decays. This will allow us to determine $d_{22}^{\pm}$ and $d_{21}^{\pm}$ in Eq.~\eqref{eq:sh_log_form}. 

\subsection{Kinematics and phase space}
\label{sec:nlo_kin}

We work in the Higgs rest frame, with total momentum $Q^\mu$ normalized to $Q^2=1$, and label
the four massless final-state partons $p_1,\dots,p_4$. After momentum conservation, the
on-shell conditions and a choice of frame, the four-parton phase space has five independent
variables. Following the light-cone parametrization of Ref.~\cite{Bhattacharya:2022dtm} we
introduce back-to-back light-like directions $n^\mu=(1,0,0,1)$ and $\bar n^\mu=(1,0,0,-1)$, with
$n\cdot\bar n=2$, and write the candidate collinear pair $(p_3,p_4)$ and the recoiling pair
$(p_1,p_2)$ as
\begin{align}
p_3 &= z\omega\,n + \frac{q_\perp^2}{4z\omega}\,\bar n + q_\perp\,, &
p_4 &= (1-z)\omega\,n + \frac{q_\perp^2}{4(1-z)\omega}\,\bar n - q_\perp\,,\nn\\
p_1 &= z_1\,n + \frac{p_\perp^2}{4z_1}\,\bar n + p_\perp\,, &
p_2 &= z_2\,n + \frac{p_\perp^2}{4z_2}\,\bar n - p_\perp\,,
\end{align}
where $q_\perp$ fixes a reference transverse direction and $p_\perp$ lies at azimuth $\phi$ to
it, $p_\perp^\mu=(0,p_T\cos\phi,p_T\sin\phi,0)$.

For convenience, we take the following independent variables
\begin{equation}
\{\,s_{234},\ s_{34},\ z,\ \omega,\ \phi\,\}\,,\qquad\text{with }\quad
s_{234}\equiv(p_2+p_3+p_4)^2\,,\quad s_{34}\equiv(p_3+p_4)^2\,,
\end{equation}
which separate into two hard and three jet-resolving variables. The hard variables are the
large light-cone momentum of the light jet, $\omega=\frac{1}{2}\,\bar n\!\cdot\!(p_3+p_4)$, and
$s_{234}$; both equal $\frac{1}{3}$ at the symmetric trijet, so their offsets
\begin{equation}
x\equiv\omega-\frac{1}{3}\,,\qquad y\equiv s_{234}-\frac{1}{3}\,
\end{equation}
measure the distance from it. The jet-resolving variables describe the light hemisphere:
$s_{34}$ is its squared mass, $z$ the light-cone momentum fraction of $p_3$ within the $(3,4)$
jet, and $\phi$ the orientation of the splitting plane.
The on-shell and momentum-conservation conditions then fix the remaining components. The transverse momentum in the
collinear pair satisfies
\begin{equation}
q_\perp^2=z(1-z)\,s_{34}\,,
\end{equation}
while the recoiling pair is given by
\begin{align}
z_1&=\frac{(1-2\omega)(2 s_{234}\omega-s_{34})}{2\,(4\omega^2-s_{34})}\,,\qquad
z_2=\frac{\omega(1-2\omega)(2\omega-s_{234})}{4\omega^2-s_{34}}\,,\nn\\
p_\perp^2&=\frac{(2\omega-s_{34})(1-2\omega)(2\omega-s_{234})(2 s_{234}\omega-s_{34})}{(4\omega^2-s_{34})^2}\,,
\end{align}
with recoil mass $s_{12}\equiv(p_1+p_2)^2=1-2\omega+s_{34}\big(1-\frac{1}{2\omega}\big)$. In the symmetric limit, the light-jet variables vanish
($s_{34},q_\perp^2\to0$), while the recoil stays hard, $z_1=z_2=p_\perp^2\to\frac1{12}$ as
$\omega,s_{234}\to\tfrac13$, consistent with $z_1+z_2=\tfrac12-\omega$.

For the integration, we trade the azimuth for the invariant $s_{23}$,
\begin{multline}
s_{23} = \frac{s_{34}^2 z + 2 s_{34}(1+2z\omega-2z-2\omega)\omega
 + 4z\,s_{234}\,\omega^2 + s_{34}s_{234}(z-1+2\omega-4z\omega)}{4\omega^2-s_{34}}\\
 + \frac{2\sqrt{s_{34}(1-z)z(2\omega-s_{34})(1-2\omega)(2\omega-s_{234})(2\omega s_{234}-s_{34})}}{4\omega^2-s_{34}}\,\cos\phi\,,
\label{eq:s23}
\end{multline}
which is linear in $\cos\phi$, the physical range $-1\le\cos\phi\le1$ bounding the $s_{23}$
integration.
With this substitution the four-parton phase space,
$d\mathrm{PS}_4=\prod_{i=1}^{4}\frac{d^3 p_i}{(2\pi)^3\,2E_i}\,(2\pi)^4\,\delta^4\big(Q-\textstyle\sum_i p_i\big)$,
reduces to a five-fold differential and the shoulder distribution becomes
\begin{equation}
\frac{1}{\sigma_0}\frac{d\sigma}{d\rho}
=\frac{1}{256\pi^5}\int ds_{234}\,ds_{34}\,d\omega\,dz\,
\frac{ds_{23}}{\sqrt{(s_{23}^{+}-s_{23})(s_{23}-s_{23}^{-})}}\;
|\mathcal M|^2\;\delta\big(\rho-\rho(s_{234},s_{34},\omega,z,s_{23})\big)\,,
\label{eq:master}
\end{equation}
and identically for $\tau$. Here $|\mathcal M|^2$ is the Born-normalized four-parton
squared matrix element, and the square-root Jacobian
$J=[(s_{23}^{+}-s_{23})(s_{23}-s_{23}^{-})]^{-1/2}$ follows from Eq.~\eqref{eq:s23} through
$\int_0^{2\pi}d\phi=2\int ds_{23}\,J$, with $s_{23}^{\pm}$ the values reached at
$\cos\phi=\pm1$. The measurement $\delta$-function removes one of the five integrations, leaving
a four-fold integral dominated by the soft and collinear regions identified below.
Identical particle symmetry factors are restored channel by channel.

The shoulder logarithms come entirely from soft and collinear divergences. Writing $r\equiv\tfrac13-\rho$ and
counting $r\sim\lambda\ll1$, the collinear region $p_3\parallel p_4$ scales as
\begin{equation}
s_{34}\sim\lambda\,,\quad x\sim\lambda\,,\quad y\sim\lambda\,,\quad z\sim\lambda^0\,,\quad
\phi\sim\lambda^0\,,
\label{eq:collscaling}
\end{equation}
for which the endpoints reduce to
$s_{23}^{\pm}=\big(\tfrac{\sqrt{s_{34}}}{2}\pm\tfrac{\sqrt z}{\sqrt3}\big)^2$; the soft region,
in which $p_3$ becomes soft, has the same scaling but with $z\sim\lambda$.

\begin{figure}[!htbp]
\centering
\includegraphics[scale=0.85]{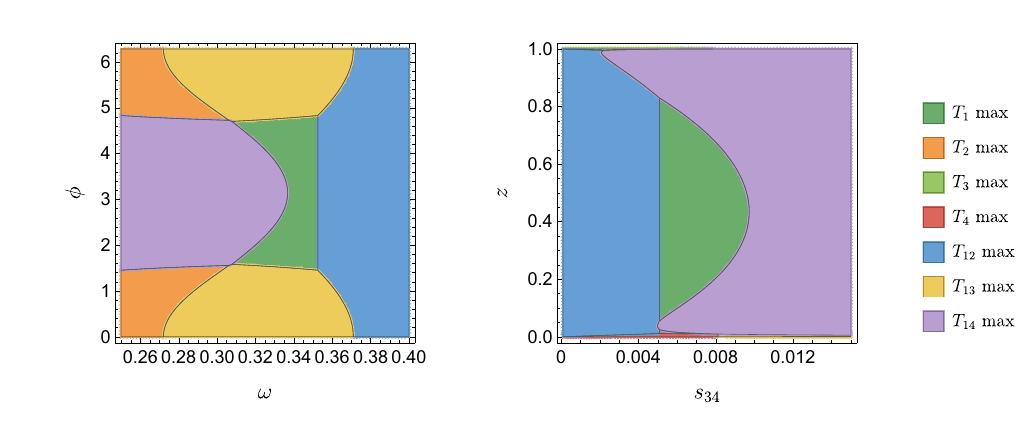}
\caption{Two slices of the five-dimensional massless four-parton phase space near the
symmetric point. Each region is colored by which hemisphere partition wins the thrust
maximization.
Left panel: the $(\omega,\phi)$ plane at $z=0.06$, $s_{34}=0.02$, $s_{234}=\frac{1}{3}-0.01$. For example,
in the green ($T_1$) region the heavy hemisphere is $\{234\}$, so $\rho=s_{234}=\frac{1}{3}-0.01$
and this region contributes to the HJM left shoulder at $r=\frac{1}{3}-\rho=0.01$. Similar analysis can be applied to other regions.
Right panel: the $(s_{34},z)$ plane at $\phi=\pi$, $s_{234}=\frac{1}{3}$, $s_{12}=\frac{1}{3}-0.01$. The blue region with $T_{12}$ max also contributes to $r=\frac{1}{3}-\rho=0.01$.}
\label{fig:shoulder_ps}
\end{figure}

Both observables require the thrust axis. By Eq.~\eqref{eq:thrustdef} the optimal axis aligns with
the vector sum of the momenta in each hemisphere, so the candidates are labeled by the
partition of $\{p_1,\dots,p_4\}$ into two hemispheres --- three partons against one ($3{+}1$) or
two against two ($2{+}2$). For a partition $J$ the squared thrust is
$T_J^2\equiv4\,\big|\!\sum_{j\in J}\vec p_j\big|^2$ and the thrust is $T=\max_J T_J$. Because the
total three-momentum vanishes, a hemisphere and its complement share the same axis,
$T_J^2=T_{\bar J}^2$, leaving seven independent candidates. The four $3{+}1$ partitions are
\begin{align}
T_1^2&=T_{234}^2=(1-s_{234})^2\,, &
T_2^2&=T_{134}^2=\left[\frac{2\omega(1+s_{234}-2\omega)-s_{34}}{2\omega}\right]^2\,,\nn\\
T_3^2&=T_{124}^2=\left[\frac{4z\omega^2+s_{34}(1-z)}{2\omega}\right]^2\,, &
T_4^2&=T_{123}^2=\left[\frac{4(1-z)\omega^2+s_{34}z}{2\omega}\right]^2\,,
\end{align}
and the three $2{+}2$ partitions are
\begin{align}
T_{12}^2&=T_{34}^2=\left(\frac{4\omega^2-s_{34}}{2\omega}\right)^2\,,\qquad
T_{14}^2=T_{23}^2=\left[\frac{s_{34}z-2\omega\big(1+s_{234}-2\omega(1-z)\big)}{2\omega}\right]^2-4s_{23}\,,\nn\\
T_{13}^2&=T_{24}^2=\frac{1}{4\omega^2}\Big[s_{34}^2(1-z)^2
+4\omega^2\big(4s_{23}+s_{234}^2+(1-2\omega z)^2-2s_{234}(1+2\omega z)\big)\nn\\
&\hspace{20mm}-4s_{34}\omega\big(1+s_{234}-z-s_{234}z+2\omega(z^2-z-2)\big)\Big]\,.
\end{align}
Which partition is thrust-maximal selects the region feeding each shoulder.
In Fig.~\ref{fig:shoulder_ps}, following Ref.~\cite{Bhattacharya:2022dtm}, we provide two slices of the five-dimensional phase space for illustration.
For the left shoulder in HJM ($\rho<\frac{1}{3}$), the heavy hemisphere can be either $3+1$ case (e.g. $T_{234}$, giving $\rho=s_{234}$) or $2+2$ case (e.g. $T_{12}$, giving $\rho=s_{12}$). The left shoulder logarithms turn out to only come from $2+2$ case, in which the light hemisphere collapses to a narrow jet of mass $s_{34}\simeq Q^2 r$. On the other hand, the right shoulder logs in HJM and shoulder logs in thrust arise from $3+1$ regions.

\subsection{Matrix elements}

\begin{figure}[!htbp]
	\centering
	% --- center: H -> ggg  via the HEFT Hggg vertex ---
	\begin{tikzpicture}[scale=0.85]
		\draw[semithick,color=blue,dashed] (0,0) -- (-1.3,0) node[left]{$H$};
		\draw[semithick,color=darkgreen,decorate,decoration={gluon,amplitude=2.1pt,segment length=2.8pt,aspect=0.6}] (0,0) -- (1.95,1.05);
		\draw[semithick,color=darkgreen,decorate,decoration={gluon,amplitude=2.1pt,segment length=2.8pt,aspect=0.6}] (0,0) -- (1.95,0);
		\draw[semithick,color=darkgreen,decorate,decoration={gluon,amplitude=2.1pt,segment length=2.8pt,aspect=0.6}] (0,0) -- (1.95,-1.05);
		\node at (2.2,1.1) {$g$}; \node at (2.25,0) {$g$}; \node at (2.2,-1.1) {$g$};
	\end{tikzpicture}
	\hspace{0.8cm}
	% --- right: H -> g q qbar  via the HEFT Hgg vertex (one gluon splits to q qbar) ---
	\begin{tikzpicture}[scale=0.85]
		\draw[semithick,color=blue,dashed] (0,0) -- (-1.3,0) node[left]{$H$};
		\draw[semithick,color=darkgreen,decorate,decoration={gluon,amplitude=2.1pt,segment length=2.8pt,aspect=0.6}] (0,0) -- (1.4,0.95);
		\draw[semithick,color=darkgreen,decorate,decoration={gluon,amplitude=2.1pt,segment length=2.8pt,aspect=0.6}] (0,0) -- (1.0,-0.6);
		\draw[semithick,->-=0.55] (2.0,-1.2) -- (1.0,-0.6);
		\draw[semithick,->-=0.55] (1.0,-0.6) -- (2.0,-0.2);
		\node at (1.65,1.05) {$g$}; \node at (2.25,-0.2) {$q$}; \node at (2.25,-1.2) {$\bar q$};
	\end{tikzpicture}
	\hspace{0.8cm}
	% --- left: H -> q qbar g  via the Hbb Yukawa vertex (one quark radiates a gluon) ---
	\begin{tikzpicture}[scale=0.85]
		\draw[semithick,color=blue,dashed] (0,0) -- (-1.3,0) node[left]{$H$};
		\draw[semithick,->-=0.55] (1.95,-1.05) -- (0,0);
		\draw[semithick,->-=0.55] (0,0) -- (0.9,0.5);
		\draw[semithick,->-=0.55] (0.9,0.5) -- (1.95,1.05);
		\draw[semithick,color=darkgreen,decorate,decoration={gluon,amplitude=2.1pt,segment length=2.8pt,aspect=0.6}] (0.9,0.5) -- (1.95,0.05);
		\node at (2.2,1.1) {$q$}; \node at (2.2,0.05) {$g$}; \node at (2.2,-1.1) {$\bar q$};
	\end{tikzpicture}
	\caption{Representative LO diagrams: $H\to ggg$ (left) and $H\to g q\bar q$ (center), contributing to $Hgg$;
		 $H\to q\bar q g$ (right), contributing to $Hq\bar{q}$.}
	\label{fig:diagrams_LO}
\end{figure}
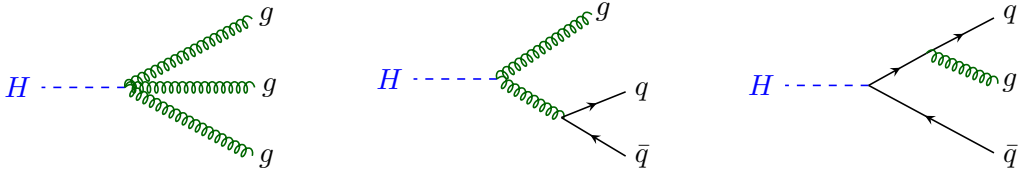

At leading order, the trijet final states are 
\begin{equation}
	H\to ggg\,,\quad H\to g q\bar q\,,
\end{equation}
for $Hgg$ process and
\begin{equation}
H\to q\bar q g\,,
\end{equation}
for $Hq\bar{q}$ process. Example diagrams are drawn in Fig.~\ref{fig:diagrams_LO}. The gluon-initiated decay $Hgg$ receives
contributions from two distinct topologies: $H\to ggg$ and $H\to g q\bar q$, where the latter case comes from a gluon splitting to a quark pair. On the other hand, $Hq\bar{q}$ contains a single topology, $H\to q\bar q g$.
To avoid potential confusion, we will always use $Hggg$ and $Hgq\bar{q}$ to refer to two distinct channels in $Hgg$ process and $Hq\bar{q}g$ with a different label order for the single channel in $Hq\bar{q}$ process.

At LO, it is straightforward to evaluate the phase space integral for HJM analytically. We find
\begin{align}
\frac{1}{\sigma_0}\frac{d\sigma^{\rm LO}_{Hgg}}{d\rho}
&=\frac{\alpha_s}{4\pi}\,\frac{2}{3\,\rho(1-\rho)}\bigg\{(1-4\rho+3\rho^2)\Big[C_A\big(-11+24\rho-15\rho^2\big)+n_f\big(2-15\rho+15\rho^2\big)\Big]\nn\\
&\hspace{3.4em}+6\Big[2C_A\big(1-\rho+\rho^2\big)^2+n_f\,\rho\big(1-3\rho+4\rho^2-2\rho^3\big)\Big]\ln\frac{1-2\rho}{\rho}\bigg\}\,,\\[5pt]
\frac{1}{\sigma_0}\frac{d\sigma^{\rm LO}_{Hq\bar q}}{d\rho}
&=\frac{\alpha_s}{4\pi}\,\frac{2\,C_F}{\rho(1-\rho)}\Big[3(1-\rho)^2(3\rho-1)+2\big(2-3\rho+3\rho^2\big)\ln\frac{1-2\rho}{\rho}\Big]\,,
\label{eq:LO_HJM}
\end{align}
and at this order, thrust and HJM are identical. Expanding at $\rho\to\frac{1}{3}$, we obtain
\begin{align}
	\frac{1}{\sigma_0}\frac{d\sigma^{\rm LO}}{d\rho}\bigg|_{\rho\to\frac13^-}
	&=\frac{\alpha_s}{4\pi}\left(\frac13-\rho\right)\times
	\begin{cases}
		\,168\,C_A+12\,n_f\,, & H\to gg\,,\\[2pt]
		\,180\,C_F\,, & H\to q\bar q\,,
	\end{cases}
	\label{eq:LOkink}
\end{align}
which is linear in the distance to $1/3$ boundary. The LO distribution has a discontinuous slope (a kink) but no logarithm. For later convenience, we will define the following notations for LO coefficients
\begin{equation}\label{eq:sigmaLO_value}
	\sigma_{\text{LO}, Hggg}= \frac{\alpha_s}{4\pi} \,56 \,C_A\,,\qquad
	\sigma_{\text{LO}, Hgq\bar{q}}= \frac{\alpha_s}{4\pi} \,4 \,n_f\,,\qquad
	\sigma_{\text{LO}, Hq\bar{q}g}= \frac{\alpha_s}{4\pi} \,60 \,C_F\,.
\end{equation}

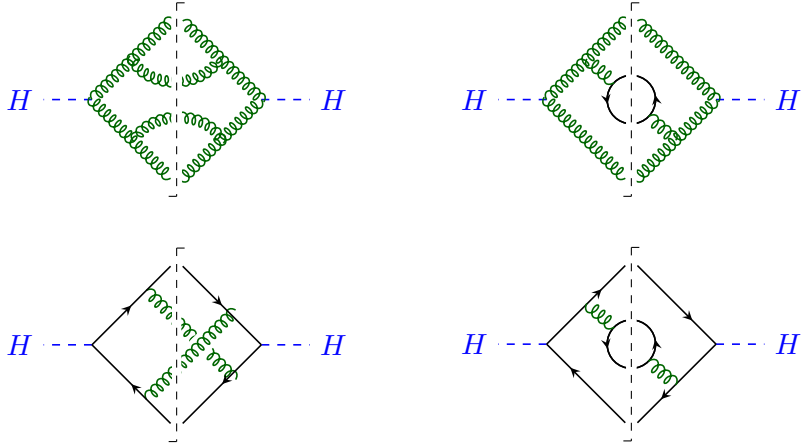
\begin{figure}[!hbtp]
	\centering
	% ===== top row: gluon channel H->gg =====
	% (a) gg loop dressed by two gluon exchanges  (C_A^2)
	\begin{tikzpicture}[scale=0.8]
		\draw [semithick][color=blue][dashed](-1.4,.0)--(-2.2,.0)node[left]{$H$};
		\draw [semithick][color=blue][dashed](1.4,.0)--(2.2,.0)node[right]{$H$};
		\draw[semithick,color=darkgreen,decorate,decoration={gluon, amplitude=2.1pt,segment length=2.8pt, aspect=0.6}](-1.4,-0)-- (-0.1,1.3);
		\draw[semithick,color=darkgreen,decorate,decoration={gluon, amplitude=2.1pt,segment length=2.8pt, aspect=0.6}](0.1,1.3)-- (1.4,0);
		\draw[semithick,color=darkgreen,decorate,decoration={gluon, amplitude=2.1pt,segment length=2.8pt, aspect=0.6}](1.4,-0)-- (0.1,-1.3);
		\draw[semithick,color=darkgreen,decorate,decoration={gluon, amplitude=2.1pt,segment length=2.8pt, aspect=0.6}](-0.1,-1.3)-- (-1.4,0);
		\draw[semithick,color=darkgreen,decorate,decoration={gluon, amplitude=2.1pt,segment length=2.8pt, aspect=0.6}] (0.7,0.7) to [bend left=61.5] (-0.7,0.63);
		\draw[semithick,color=darkgreen,decorate,decoration={gluon, amplitude=2.1pt,segment length=2.8pt, aspect=0.6}] (-0.7,-0.7) to [bend left=61.5] (0.7,-0.63);
		\draw[white] (0,1.6) edge [-,line width =4pt,draw=white] (0,-1.6) edge (0,-1.6);
		\draw[dashed] (0,1.6)-- (0,-1.6);
		\draw[dashed] (0,1.6) -- (0.2,1.6);
		\draw[dashed] (0,-1.6) -- (-0.2,-1.6);
%		\node at (0,-2.0) {(a) $H\to gg$};
	\end{tikzpicture}
	\hspace{1.0cm}
	% (b) gg loop with an internal quark loop  (n_f)
	\begin{tikzpicture}[scale=0.8]
		\draw [semithick][color=blue][dashed](-1.4,.0)--(-2.2,.0)node[left]{$H$};
		\draw [semithick][color=blue][dashed](1.4,.0)--(2.2,.0)node[right]{$H$};
		\draw[semithick,color=darkgreen,decorate,decoration={gluon, amplitude=2.1pt,segment length=2.8pt, aspect=0.6}](-1.4,-0)-- (-0.1,1.3);
		\draw[semithick,color=darkgreen,decorate,decoration={gluon, amplitude=2.1pt,segment length=2.8pt, aspect=0.6}](0.1,1.3)-- (1.4,0);
		\draw[semithick,color=darkgreen,decorate,decoration={gluon, amplitude=2.1pt,segment length=2.8pt, aspect=0.6}](1.4,-0)-- (0.1,-1.3);
		\draw[semithick,color=darkgreen,decorate,decoration={gluon, amplitude=2.1pt,segment length=2.8pt, aspect=0.6}](-0.1,-1.3)-- (-1.4,0);
		\draw [->-=0.03,semithick] (0,0) circle (0.4);
		\draw [->-=0.53,semithick] (0,0) circle (0.4);
		\draw[semithick,color=darkgreen,decorate,decoration={gluon, amplitude=2.3pt,segment length=3.1pt, aspect=0.6}] (-0.282843,0.282843) to [bend left=0] (-0.685,0.63);
		\draw[semithick,color=darkgreen,decorate,decoration={gluon, amplitude=2.3pt,segment length=3.1pt, aspect=0.6}] (0.282843,-0.282843) to [bend left=0] (0.685,-0.63);
		\draw[white] (0,1.6) edge [-,line width =4pt,draw=white] (0,-1.6) edge (0,-1.6);
		\draw[dashed] (0,1.6)-- (0,-1.6);
		\draw[dashed] (0,1.6) -- (0.2,1.6);
		\draw[dashed] (0,-1.6) -- (-0.2,-1.6);
%		\node at (0,-2.0) {(b) $H\to gg$, $n_f$};
	\end{tikzpicture}
	\\[0.5cm]
	% ===== bottom row: quark channel H->bb =====
	% (c) bb loop dressed by two crossed gluon exchanges  (C_F^2, C_F C_A)
	\begin{tikzpicture}[scale=0.8]
		\draw [semithick][color=blue][dashed](-1.4,.0)--(-2.2,.0)node[left]{$H$};
		\draw [semithick][color=blue][dashed](1.4,.0)--(2.2,.0)node[right]{$H$};
		\draw[semithick,color=darkgreen,decorate,decoration={gluon, amplitude=2.1pt,segment length=3.1pt, aspect=0.6}] (-0.5,0.9) to [bend left=0] (0.9,-0.5);
		\draw[white] (-0.5,-0.9) edge [-,line width =8pt,draw=white] (0.9,0.5) edge (0.9,0.5);
		\draw[semithick,color=darkgreen,decorate,decoration={gluon, amplitude=2.1pt,segment length=3.1pt, aspect=0.6}] (-0.5,-0.9) to [bend left=0] (0.90,0.5);
		\draw[->-=0.5,semithick](-1.4,-0)-- (-0.1,1.3);
		\draw[->-=0.5,semithick](0.1,1.3)-- (1.4,0);
		\draw[->-=0.5,semithick](1.4,-0)-- (0.1,-1.3);
		\draw[->-=0.5,semithick](-0.1,-1.3)-- (-1.4,0);
		\draw[white] (0,1.6) edge [-,line width =3.5pt,draw=white] (0,-1.6) edge (0,-1.6);
		\draw[dashed] (0,1.6) --(0,-1.6);
		\draw[dashed] (0,1.6) -- (0.2,1.6);
		\draw[dashed] (0,-1.6) -- (-0.2,-1.6);
%		\node at (0,-2.0) {(c) $H\to q\bar q$};
	\end{tikzpicture}
	\hspace{1.0cm}
	% (d) bb loop with a secondary quark loop  (n_f)
	\begin{tikzpicture}[scale=0.8]
		\draw [semithick][color=blue][dashed](-1.4,.0)--(-2.2,.0)node[left]{$H$};
		\draw [semithick][color=blue][dashed](1.4,.0)--(2.2,.0)node[right]{$H$};
		\draw[->-=0.7,semithick](-1.4,-0)-- (-0.1,1.3);
		\draw[->-=0.7,semithick](0.1,1.3)-- (1.4,0);
		\draw[->-=0.7,semithick](1.4,-0)-- (0.1,-1.3);
		\draw[->-=0.7,semithick](-0.1,-1.3)-- (-1.4,0);
		\draw [->-=0.03,semithick] (0,0) circle (0.4);
		\draw [->-=0.53,semithick] (0,0) circle (0.4);
		\draw[semithick,color=darkgreen,decorate,decoration={gluon, amplitude=2.3pt,segment length=3.1pt, aspect=0.6}] (-0.282843,0.282843) to [bend left=0] (-0.685,0.63);
		\draw[semithick,color=darkgreen,decorate,decoration={gluon, amplitude=2.3pt,segment length=3.1pt, aspect=0.6}] (0.282843,-0.282843) to [bend left=0] (0.685,-0.63);
		\draw[white] (0,1.6) edge [-,line width =4pt,draw=white] (0,-1.6) edge (0,-1.6);
		\draw[dashed] (0,1.6)-- (0,-1.6);
		\draw[dashed] (0,1.6) -- (0.2,1.6);
		\draw[dashed] (0,-1.6) -- (-0.2,-1.6);
%		\node at (0,-2.0) {(d) $H\to q\bar q$, $n_f$};
	\end{tikzpicture}
	\caption{Representative NLO cut diagrams for the two channels. The vertical dashed line is the final-state cut. The top panel shows the gluon channel $H\to gg$ and the bottom panel shows the quark channel $H\to q\bar q$.}
	\label{fig:diagrams_NLO}
\end{figure}

At NLO, the HJM distributions receive contribution from both $H\to 3$ partons at one-loop and $H\to 4$ partons at tree-level. Since we are only interested in the leading power expansion near $\rho=\frac{1}{3}$, the phase space parameterization and sector decomposition discussed in the previous section allow us to derive the logs purely from tree-level $H\to 4$ partons. For $Hgg$ process, we have the following subprocesses,
\begin{align}\label{eq:Hgg_process}
	H&\to g(p_1)\, g(p_2)\, g(p_3)\, g(p_4)\,,\nn\\
	H&\to g(p_1)\, g(p_2)\, q(p_3)\, \bar q(p_4)\,,\nn\\
	H&\to q(p_1)\, \bar q(p_2)\, q(p_3)\, \bar q(p_4)\,,
\end{align}
where the last one includes both non-identical and identical quark contributions.
The $Hq\bar{q}$ process contains
\begin{align}\label{eq:Hqq_process}
	H&\to q(p_1)\, \bar q(p_2)\, g(p_3)\, g(p_4) \,,\nn\\
	H&\to q(p_1)\, \bar q(p_2)\, q(p_3)\, \bar q(p_4) \,. 
\end{align}
In Fig.~\ref{fig:diagrams_NLO}, we show some example diagrams at NLO. To obtain the matrix elements, we use \textsc{Qgraf}~\cite{NOGUEIRA1993279} and \textsc{Form}~\cite{Vermaseren:2000nd,Kuipers:2012rf,Ruijl:2017dtg}, and evaluate the color algebra with \textsc{Color} package~\cite{vanRitbergen:1998pn}. Regarding the polarization in both processes, we choose the axial gauge for gluon polarization,
\begin{equation}
	\sum_{\lambda=1}^2 \epsilon^\mu (p_i,\lambda)\epsilon^{\ast \nu}(p_i,\lambda)=-g^{\mu\nu} +\frac{\bar q^\mu p_i^\nu+\bar q^\nu p_i^\mu}{\bar q\cdot p_i}-\frac{\bar q^2 p_i^\mu p_i^\nu}{(p_i\cdot \bar q)^2}\,, 
\end{equation}
where $\bar{q}$ is a reference vector. In practice, we choose $\bar{q}$ to be $p_j$ with $i\neq j$ during the summation of the gluon polarization with momentum $p_i$. Eventually, all the squared matrix elements are expressed in terms of invariant quantities $s_{ij}$.

\subsection{Leading-power expansion}

Using the matrix elements obtained in the previous section, we can expand them to leading power with the scalings in Eq.~\eqref{eq:collscaling}, separating the collinear (C), soft (S), and soft-collinear (SC) regions. 
We illustrate the procedure on one representative subprocess, $H\to q\bar q gg$ in Eq.~\eqref{eq:Hqq_process}; the others are treated in similar ways.

\paragraph{Singular limits:} With the light jet formed by $(p_3,p_4)$, the regions of
$H\to q\bar qgg$ that generate shoulder logarithms are
\begin{itemize}
	\item[(C)] $s_{34}\to0$: the light-jet pair becomes collinear, realized either
	as the quark--gluon splitting $q\to qg$ or the gluon splitting $g\to gg$;
	\item[(S)] $z\to0$: one of the two final-state gluons becomes soft;
	\item[(SC)] $s_{34}\to0$ and $z\to0$ together: the gluon is simultaneously
	soft and collinear --- the overlap of (C) and (S).
\end{itemize}

\paragraph{Expanded matrix elements:} Using the variables in Sec.~\ref{sec:nlo_kin}
($x=\omega-\frac13$, $y=s_{234}-\frac13$, $s_{34}$, $z$ and $\phi$), the leading power expansions in these limits can be written as
\begin{align}
	|\mathcal M|^2_{\rm C}\Big|_{C_F^2}
	&=\frac{128\,\pi^2\alpha_s^2\,C_F^2}{s_{34}\,z}\,(5+36x-18y)(2-2z+z^2)\,,
	\label{eq:exp_C}\\[2pt]
	|\mathcal M|^2_{\rm S}
	&=\frac{5120\,\pi^2\alpha_s^2\,C_F\big[\,6C_A s_{34}+3C_F s_{34}+4C_F z-4\sqrt3\,(C_A-C_F)\sqrt{s_{34}z}\,\cos\phi\,\big]}
	{s_{34}\,\big(9s_{34}^2+16z^2-24\,s_{34}z\cos2\phi\big)}\,,
	\label{eq:exp_S}\\[2pt]
	|\mathcal M|^2_{\rm SC}\Big|_{C_F^2}
	&=\frac{256\,\pi^2\alpha_s^2\,C_F^2}{s_{34}\,z}\,(5+36x-18y)\,.
	\label{eq:exp_SC}
\end{align}
Eq.~\eqref{eq:exp_C} is the $q\to qg$ collinear splitting ($C_F^2$), and the one from $g\to gg$ collinear splitting ($C_F C_A$) can be obtained with similar ways. Eq.~\eqref{eq:exp_S} is the leading soft expansion, and it contains $\cos\phi$ dependence. 
Eq.~\eqref{eq:exp_SC} is the soft ($z\to0$) limit of Eq.~\eqref{eq:exp_C}, equivalently the collinear ($s_{34}\ll z$) limit of Eq.~\eqref{eq:exp_S}. 
We automate these expansions and apply them to all matrix elements.

\paragraph{Phase-space integration:} At leading power, the shoulder contribution can be written as the non-overlapping sum of soft and collinear integrals, i.e.
\begin{equation}
	\frac{1}{\sigma_0}\frac{d\sigma}{dr}=\mathcal S_{\rm C}+\mathcal S_{\rm S}-\mathcal S_{\rm SC}\,,
	\qquad
	\mathcal S_{\rm X}=\int\! d\text{PS}\;|\mathcal M|^2_{\rm X}\,,
\end{equation}
$|\mathcal{M}|^2_{\rm X}$ is the power-expanded matrix elements with $\rm{X}=\rm{S}, \rm{C}, \rm{SC}$. The phase space restricted to $2+2$ or $3+1$ regions from the thrust axis takes the form in Eq.~\eqref{eq:master}. Then the remaining task is to perform the integrations over $s_{234}, s_{34},\omega, z, s_{23}$, with one of them fixed by the delta function. Note that the azimuthal $\cos\phi$ and $\cos2\phi$
spin-correlation terms average to zero in the $\ln^2 r$ and $\ln r$ coefficients.

\subsection{Results}
\label{sec:NLOResults}

Finally, we list the analytic expressions of shoulder logarithms of thrust and HJM in both decay processes. We use the following notations: $r=\frac13-\rho$ (for HJM left shoulder), $s=\rho-\frac13$ (for HJM right shoulder) and $t=\tau-\frac13$. We omit the LO terms, i.e. $\mathcal{O}(\alpha_s)$ and only show the pure NLO expressions.

For $Hgg$ process, we find HJM left and right shoulders,
\begin{align}
	\frac{1}{\sigma_0}\frac{d\sigma_{Hgg}}{d\rho}\bigg|_{\rho\to\frac13^-}
	&= \left(\frac{\alpha_s}{4\pi}\right)^2\,r\Bigg\{C_A^2\bigg[-\!336\ln^2 r+\left(56+1344\ln2-672\ln3\right)\ln r\bigg]\nn\\
	&\quad+C_A n_f\bigg[-8\ln^2 r+\left(\frac{340}{3}+32\ln2-16\ln3\right)\ln r\bigg]\nn\\
	&\quad+C_F n_f\bigg[-16\ln^2 r+\left(8+64\ln2-32\ln3\right)\ln r\bigg]+\frac83 n_f^2\ln r\,\Bigg\}\,,\\
	\frac{1}{\sigma_0}\frac{d\sigma_{Hgg}}{d\rho}\bigg|_{\rho\to\frac13^+}
	&= \left(\frac{\alpha_s}{4\pi}\right)^2\,s\Bigg\{\,C_A^2\bigg[-672 \ln^2 s+\left(112-1344\ln6\right)\ln s\bigg]\nn\\
	&\quad+C_A n_f\bigg[-16\ln^2 s+\left(\frac{680}{3}-32\ln6\right)\ln s\bigg]\nn\\
	&\quad+C_F n_f\bigg[-32\ln^2 s+\left(16-64\ln6\right)\ln s\bigg]+\frac{16}3 n_f^2\ln s\,\Bigg\}\,,
\end{align}
and thrust right shoulder,
\begin{align}
	\frac{1}{\sigma_0}\frac{d\sigma_{Hgg}}{d\tau}\bigg|_{\tau\to\frac13^+}
	&= \left(\frac{\alpha_s}{4\pi}\right)^2\,t\Bigg\{C_A^2\bigg[-1008\ln^2 t+(168-2016\ln3)\ln t\bigg]\nn\\
	&\quad+C_A n_f\bigg[-24\ln^2 t+(340-48\ln3)\ln t\bigg]\nn\\
	&\quad+C_F n_f\bigg[-48\ln^2 t+(24-96\ln3)\ln t\bigg]+8\,n_f^2\ln t\,\Bigg\}\,.
\end{align}

For $Hq\bar{q}$ process, we find HJM left and right shoulders,
\begin{align}
	\frac{1}{\sigma_0}\frac{d\sigma_{Hq\bar q}}{d\rho}\bigg|_{\rho\to\frac13^-}
	&= \left(\frac{\alpha_s}{4\pi}\right)^2\,r\Bigg\{C_F^2\bigg[-240\ln^2 r+\left(120+960\ln2-480\ln3\right)\ln r\bigg]\nn\\
	&\quad+C_AC_F\bigg[-120\ln^2 r+\left(20+480\ln2-240\ln3\right)\ln r\bigg]+40\,C_F n_f\ln r\,\Bigg\}\,,\\
	\frac{1}{\sigma_0}\frac{d\sigma_{Hq\bar q}}{d\rho}\bigg|_{\rho\to\frac13^+}
	&= \left(\frac{\alpha_s}{4\pi}\right)^2\,s\Bigg\{\,C_F^2\bigg[-480\ln^2 s+\left(240-960\ln6\right)\ln s\bigg]\nn\\
	&\quad+C_AC_F\bigg[-240\ln^2 s+\left(40-480\ln6\right)\ln s\bigg]+80\,C_F n_f\ln s\,\Bigg\}\,,
\end{align}
and thrust right shoulder,
\begin{align}
	\frac{1}{\sigma_0}\frac{d\sigma_{Hq\bar q}}{d\tau}\bigg|_{\tau\to\frac13^+}
	&= \left(\frac{\alpha_s}{4\pi}\right)^2\,t\Bigg\{\,C_F^2\bigg[-720\ln^2 t+(360-1440\ln3)\ln t\bigg]\nn\\
	&\quad+C_AC_F\bigg[-360\ln^2 t+(60-720\ln3)\ln t\bigg]+120\,C_F n_f\ln t\,\Bigg\}\,.
\end{align}

These analytic results will serve as non-trivial checks for the shoulder factorization theorem discussed in the next section.

\section{Sudakov shoulder resummation}
\label{sec:shoulder}

Now we turn to study the Sudakov shoulder logarithms for HJM in the SCET framework. The shoulder factorization formula is derived in Ref.~\cite{Bhattacharya:2022dtm} and applied to $\gamma^\star\to q\bar{q}$ in Ref.~\cite{Bhattacharya:2023qet}. In the following, we extend it to both $Hgg$ and $Hq\bar{q}$.

\subsection{Shoulder factorization}
\label{sec:shoulder_factorization}

In this subsection, we discuss the trijet kinematics around $\rho=\frac{1}{3}$ further and review the shoulder factorization theorem.
At LO, the edge $\rho=\frac13$ is reached only at the symmetric
configuration, where three massless partons of energy $Q/3$ are separated by $2\pi/3$, two clustered in the heavy hemisphere and one in the light hemisphere. Collinear splittings turn the three partons into narrow jets of invariant masses $m_i^2$, while soft radiation is emitted at wide angles between them. Near the shoulder regime, the thrust
axis aligns with one of the jets. Taking jet $1$ to lie in the light hemisphere and jets $2,3$ in
the heavy one, the configuration has $\rho<\frac13$ as long as
\begin{equation}
	m_1^2 < r + m_2^2 + m_3^2\,,\qquad r=\frac13-\rho\,,
	\label{eq:sh_axisbound}
\end{equation}
and the thrust axis flips to a different parton once Eq.~\eqref{eq:sh_axisbound} is violated.
In general, multiple collinear emissions in each hemisphere combine into effective light- and heavy-hemisphere
mass-shift variables, $m_\ell^2\simeq m_1^2$ and $m_h^2\simeq m_2^2+m_3^2$ and the bound is upgraded to
\begin{equation}
	m_\ell^2 < r + m_h^2\,.
\end{equation}
Similar arguments can be drawn for soft emissions.
Note that this inequality only indicates the difference of hemisphere masses is $\mathcal{O}(\lambda)$; while the derivation implicitly assumes both masses are small, $\mathcal{O}(\lambda)$. On the other hand, hard configurations that reach $\rho\simeq 1/3$, where $m_\ell,m_h\sim\mathcal{O}(1)$, can also have very small difference. However, they contribute equally for $r>0$
and $r<0$ and vary smoothly across the edge, so they can not produce a logarithm of $r$.

Based on the shoulder kinematics, the factorized cross-section at $\rho\simeq 1/3$ can be written in terms of the double differential hemisphere mass distribution,
\begin{equation}
	\frac{1}{\sigma_0}\frac{d\sigma_\text{sh}}{d\rho}=\int_0^\infty\! dm_h^2\int_0^\infty\! dm_\ell^2\,
	\frac{d^2\sigma_\text{sh}}{dm_\ell^2\,dm_h^2}\,
	\big(r+m_h^2-m_\ell^2\big)\,\theta\big(r+m_h^2-m_\ell^2\big)\,,
	\label{eq:sh_massshift}
\end{equation}
which holds on both sides of the edge ($r>0$ on the left shoulder, $r<0$ on the right), and
similarly for thrust with $r+m_h^2-m_\ell^2\to m_h^2+m_\ell^2-t$ and $t=\tau-\frac13$. The single
power of $(r+m_h^2-m_\ell^2)$ is the remnant of the linear phase space we observed in the fixed-order calculations. As derived in Refs.~\cite{Bhattacharya:2022dtm,Bhattacharya:2023qet}, the double differential cross-section can be factorized as
\begin{multline}
	\frac{d^2\sigma_\text{sh}}{dm_\ell^2\,dm_h^2}=\sigma_{\text{LO}}\,H^{\text{sh}}(Q,\mu)
	\int\!\Big[\prod_{i=1}^3 dm_i^2\,J_i(m_i^2,\mu)\Big]\,dk_\ell\,dk_h\,S^{\text{sh}}(k_\ell,k_h,\mu)\\
	\times\,\delta\!\big(m_\ell^2-m_1^2-k_\ell\big)\,
	\delta\!\big(m_h^2-m_2^2-m_3^2-k_h\big)\,,
	\label{eq:sh_fact}
\end{multline}
where $J_i$ is the inclusive quark or gluon jet function of jet $i$, $S^\text{sh}(k_\ell,k_h)$ is the trijet
hemisphere soft function for the soft radiation projected onto the light and heavy hemispheres
(with $k_\ell,k_h$ its contributions to the two hemisphere masses), and the two $\delta$-functions
decompose each hemisphere mass-shift into its collinear ($m_i^2$) and soft ($k_{\ell,h}$) pieces.
The hard function $H^{\text{sh}}(Q,\mu)$ encodes the virtual corrections to trijet production at the symmetric
point, obtained by matching the relevant $H\to 3$ partons currents onto the SCET
operators at the hard scale $\mu_h\sim Q$.

The decay-channel dependence resides entirely in the hard function and in the color
representation of the Wilson lines of the soft function; the jet functions are
process-independent. In the $Hgg$ process, we have two subprocesses, $Hgq\bar q$ and $H ggg$. For $Hgq\bar q$, we can have either a gluon or quark in the light hemisphere, and it will change the soft function expression. For convenience, we will denote the former as  $H{\color{blue}g}{\color{red} q} {\color{red} \bar q}$ and the latter as $H{\color{red}g}{\color{blue} q} {\color{red} \bar q}$ and $H{\color{red}g}{\color{red} q} {\color{blue} \bar q}$, with the blue color for light hemisphere and red color for heavy hemisphere. Putting them together, we will have the $Hgg$ distribution as a sum over all different channels,
\begin{equation}
	\frac{d\sigma_{Hgg}}{d\rho}=\frac{d\sigma_{H{\color{blue}g}{\color{red} q} {\color{red} \bar q}}}{d\rho}+2\,\frac{d\sigma_{H{\color{red}g}{\color{blue} q} {\color{red} \bar q}}}{d\rho}\,+3 \frac{d\sigma_{H{\color{blue}g}{\color{red}g}{\color{red}g}}}{d\rho}\,.
	\label{eq:sh_channels_Hgg}
\end{equation}
Similarly, for $Hq\bar{q}$ channel, we can have either quark or gluon in the light hemisphere, namely $H{\color{blue}q}{\color{red} \bar q} {\color{red} g}$ and $H{\color{red}q}{\color{red} \bar q} {\color{blue} g}$.
In addition, it also receives the loop-induced contribution from $H\to ggg$, though it is suppressed by $\mathcal{O}(\alpha_s^2)$. Explicitly, we write the $Hq\bar{q}$ distribution as
\begin{equation}
	\frac{d\sigma_{Hq\bar q}}{d\rho}=2\frac{d\sigma_{H{\color{blue}q}{\color{red} \bar q} {\color{red} g}}}{d\rho}+\frac{d\sigma_{H{\color{red}q}{\color{red} \bar q} {\color{blue} g}}}{d\rho}\,+3 \frac{d\sigma^{\text{loop}}_{H{\color{blue}g}{\color{red}g}{\color{red}g}}}{d\rho}\,,
	\label{eq:sh_channels_Hqq}
\end{equation}
In this work, we will neglect the loop-induced contribution since the hard function is suppressed by $\mathcal{O}(\alpha_s^2)$.

Regarding the resummation, one should apply Eq.~\eqref{eq:sh_fact} to all six individual distributions, with proper hard, jet and soft functions. In particular, $\sigma_\text{LO}$ will take the values in Eq.~\eqref{eq:sigmaLO_value}. Similar to the dijet case, the total decay rate contains renormalization scale $\mu$ and we convert it to $Q$ as in Eq.~\eqref{eq:lambda_res} and~\eqref{eq:yukawa_res}. For simplicity, we absorb the resulting kernel and boundary into those of hard functions, and from now on, we do not write them explicitly.

\subsection{Ingredients}
\label{sec:ingredients}

\subsubsection{Trijet hard functions}

The shoulder hard function is the symmetric-point value of a more general trijet hard function $H_i(u,v,w)$, the squared IR-finite amplitude to produce three energetic partons,
written in terms of the normalized pairwise invariants
\begin{equation}
	u=\frac{s_{23}}{Q^2}\,,\qquad v=\frac{s_{13}}{Q^2}\,,\qquad w=\frac{s_{12}}{Q^2}\,,\qquad u+v+w=1\,,
\end{equation}
with $s_{jk}=(p_j+p_k)^2$. The shoulder hard function is evaluated at $u=v=w=1/3$.
The trijet hard function is obtained by matching the UV-renormalized QCD
amplitude of the relevant $1\to3$ current onto the SCET trijet operators.
In all individual channels from $Hgg$ and $Hq\bar{q}$ processes, the hard functions satisfy the following RGE,
\begin{align}
	\frac{d\ln H_i}{d\ln\mu^2}
	&=-\Gamma^{\rm cusp}(\alpha_s)\sum_{1\le j<k\le3}\mathbf T_j\!\cdot\!\mathbf T_k\,\ln\frac{s_{jk}}{\mu^2}
	+\sum_{j=1}^{3}\gamma^{j}(\alpha_s)\nn\\
	&\quad+\gamma_{\rm quad}(\alpha_s)\;f^{abe}f^{cde}\sum_{l=1}^{3}\sum_{\substack{1\le j<k\le3\\ j,k\neq l}}
	\big\{\mathbf T_l^{a},\mathbf T_l^{d}\big\}\,\mathbf T_j^{b}\,\mathbf T_k^{c}+\mathcal{O}(\alpha_s^4)\,.
	\label{eq:sh_hard_rge}
\end{align}
with $i=q,g$ on the LHS corresponding to $\{q,\bar{q},g\}$ or $\{g,g,g\}$ trijet final-state. Note that $\{q,\bar{q},g\}$ in either $Hgg$ or $Hq\bar{q}$ shares the same RG structure, but different constants.
The first line is the color-dipole contribution. $\mathbf T_i$ is the color-charge operator of parton $i$ and $\gamma^{j}$ its quark or gluon collinear anomalous dimension. 
Here we use the notation $\Gamma^{\rm cusp}_i=C_i \Gamma^{\rm cusp}$, with $C_q=C_F$ and $C_g=C_A$ to make the color explicit. But we emphasize again that at four loops, the Casimir scaling is broken and one should read the expression below in terms of $\Gamma^{\rm cusp}_i$.
The second line is the quadrupole term~\cite{Almelid:2015jia,Almelid:2017qju}, built from four color charges with the structure
constants $f^{abe}f^{cde}$, which first appears at three
loops with~\cite{Gao:2024wcg}
\begin{equation}
	\gamma_{\rm quad}(\alpha_s)=\Big(\frac{\alpha_s}{4\pi}\Big)^{3}\,16\,(\zeta_5+2\zeta_2\zeta_3)+\mathcal O(\alpha_s^4)\,.
	\label{eq:sh_gammaquad}
\end{equation}
For a configuration of only three lightlike directions one cannot build a nontrivial conformal cross ratio, so the quadrupole carries no kinematic dependence and reduces to a constant.

Color conservation $\sum_j\mathbf T_j=0$ gives for $\{q,\bar{q},g\}$ trijet, $\mathbf T_q\!\cdot\!\mathbf T_g=\mathbf T_{\bar q}\!\cdot\!\mathbf T_g=-\frac{C_A}{2}$
and $\mathbf T_q\!\cdot\!\mathbf T_{\bar q}=\frac{C_A-2C_F}{2}$, leading to 
\begin{align}
	\frac{d\ln H_{q}}{d\ln\mu^2}
	&=\frac{C_A+2C_F}{2}\,\Gamma^{\rm cusp}\ln\frac{Q^2}{\mu^2}\nn\\
	&\hspace{1cm}+\Gamma^{\rm cusp}\Big[\frac{C_A}{2}\ln(vw)+\frac{2C_F-C_A}{2}\ln u\Big]+2\gamma^q+\gamma^g+\frac{3}{2} C_A\,\gamma_{\rm quad}\,,
\end{align}
while for the all-gluon $\{g,g,g\}$ trijet every dipole is $\mathbf T_j\!\cdot\!\mathbf T_k=-\frac{C_A}{2}$, so
\begin{align}
	\frac{d\ln H_{g}}{d\ln\mu^2}
	&=\frac{3C_A}{2}\,\Gamma^{\rm cusp}\ln\frac{Q^2}{\mu^2}\nn\\
	&\hspace{1cm}+\frac{C_A}{2}\,\Gamma^{\rm cusp}\ln(uvw)+3\gamma^g+9\,C_A\,\gamma_{\rm quad}\,.
\end{align}
We denote the second line of these equations as $\gamma_{H,q}^\text{sh}$ and $\gamma_{H,g}^\text{sh}$ respectively.
Solving the RGEs leads to closed-form solutions in terms of the RG kernels introduced in
Eq.~\eqref{eq:SAfunctions},
\begin{equation}
H_i\big(u,v,w,\mu\big)\,=H_i\big(u,v,w,\mu_h\big)\,
\exp\!\Big[2\,\mathcal C_i\,S_{i}(\mu_h,\mu)-2\,A^\text{sh}_{H,i}(\mu_h,\mu)\Big]
\Big(\frac{Q^2}{\mu_h^2}\Big)^{-\mathcal C_i\,A_{\Gamma}(\mu_h,\mu)}\,,
\label{eq:sh_hard_sol}
\end{equation}
where $\mathcal C_{q\bar q g}=\mathcal C_{g q\bar q}=C_A+2C_F$ and $\mathcal C_{ggg}=3C_A$. The RG solutions with proper hard anomalous dimensions predict the scale-dependent part. 
The constants in the boundary conditions $H_i\big(u,v,w,\mu_h\big)$ are instead process dependent. To obtain the hard boundary up to two-loops, we take the IR-subtracted helicity amplitudes from Refs.~\cite{Ahmed:2014pka,Mondini:2019vub,Gehrmann:2022vuk,Gehrmann:2023etk,Gehrmann:2023zpz}, square them and perform the helicity sums. For $Hggg$ in the $Hgg$ process, we also cross-check the results with Refs.~\cite{Chen:2025utl,Guan:2026igf}. Note that some of the three-loop results are also available~\cite{Gehrmann:2023jyv,Chen:2025utl,Guan:2026igf}.

\subsubsection{Trijet hemisphere soft functions}
\label{sec:trijet_soft_func}

The shoulder soft function is the trijet hemisphere soft function, which is defined as the vacuum matrix element of three soft Wilson lines $Y_{n_1},Y_{n_2},Y_{n_3}$ along the jet directions. 
Explicitly, we have
\begin{equation}
S_{i}^\text{sh}(k_\ell,k_h,\mu)=\frac{1}{\mathcal N_i}\sum_{X_s}
\Big|\big\langle X_s\big|\,\mathcal Y_i\,\big|0\big\rangle\Big|^2\,
\widehat M(k_\ell,k_h)\,,
\label{eq:sh_soft_def}
\end{equation}
where the sum runs over all soft final states $X_s$ and $\widehat M$ is the hemisphere measurement. 
$\mathcal Y_i$ is the color-singlet Wilson-line operator fixed by the trijet color flow, with all open color indices (the external gluon index and the fundamental indices of the quark) summed against the conjugate in the square.

For the quark channel $\{q,\bar q, g\}$ with the gluon jet in the light hemisphere, $\mathcal Y_q$ can be written as (with $n_1$ along the thrust axis pointing into the light hemisphere) 
% \begin{equation}
% \big(\mathcal Y_{q}\big)^a
% =Y^{\dagger}_{n_2}\,Y_{n_1}\,T^{a}\,Y^{\dagger}_{n_1}\,Y_{n_3}\,,
% \qquad \mathcal N_q=N_c\,C_F\,,
% \end{equation}
\begin{equation}
\big(\mathcal Y_{q}\big)^a_{ij}
=\big(Y^{\dagger}_{n_2}\,Y_{n_1}\,T^{a}\,Y^{\dagger}_{n_1}\,Y_{n_3}\big)_{ij}\,,
\qquad \mathcal N_q=N_c\,C_F\,,
\end{equation}
with $T^{a}$ the color generator associated to the gluon in the light hemisphere. The other two channels for $q\bar q g$ follow by permuting which jet occupies the light hemisphere. For the gluon channel $\{g,g,g\}$ all three legs carry adjoint color, and the singlet is the antisymmetric contraction of three adjoint Wilson lines $\mathcal Y^{\rm adj}_{n}$. Writing $\big(\mathcal Y^{\rm adj}_{n}\big)^{ab}\equiv 2\,\mathrm{Tr}\big(T^{a}\,Y_{n}\,T^{b}\,Y^{\dagger}_{n}\big)$, the gluon channel operator is
\begin{equation}
\big(\mathcal Y_{g}\big)^{a'b'c'}=f^{abc}\,\big(\mathcal Y^{\rm adj}_{n_1}\big)^{aa'}
\big(\mathcal Y^{\rm adj}_{n_2}\big)^{bb'}\big(\mathcal Y^{\rm adj}_{n_3}\big)^{cc'}\,,
\qquad \mathcal N_g=f^{abc}f^{abc}=N_c\,N_A\,.
\end{equation}
In both cases, the color indices are summed over in the matrix element square.

The hemisphere measurement can be defined as 
\begin{align}
&\widehat M(k_\ell,k_h)\nn\\
&=\delta\!\left(k_\ell-\sum_{k_i\in|X_s\rangle}\left[\theta_{\mathbf{1}}(k_i)\left(\frac{2}{3}\,n_1\!\cdot\!k_i\right)
+\theta_{\mathbf{\bar 2}}(k_i)\left(\frac{2}{3}\,N_2\!\cdot\!k_i\right)
+\theta_{\mathbf{\bar 3}}(k_i)\left(\frac{2}{3}\,N_3\!\cdot\!k_i\right)\right]\right)\nn\\
&\times\delta\!\left(k_h-\sum_{k_m\in|X_s\rangle}\left[\theta_{\mathbf{\bar 1}}(k_m)\left(\frac{2}{3}\,n_{\bar 1}\!\cdot\!k_m\right)
+\theta_{\mathbf{2}}(k_m)\left(\frac{2}{3}\,n_2\!\cdot\!k_m\right)
+\theta_{\mathbf{3}}(k_m)\left(\frac{2}{3}\,n_3\!\cdot\!k_m\right)\right]\right)\,.
\label{eq:sh_hemi_measure}
\end{align}
Here we use short-hand notations $\theta_{\mathbf{i}}$ ($\theta_{\bar{\mathbf{i}}}$) to project onto the sextant around (opposite) jet $i$. For example, $\theta_{\mathbf{1}}(k)=\theta(n_2\cdot k-\bar{n}_2\cdot k)\theta(n_3\cdot k-\bar{n}_3\cdot k)$ and $\theta_{\bar{\mathbf{1}}}(k)=\theta(\bar{n}_2\cdot k-n_2\cdot k)\theta(\bar{n}_3\cdot k-n_3\cdot k)$.
$N_i$ are some directions between the jets determined by kinematics. For thrust, we have $N_2=\bar{n}_2$ and $N_3=\bar{n}_3$. For HJM, we get $N_2=\bar{n}_3+\frac{1}{2}(n_1-\bar{n}_1)$ and $N_3=\bar{n}_2+\frac{1}{2}(n_1-\bar{n}_1)$. 

The one-loop trijet hemisphere quark soft function is computed in Ref.~\cite{Bhattacharya:2022dtm}, and we can obtain the gluon soft function by Casimir scaling, $C_F\to C_A$. This is sufficient to perform the shoulder resummation to NNLL accuracy. Here let us quote the anomalous dimensions. To define the convention, we write the one-loop renormalized soft function as
\begin{equation}
S_i^\text{sh}(k_\ell,k_h,\mu)=\delta(k_\ell)\delta(k_h)
+\frac{\alpha_s}{4\pi}\bigg\{\delta(k_\ell)\Big[\frac{-2 C_h\Gamma_0\ln\frac{k_h}{\mu}+2\gamma_s^{h}}{k_h}\Big]_{+}
+\delta(k_h)\Big[\frac{-2 C_\ell\Gamma_0\ln\frac{k_\ell}{\mu}+2\gamma_s^{\ell}}{k_\ell}\Big]_{+}\bigg\}\,,
\label{eq:sh_soft_oneloop}
\end{equation}
with $C_\ell$ and $C_h$ the total color charge in each hemisphere. For the quark channel $\{q,\bar q, g\}$ with the gluon jet in the light hemisphere, we have $C_\ell=C_A$ and $C_h=2C_F$, and the non-cusp anomalous dimensions are 
\begin{equation}
\gamma_s^{g}=-2C_A\ln3+4C_F\ln2\,,\qquad \gamma_s^{q\bar q}=-4C_F\ln6\,,
\end{equation}
while with a quark jet in the light hemisphere, $C_\ell=C_F$ and $C_h=C_F+C_A$, and
\begin{equation}
\gamma_s^{q}=-2C_F\ln\frac{3}{2}+2C_A\ln2\,,\qquad \gamma_s^{qg}=-2(C_F+C_A)\ln6\,.
\end{equation}
For the gluon channel $\{g,g,g\}$ with $C_\ell=C_A$ and $C_h=2C_A$, we have
\begin{equation}
\gamma_s^{g}=-2C_A\ln3+4C_A\ln2\,,\qquad \gamma_s^{gg}=-4C_A\ln6\,.
\end{equation}
Note that while we need the split of soft anomalous dimensions in the soft function boundary, the RG kernels only require the sum, which is also uniquely fixed by RG consistency. Therefore, at NNLL, we can use hard and jet anomalous dimensions to fix the sum of the two-loop soft anomalous dimensions.

\subsection{Fixed-order expansion}
\label{sec:shoulder_fixed_order}

\begin{figure}[!htbp]
\centering
\includegraphics[scale=0.65]{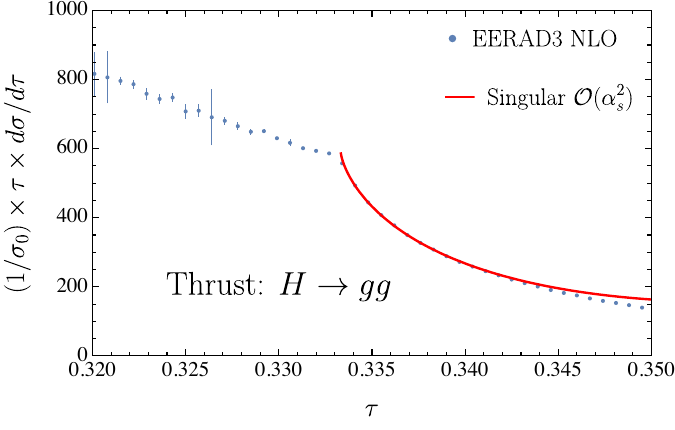}
\includegraphics[scale=0.65]{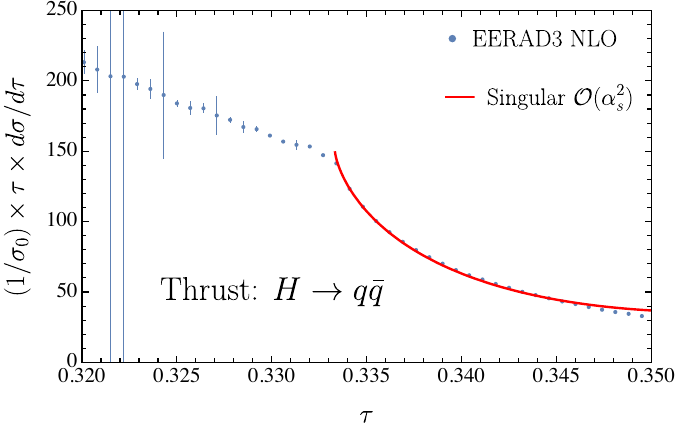}\\
\includegraphics[scale=0.65]{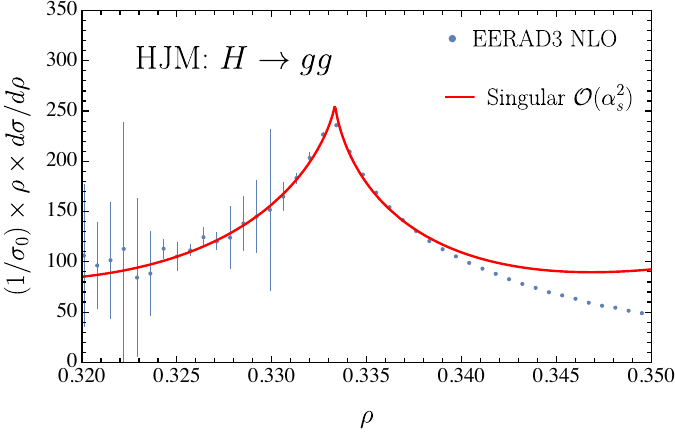}
\includegraphics[scale=0.65]{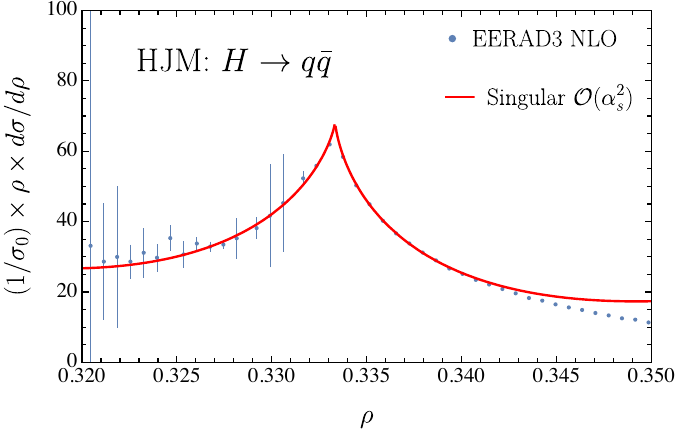}
\caption{Comparison of $\mathcal{O}(\alpha_s^2)$ expansion of shoulder factorization formula with NLO data from \textsc{Eerad3}. In all four cases, we find good agreement.}
\label{fig:shoulder_singular_compare}
\end{figure}

With all the ingredients, now we can substitute into the factorization formula in Eq.~\eqref{eq:sh_fact}. As a cross-check, we expand the factorization theorem in $\alpha_s$ and compare with fixed-order results. Since LO is just the normalization $\sigma_{\text{LO}}$, the first non-trivial order is NLO, $\mathcal{O}(\alpha_s^2)$.

First of all, we verified that the $\mathcal{O}(\alpha_s^2)$ expansion of the factorization theorem gives rise to identical results for HJM in Sec.~\ref{sec:NLOResults}. Secondly, we also compare the analytic expressions with NLO numerical results from \textsc{Eerad3}. Taking the HJM left shoulder as an example, note that in addition to $r\ln^2 r$ and $r\ln r$ that we predict or calculate, there are also linear terms in $r$ and constant terms near $\rho\to \frac{1}{3}$. Thus, instead of a direct comparison, we fit the \textsc{Eerad3} data with the following HJM ansatz
\begin{multline}
	\frac{d\sigma}{d\rho} = \theta\left(\frac{1}{3}-\rho\right) \left[\frac{d\sigma}{d\rho}\bigg|_{\rho\to\frac13^-}+c_\text{left}\left(\frac{1}{3}-\rho\right)\right]\\
	+ \theta\left(\rho-\frac{1}{3}\right) \left[\frac{d\sigma}{d\rho}\bigg|_{\rho\to\frac13^+}+c_\text{right}\left(\rho-\frac{1}{3}\right)\right]+\frac{d\sigma}{d\rho} \left(\rho=\frac{1}{3}\right)
\end{multline}
and a similar ansatz with only the right shoulder for thrust. We fit the parameters $c_\text{left},\, c_\text{right}$ and 
$\frac{d\sigma}{d\rho} \left(\rho=\frac{1}{3}\right)$ with multiple fit ranges around $\rho=\frac{1}{3}$. We found that different fit ranges lead to stable values, and we show the comparison in Fig.~\ref{fig:shoulder_singular_compare}. For both thrust and HJM in $Hgg$ and $Hq\bar{q}$, we find good agreement with the NLO fit form\footnote{Since the NNLO data from \textsc{Eerad3} is not accurate enough to resolve the Sudakov shoulder, we do not compare them.}. 
These two tests provide non-trivial checks on our factorization theorem and the perturbative ingredients.

\subsection{Shoulder resummation}
\label{sec:shoulder_resummation}

Since HJM left shoulder is phenomenologically important, we will only focus on the shoulder resummation for HJM.
With the hard, jet and soft functions explained in Sec.~\ref{sec:ingredients}, we can solve their RGEs and obtain the solutions in terms of RG kernels. Plugging them into Eq.~\eqref{eq:sh_massshift}, we obtain the resummed HJM formula. Systematically, the resummed distribution takes the following form, 
\begin{align}
    \frac{1}{\sigma_{\text{LO}}}\frac{d \sigma_\text{sh}}{d \rho}
  &=  \Pi(\mu_h,\mu_{j\ell}, \mu_{jh},\mu_{s\ell},\mu_{sh})\,
    \int_0^\infty d m_\ell^2 \int_0^\infty  d m_h^2\,
    \sjt \cdot
    \left( \frac{Q m_\ell^2}{\mu_{s\ell}} \right)^{a}
    \left( \frac{Q m_h^2}{\mu_{sh}} \right)^{b} 
    \nn \\
&\qquad \left.  \times\frac{1}{m_l^2 m_h^2} \frac{e^{-\gamma_E (a+b)}}{\Gamma(a) \Gamma(b)}  \right|_{\evenoverset{a=\eta_\ell}{b=\eta_h}}   (r + m_h^2 - m_{\ell}^2) \theta (r +
  m_h^2 - m_{\ell}^2) \,,
\end{align}
where the evolution block $\Pi(\mu_h,\mu_{j\ell}, \mu_{jh},\mu_{s\ell},\mu_{sh})$ takes the form
\begin{align}
   \Pi
    &=  \exp \Big[ 
    2C_h \left(S (\mu_h, \mu_{jh} )
    + S (\mu_{sh}, \mu_{jh})\right)
    + 2 C_\ell \left(S (\mu_h, \mu_{j\ell}) 
    +  S (\mu_{s\ell}, \mu_{j\ell})\right)\Big]
    \nn\\
   & 
   \times
    \exp \Big[ 
    2 A_{S_\ell}^\text{sh} (\mu_{s\ell}, \mu_h) 
    + 2 A_{S_h}^\text{sh}  (\mu_{sh}, \mu_h) 
    + 2 A_{J_\ell}  (\mu_{j\ell}, \mu_h) 
    + 2 A_{J_{h_1}} (\mu_{jh}, \mu_h) + 2 A_{J_{h_2}} (\mu_{jh}, \mu_h)\Big]
   \nn\\
   & \times H^\text{sh} (Q, \mu_h)   \left(\frac{Q^2}{\mu_h^2}\right)^{-C_h A_\Gamma (\mu_h,\mu_{jh})-C_\ell A_\Gamma (\mu_h,\mu_{j\ell})} \label{Pig}
  \,,
\end{align}
and the boundary function $\sjt$ is
\begin{equation}
\sjt =     \widetilde{j}_{h_1}\left( \partial_b + \ln \frac{Q \mu_{sh}}{\mu_{jh}^2} \right) 
     \widetilde{j}_{h_2} \left( \partial_b+ \ln \frac{Q \mu_{sh}}{\mu_{jh}^2} \right) \widetilde{j}_\ell \left( \partial_a + \ln \frac{Q \mu_{s\ell}}{\mu_{j\ell}^2} \right) \widetilde{s}_{\ell; h_1 h_2} (\partial_a,\partial_b) 
  \,,
\end{equation}
as well as
\begin{equation}
a = \eta_\ell = 2 C_\ell A_{\Gamma} (\mu_{j\ell}, \mu_{s\ell}),\qquad
b = \eta_h= 2 C_h A_{\Gamma} (\mu_{jh}, \mu_{sh})
  \,.
\end{equation}
Here $C_\ell$ and $C_h$ are the color factors associated with the light and heavy hemispheres; $\gamma_{s\ell}$ and $\gamma_{sh}$ are the split of soft anomalous dimensions; $\gamma_{j\ell}$, $\gamma_{jh_1}$ and $\gamma_{jh_2}$ are the jet anomalous dimensions, where we use $h_1$ and $h_2$ to label the two particles in the heavy hemisphere. $\widetilde{j}_{\ell}, \widetilde{j}_{h_1}, \widetilde{j}_{h_2}$ are the jet function boundary in Laplace space and $\widetilde{s}_{\ell; h_1 h_2}$ is the soft function boundary. 
For example, for $H{\color{blue}g}{\color{red} q} {\color{red} \bar q}$ in $Hgg$ process, we have
\begin{equation}
	C_\ell =C_A,\quad C_h=2 C_F,\quad \gamma_{s\ell}=\gamma_{sg},\quad \gamma_{sh}=\gamma_{sq\bar{q}}, \quad \gamma_{j\ell}=\gamma_{jg},\quad \gamma_{jh_1}=\gamma_{jh_2}=\gamma_{jq}\,,
\end{equation}
and
\begin{equation}
	\widetilde{j}_{\ell}=\widetilde{j}_{g},\quad \widetilde{j}_{h_1}=\widetilde{j}_{q},\quad \widetilde{j}_{h_2}=\widetilde{j}_{\bar{q}},\quad \widetilde{s}_{\ell; h_1 h_2}=\widetilde{s}_{g;q\bar{q}}\,.
\end{equation}
Similar substitutions can be derived for all the other five channels.

The RG kernels are again expressed in terms of $S$ and $A$ functions defined in Eq.~\eqref{eq:SAfunctions}, and the last step for resummation is the integration over $m_\ell^2$ and $m_h^2$. However, as observed in Ref.~\cite{Bhattacharya:2022dtm}, a naive integration will give rise to an infinite set of spurious poles in the physical region, referred to as Sudakov Landau poles. Similar issues were found in the transverse momentum dependent (TMD) resummation and people perform the resummation in position space~\cite{Frixione:1998dw,Ebert:2016gcn}.
HJM shoulder is the first non-TMD (or SCET${}_\text{I}$) example that requires similar treatment. As illustrated in Ref.~\cite{Bhattacharya:2023qet} (cf. Eq.(3.21)), the position space will bring higher logarithmic order terms in an exponential form and suppress the spurious poles. The physical interpretation is that the shoulder logs arise from the region where both hemisphere masses are small and the difference between them is also small. However, the measurement function $\theta (r + m_h^2 - m_{\ell}^2)$ in the factorization theorem only constrains their difference, and we include some divergent contributions from regions where EFT is not valid.

To proceed, we first take the second derivative of the distribution, namely $d^3\sigma/d\rho^3$, and turn the resummed formula into 
\begin{align}
    \frac{1}{\sigma_{\text{LO}}}\frac{d^3 \sigma_\text{sh}}{d \rho^3}
  &=  \Pi(\mu_h,\mu_{j\ell}, \mu_{jh},\mu_{s\ell},\mu_{sh})\,
    \int_0^\infty d m_\ell^2 \int_0^\infty  d m_h^2\,
    \sjt \cdot
    \left( \frac{Q m_\ell^2}{\mu_{s\ell}} \right)^{a}
    \left( \frac{Q m_h^2}{\mu_{sh}} \right)^{b} 
    \nn \\
&\qquad \left.  \times\frac{1}{m_l^2 m_h^2} \frac{e^{-\gamma_E (a+b)}}{\Gamma(a) \Gamma(b)}  \right|_{\evenoverset{a=\eta_\ell}{b=\eta_h}}    \delta (r +m_h^2 - m_{\ell}^2) \,.
\end{align}
Then we perform the Fourier transformation,
\begin{equation}
    \frac{1}{\sigma_{\text{LO}}}\frac{d^3 \sigma_\text{sh}(\mu_\text{res})}{d\rho^3}=  
    \int_{-\infty}^\infty \frac{dz}{2\pi}\, e^{-i z r}\, \widetilde\sigma_\text{sh}(z,\mu_\text{res})
  \,,
\end{equation}
with $\widetilde\sigma_\text{sh}(z,\mu_\text{res})$ the resummed formula in the position (Fourier) space and $\mu_\text{res}$ the short-hand notation for shoulder resummation scale. Explicitly we have
\begin{equation}
\label{eq:resummed_sigmaz}
    \widetilde\sigma_\text{sh}(z,\mu_\text{res}) =\Pi(\mu_h,\mu_{j\ell}, \mu_{jh},\mu_{s\ell},\mu_{sh}) \cdot 
    \sjt \cdot
\left( -i z\frac{e^{\gamma_E} \mu_{s\ell} }{Q} \right)^{-a}
\left( i z\frac{e^{\gamma_E} \mu_{sh} }{Q} \right)^{-b}
  \,,
\end{equation}
with the same factor $\Pi(\mu_h,\mu_{j\ell}, \mu_{jh},\mu_{s\ell},\mu_{sh})$ and the following boundary
\begin{equation}
\label{eq:sjt_position}
\sjt =     
\widetilde{j}_{h_1}
\left( 
\ln \frac{Q^2 e^{-\gamma_E}}{i z \mu_{jh}^2} 
\right) 
\widetilde{j}_{h_2}
\left( 
\ln \frac{Q^2 e^{-\gamma_E}}{i z \mu_{jh}^2}
\right) 
\widetilde{j}_\ell
\left( 
\ln \frac{Q^2 e^{-\gamma_E}}{-i z \mu_{j\ell}^2} 
\right)
\widetilde{s}_{\ell;h_1 h_2} \left(
\ln \frac{Q e^{-\gamma_E}}{-i z \mu_{s\ell}},
\ln \frac{Q e^{-\gamma_E}}{i z \mu_{sh}}
\right) .
\end{equation}
The important point to fix the spurious pole is setting scale in the position space. So far we have been using $\mu_{j\ell},\mu_{jh}$ and $\mu_{s\ell}, \mu_{sh}$ to label the jet and soft scales in different hemispheres. As studied in Ref.~\cite{Bhattacharya:2023qet}, one concludes that the real scale choice
\begin{equation}
\mu_h = Q,\quad
\mu_{s \ell} = \mu_{sh} =\frac{Qe^{-\gamma_E}}{|z|}, \quad
\mu_{j \ell} = \mu_{jh} =\frac{Qe^{-\gamma_E/2}}{\sqrt{|z|}}\,,
\end{equation}
eliminates the large logarithms in Eq.~\eqref{eq:sjt_position} up to a phase. And we will use the real scale for shoulder resummation in this work.

To obtain the HJM distribution, we need to perform inverse Fourier transformation and integrate over $\rho$ (or equivalently, $r$) twice from some boundary values. This can be absorbed into a transformation kernel $K(z,r)$,
\begin{equation}
\frac{1}{\sigma_{\rm LO}}\frac{d\sigma_\text{sh}(\mu_\text{res})}{d\rho}
=\int dr'\int dr''\,
\frac{1}{\sigma_{\rm LO}}\frac{d^3\sigma_\text{sh}(\mu_\text{res})}{d\rho^3}
=2\,{\rm Re}\int_0^{\infty}\frac{dz}{2\pi}\,K(z,r)\,\widetilde\sigma_\text{sh}(z,\mu_\text{res})\,,
\label{eq:sh_int}
\end{equation}
where we use $\widetilde\sigma_\text{sh}^{\star}(-z)=\widetilde\sigma_\text{sh}(z)$ followed by the real scale choice. The kernel will take the form
\begin{equation}
K(z,r)=-\frac{e^{-i r z}}{z^2}+K_0(z)+K_1(z) r\,.
\label{eq:sh_kernel}
\end{equation}
$K_{0,1}(z)$ account for the chosen integration boundaries, effectively the offset and the slope near $r=0$. The boundary condition we adopt, as in Ref.~\cite{Bhattacharya:2023qet} is that when expanding the distribution in $\alpha_s$, one should reproduce the leading behavior of the full theory prediction near $r=0$. This leads to 
\begin{equation}
    K(z,r)=\frac{1}{z^2}\left[1-e^{-i z r}+ r(1-e^{iz})\right] \label{eq:LOkernel}
  \,.
\end{equation}
Thus we can obtain the shoulder resummation distributions by performing the numerical integration over $z$.

\subsection{Canonical scale and NNLL resummation}
\label{sec:canonical_scale}

In this subsection, we present the canonical resummation. Including the scale variation in the canonical scale choice and freezing the soft scale in the non-perturbative regime give the following canonical scale $\mu_\text{can}$: 
\begin{align} 
	\mu_{h} &= e_h Q  \notag \,, \\
	\mu_{s}^\text{sh}(z) &= \sqrt{
 \Big(e_h e_s^\text{sh} \frac{Q e^{-\gamma_E}}{|z|}\Big)^2
 +\Big(\mu_{s}^{\text{min}}\Big)^2}\notag \,,\\
	\mu_{j}^\text{sh}(z) &= \bigl[ \mu_{s}^\text{sh}(z) \bigr]^{v_j} \mu_{h}^{1- v_j}
  \,.
  \label{eq:shoulder_canonical_scale}
\end{align}
This follows the spirit of dijet canonical scale in Eq.~\eqref{eq:canonical_scales}. The hard variation $e_h$ is again correlated among all scales to preserve the hierarchy, and it will also be correlated with both the dijet hard scale and FO scale. When $|z|\to \infty$, $\mu_s^{\text{sh}}$ will converge to $\mu_s^{\text{min}}$ regardless of any variation.
The default values for these variation parameters are $e_h=1$, $e_s^\text{sh}=1$ and $v_j=1/2$. And for perturbative uncertainty estimation, we vary $e_h\in [1/2,2]$, $e_s^\text{sh} \in [1/2,2]$ and $v_j\in [0.4,0.6]$ and take the envelop.

The FO scale for shoulder is simply $\mu_\text{FO}: \mu_h=\mu_s^\text{sh}=\mu_j^\text{sh}=e_h Q$ in the position space, and it will produce the singular expansions. To test the kernel $K(z,r)$ and the inverse Fourier integration implementation, we have checked the singular expansions through setting FO scales against the analytic expression we obtain in Sec.~\ref{sec:NLOResults}.

The matching formula for shoulder resummation is
\begin{align}
\frac{d\sigma^{\text{sh,match}}}{d\rho} &= \frac{d\sigma^\text{sh,res}}{d\rho} + \left(\frac{d\sigma^\text{FO}}{d\rho}-\frac{d\sigma^\text{sh,sing}}{d\rho}\right)\nn\\
& = 2\,{\rm Re}\int_0^{\infty}\frac{dz}{2\pi}\,K(z,r)\,\Big[\widetilde\sigma_\text{sh}(z,\mu_\text{can})-\widetilde\sigma_\text{sh}(z,\mu_\text{FO})\Big]+\frac{d\sigma^\text{FO}}{d\rho}\,,
\end{align}
where in the first line, the first term is the resummed distribution and the second term is the non-singular contribution. In the second line, we can rewrite both the resummed and singular contributions in position space, and the difference between them is the scale choice. Since the trijet hard function starts at $\mathcal{O}(\alpha_s)$, the matching order is N${}^{k}$LL${}_\text{sh}+$N${}^{k-1}$LO. 

\begin{figure}[!htbp]
    \centering
    \includegraphics[scale=0.36]{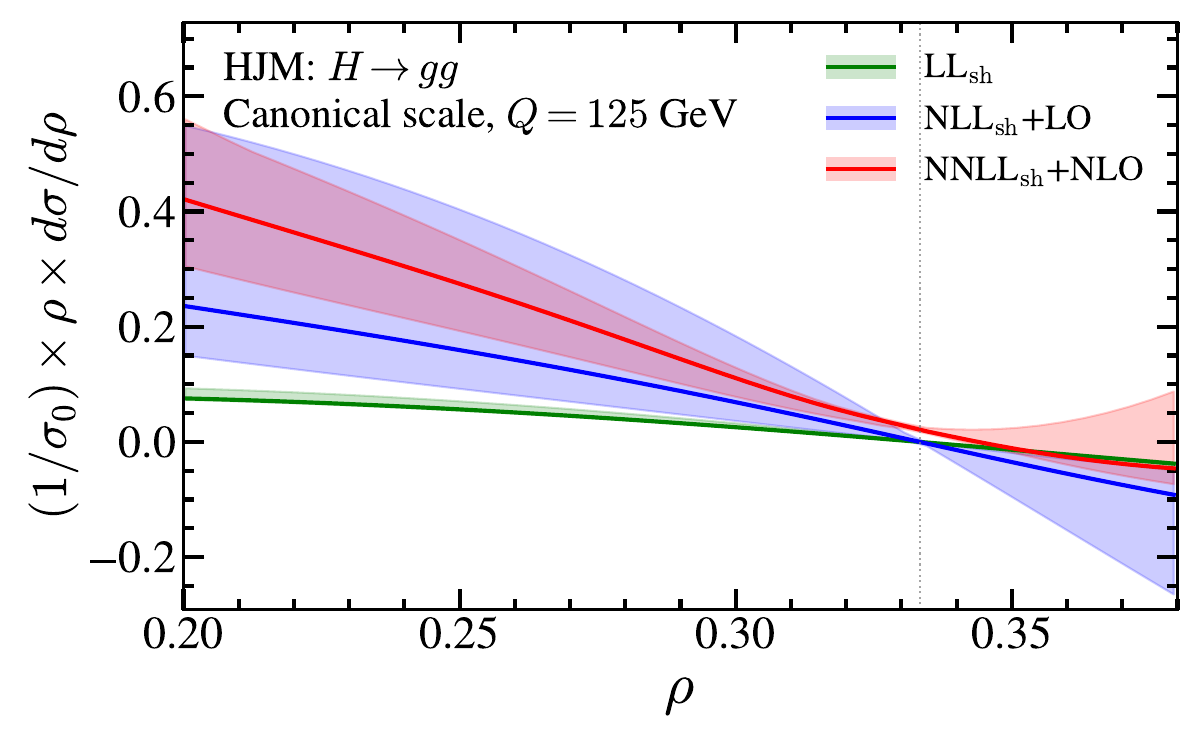}
    \includegraphics[scale=0.36]{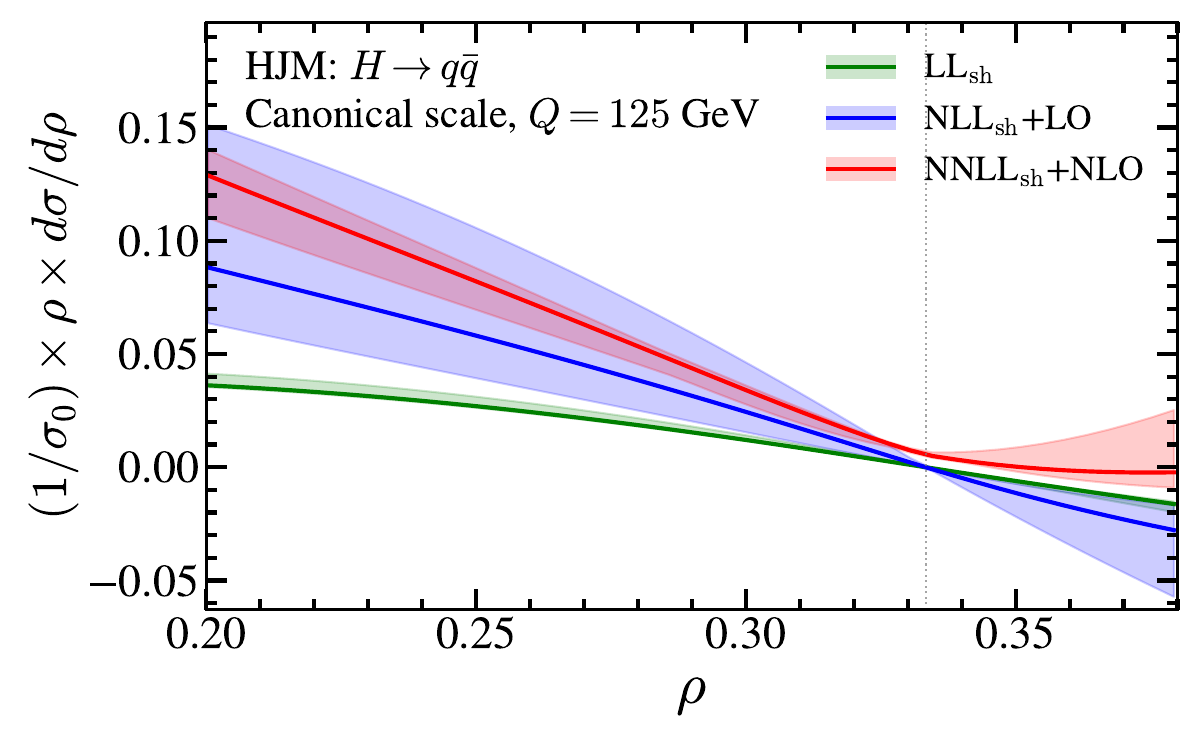}
    \caption{The shoulder resummed distributions with canonical scale and matched to fixed-order. The resummation smoothes the shoulder kink at $\rho=1/3$. We find good convergence from NLL${}_\text{sh}+$LO to NNLL${}_\text{sh}+$NLO on the left shoulder.}
    \label{fig:shoulder_canonical_res}
\end{figure}

In Fig.~\ref{fig:shoulder_canonical_res}, we show the canonical resummation results for HJM in $Hgg$ and $Hq\bar{q}$ channels, matched to fixed-order. The uncertainty bands are obtained by varying the canonical scales independently and taking the envelope. With shoulder resummation, the Sudakov shoulder kink at $\rho=1/3$ is smoothed and the distribution is now well-behaved. 
We also observe that LL is underestimating the uncertainty, however, from NLL${}_\text{sh}$+LO to NNLL${}_\text{sh}$+NLO on the left shoulder, we observe good convergence. 
The fact that right shoulder distributions go to negative very fast indicates the existence of second shoulder logarithms, which is beyond the scope of this paper.

\subsection{Shoulder profile functions}
\label{sec:shoulder_profile}

In order to combine the shoulder resummation with dijet resummation, we also need a shoulder profile scale, such that we can smoothly turn off the shoulder resummation away from $\rho=1/3$. In Ref.~\cite{Bhattacharya:2023qet}, we have used the sigmoid function to turn off the shoulder resummation. In this work, we will adopt a quadratic function as in Ref.~\cite{Benitez:2025vsp}, which is similar to the dijet profile function in Sec.~\ref{sec:dijet_profile} and will not differ much from the sigmoid function.
Explicitly, we introduce a weight function $g_{\rm sh}(\rho)\in[0,1]$ that turns the shoulder
resummation fully on around $\rho=1/3$ and smoothly off in the fixed-order region on both sides.
To achieve this, we define 
\begin{equation}
\label{eq:gdef}
g_{\rm sh}(\rho) = g_0(\rho;\rho_{L1},\rho_{L2},\rho_{L3})-g_0(\rho;\rho_{R1},\rho_{R2},\rho_{R3})\,,
\end{equation}
where $g_0$ is a three-point quadratic smoothstep function that rises from $0$ to $1$ across $a<b<c$ with continuous value and first derivative,
\begin{equation}
\label{eq:smoothstep}
g_0(\rho;a,b,c) = \begin{cases}
0\,, & \rho<a\,,\\[4pt]
\dfrac{(\rho-a)^2}{(c-a)(b-a)}\,, & a\le\rho<b\,,\\[10pt]
1-\dfrac{(\rho-c)^2}{(c-a)(c-b)}\,, & b\le\rho<c\,,\\[10pt]
1\,, & \rho\ge c\,.
\end{cases}
\end{equation}
Thus we have left transition points $\rho_{L1}<\rho_{L2}<\rho_{L3}$ and right points $\rho_{R1}<\rho_{R2}<\rho_{R3}$.
The shoulder profile scale then takes the following form,
\begin{equation}
\mu_{s}^{\rm sh,prof}(z,\rho)=\mu_h^{\,1-g_{\rm sh}(\rho)}\,\big[\mu_s^{\rm sh}(z)\big]^{g_{\rm sh}(\rho)}\,,
\qquad
\mu_{j}^{\rm sh,prof}=\left[\mu_{s}^{\rm sh,prof}(z,\rho)\right]^{v_j}\mu_h^{1-v_j}\,,
\label{eq:sh_profile_scales}
\end{equation}
with $\mu_h=e_h Q$ and $\mu_s^\text{sh}(z)$ the canonical soft scale in Eq.~\eqref{eq:shoulder_canonical_scale}. When $g_{\rm sh}(\rho)=1$, the canonical scales are recovered and the shoulder resummation is fully on; when $g_{\rm sh}(\rho)=0$, all scales become equal, i.e. become $\mu_\text{FO}$, and the resummation is off.

\begin{figure}[!htbp]
\centering
\includegraphics[width=0.48\linewidth]{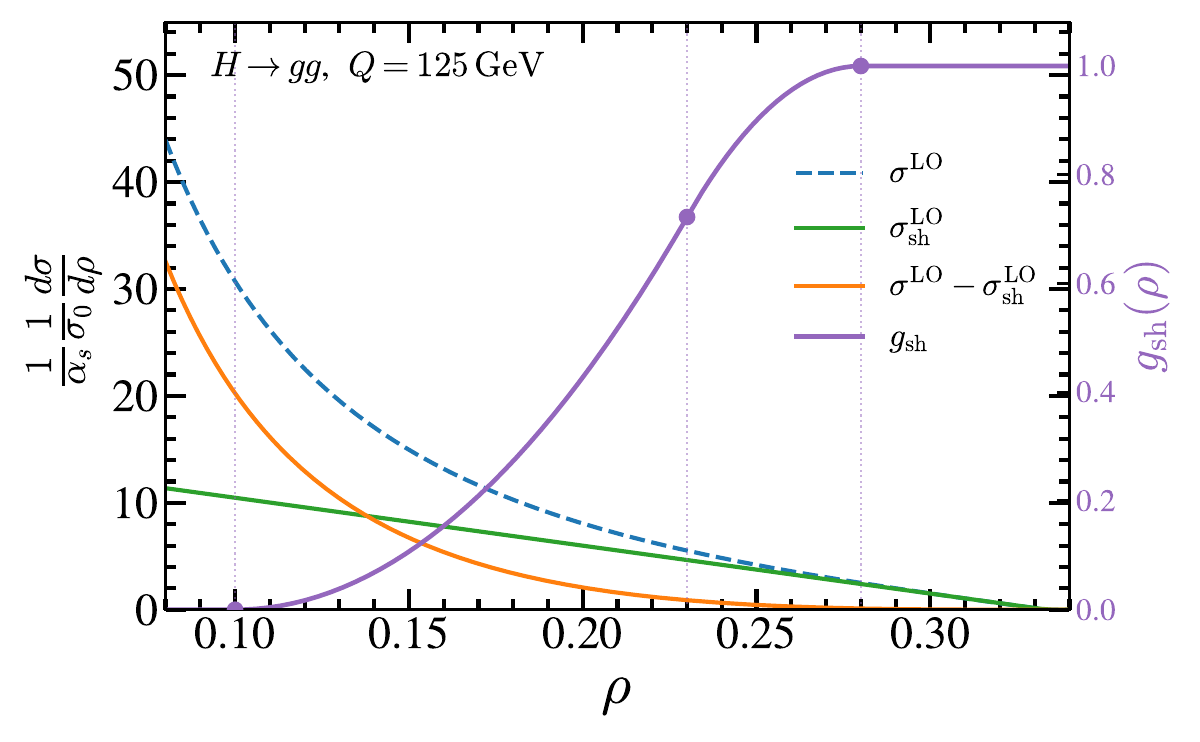}
\includegraphics[width=0.48\linewidth]{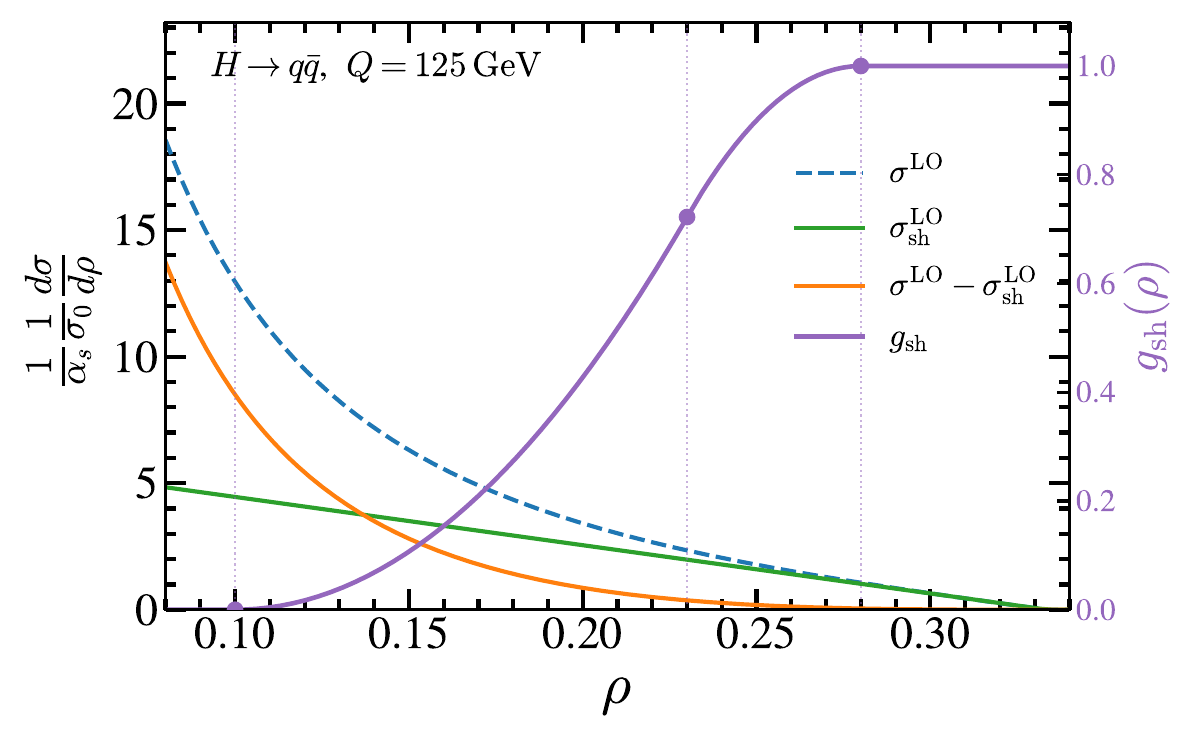}
\caption{The comparison of singular and non-singular at LO for $Hgg$ (left) and $Hq\bar q$ (right).
The left axis shows the full LO distribution
$\sigma^{\rm LO}$ (blue dashed), the shoulder singular $\sigma^{\rm LO}_{\rm sh}$ (green), and the
non-singular remainder $\sigma^{\rm LO}-\sigma^{\rm LO}_{\rm sh}$ (orange), with the overall
$\alpha_s$ factored out. Overlaid on the right axis (purple) is the shoulder weight $g_{\rm sh}$ with
transition points $\{\rho_{L1},\rho_{L2},\rho_{L3}\}=\{0.10,0.23,0.28\}$ (dots): the resummation is
switched off ($g_{\rm sh}=0$) at $\rho_{L1}$, where the non-singular remainder dominates, and fully
on ($g_{\rm sh}=1$) by $\rho_{L3}=0.28$, where the shoulder singular already saturates the full LO.
The singular reaches half and $90\%$ of the full LO at $\rho\approx 0.14$ and $\rho\approx 0.25$,
almost identically in the two channels, so a single, process-independent shoulder profile is used.}
\label{fig:sh_ns}
\end{figure}

The transition values of $g_{\rm sh}(\rho)$ are determined by comparing the singular and non-singular distribution in the shoulder region. In Ref.~\cite{Bhattacharya:2023qet}, the singular vs non-singular comparison at LO is studied for $e^+e^-\to q\bar{q}$ and is used to determine the transition points for sigmoid profile function. In Fig.~\ref{fig:sh_ns}, we plot the comparison also at LO for both $Hgg$ and $Hq\bar{q}$ decays.
The shoulder singular
accounts for $90\%$ of the full LO down to $\rho\approx 0.25$ and falls to the size of the
non-singular remainder at $\rho\approx 0.14$. The two crossovers are nearly identical in the two
figures and thus we can use a single profile for both processes. Switching to quadratic profile function, this leads to the transition values $\{\rho_{L1},\rho_{L2},\rho_{L3}\}=\{0.10,0.23,0.28\}$ in $g_{\rm sh}(\rho)$ on the left shoulder. We also plot the values of $g_{\rm sh}(\rho)$ in Fig.~\ref{fig:sh_ns} for illustration.
On the right shoulder
($\rho>\frac{1}{3}$) the singular rises more steeply, so the resummation is switched off over a
narrower window $\{\rho_{R1},\rho_{R2},\rho_{R3}\}=\{1/3, 0.338, 0.342\}$. We have listed all profile parameters in Tab.~\ref{tab:shoulder_params}.

\begin{table}[!htbp]
\centering
\setlength{\tabcolsep}{12pt}
\begin{tabular}{|c|c|c|}
\hline
parameter & default & variation range\\
\hline
$e_h$ & $1$ & $[1/2,\,2]$ \\
$e_s^\text{sh}$ & $1$ & $[1/2,\,2]$ \\
$v_j$ & $1/2$ & $\{0.4,\,0.6\}$ \\
\hline
$\rho_{L1}$ & $0.10$ & $[0.09,0.11]$ \\
$\rho_{L2}$ & $0.23$ & rescaled \\
$\rho_{L3}$ & $0.28$ & $[0.25, 0.31]$ \\
\hline
$\rho_{R1}$ & 1/3 & $-$ \\
$\rho_{R2}$ &0.338 & $-$ \\
$\rho_{R3}$ & 0.342 & $-$ \\
\hline
\end{tabular}
\caption{Shoulder scale-variation and profile-function parameters. The left shoulder profile parameters $\rho_{L1}$ and $\rho_{L3}$ are varied by $10\%$ change, while $\rho_{L2}$ is rescaled accordingly. For the right shoulder profile parameters, we do not vary them.}
\label{tab:shoulder_params}
\end{table}

When the shoulder resummation is included, the dijet resummation can no longer be maintained all the way to $\rho=0.4$. Therefore, we rescale the dijet profile and turn the dijet resummation off earlier than in the standalone case,  as in Ref.~\cite{Bhattacharya:2023qet}. Explicitly, we choose 
\begin{equation}
t_2=0.2\,,\qquad t_s=\frac{1}{3}\,,
\label{eq:sh_dijet_profile}
\end{equation}
in place of the standalone values $t_2=0.25$, $t_s=0.40$ of Tab.~\ref{tab:profile_params} and vary them with the same relative amounts; the
remaining dijet parameters ($\mu_s^{\rm min}$, $n_0$, $n_1$, $e_h$, $e_s$, $e_j$) are unchanged. Also we emphasize that both resummation and the non-singular contribution share a common hard scale as in the matching formula, i.e. $e_h$ is fully correlated in all pieces.
With this choice, the dijet distribution falls off through the shoulder region precisely where the shoulder resummation rises, producing a smooth hand-off between the two contributions.

The last step is to combine the two resummed descriptions without double counting their fixed-order limits. We use the additive matching procedure
\begin{equation}
\frac{d\sigma_i^{\rm match}}{d\rho}
=\frac{d\sigma_i^{\rm dij,res}(\mu_{\rm dij}^{\rm prof})}{d\rho}
+\frac{d\sigma_i^{\rm sh,res}(\mu_{\rm sh}^{\rm prof})}{d\rho}
+\left[\frac{d\sigma_i^{\rm FO}(\mu_{\rm FO})}{d\rho}
-\frac{d\sigma_i^{\rm dij,sing}(\mu_{\rm FO})}{d\rho}
-\frac{d\sigma_i^{\rm sh,sing}(\mu_{\rm FO})}{d\rho}\right]\,.
\label{eq:full_match}
\end{equation}
Here $d\sigma_i^{\rm dij,sing}$ and $d\sigma_i^{\rm sh,sing}$ denote the fixed-order expansions of the corresponding resummed formulae, evaluated with all profile scales set to the fixed-order scale $\mu_{\rm FO}$. Thus the bracket contains the non-singular remainder after subtracting both singular limits from the full fixed-order result. This form has the desired limiting behavior: away from the shoulder, the shoulder profile gradually sends $\mu_{\rm sh}^{\rm prof}\to\mu_{\rm FO}$ and the shoulder-resummed term cancels its singular subtraction, while near the shoulder the modified dijet profile gradually turns off the dijet resummation so that the dijet-resummed term largely cancels its singular subtraction. The hard scale is shared by all terms in Eq.~\eqref{eq:full_match}, and the remaining dijet and shoulder profile variations are treated as independent probes of the two resummation regions.

\section{Numerical results}
\label{sec:numerical_result}

We now present the final matched prediction for HJM distributions in hadronic Higgs decays. The result combines the N$^{3}$LL${}^{\prime}_{\rm dij}$ dijet resummation, the NNLL${}_{\rm sh}$ shoulder resummation, and the NNLO fixed-order distribution according to Eq.~\eqref{eq:full_match}
In principle, the consistent matching order is N${}^k$LL${}^{\prime}_\text{dij}+$N${}^k$LL$+$N${}^{k-1}$LO or N${}^k$LL${}^{\prime}_\text{dij}+$N${}^{k-1}$LL${}^\prime+$N${}^{k-1}$LO, where the shoulder resummation is one logarithmic order lower because it starts with trijet configuration. However since the current accuracy of NNLO data cannot resolve the shoulder kink, and the two-loop trijet hemisphere soft functions are not available, we only use up to NNLL${}_{\rm sh}$ in the shoulder resummation. 

\begin{figure}[!htbp]
    \centering
    \includegraphics[scale=0.38]{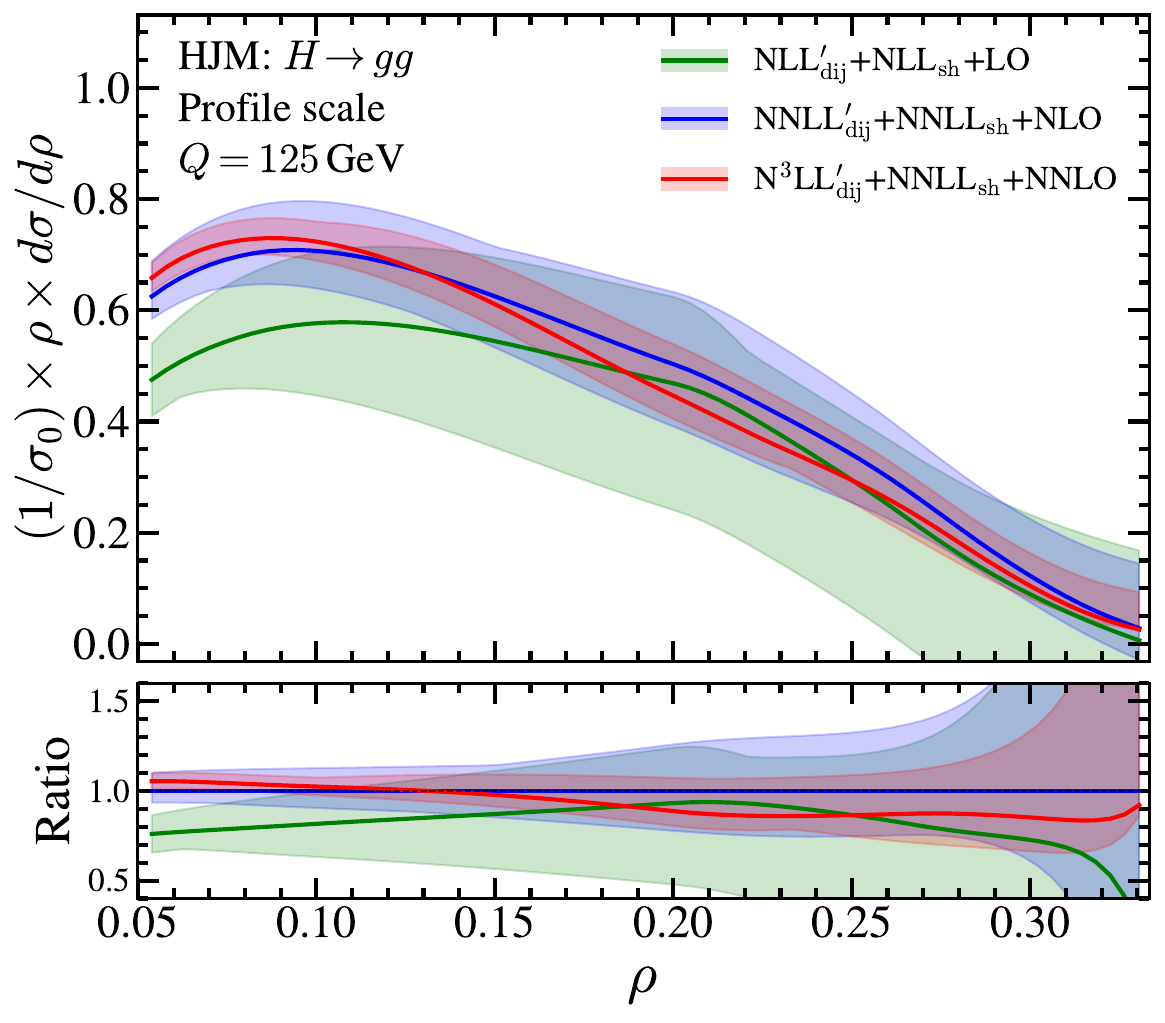}
    \includegraphics[scale=0.38]{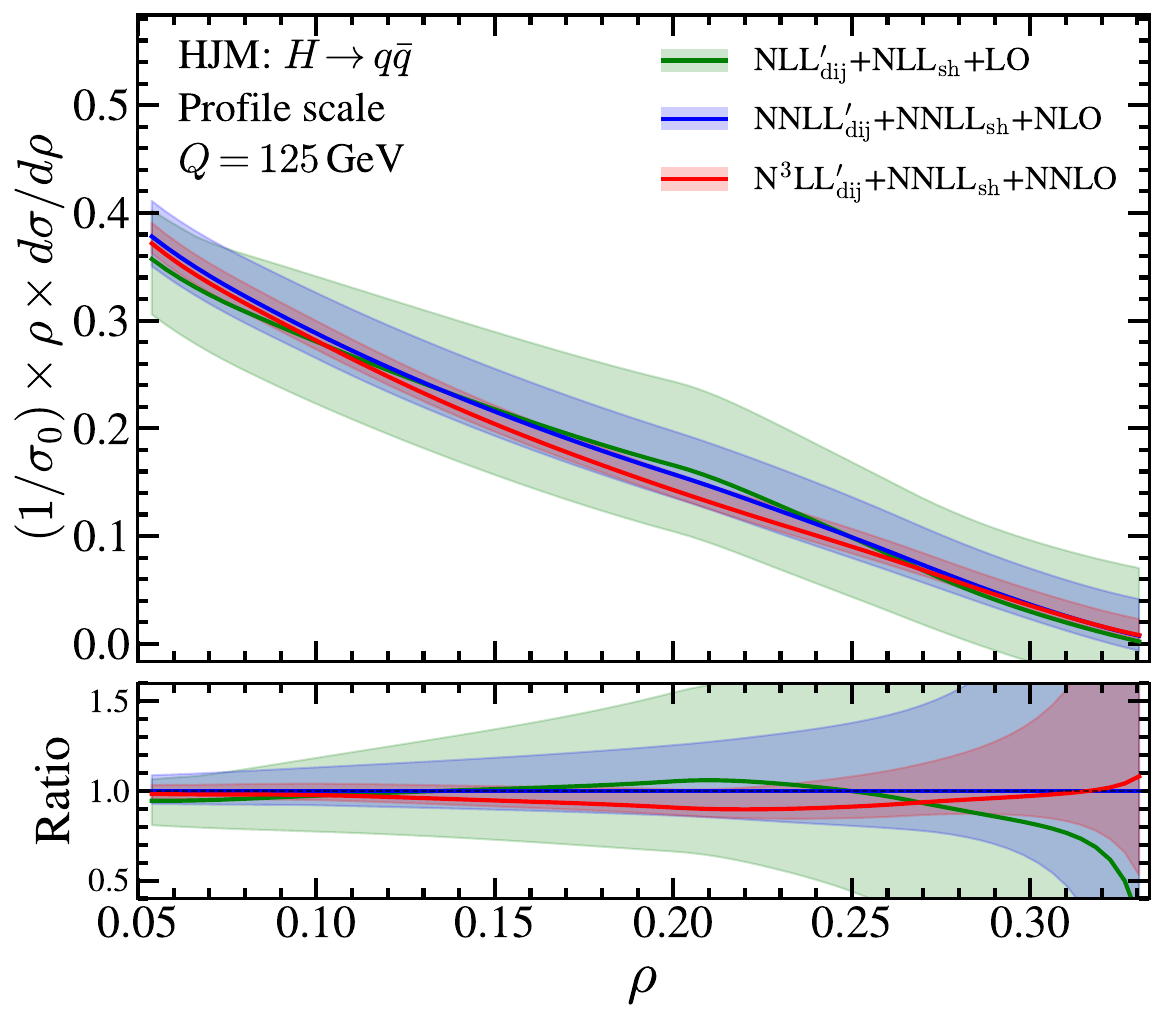}
    \caption{The matched HJM distributions for $Hgg$ (left panel) and $Hq\bar q$ (right panel), including dijet resummation, shoulder resummation, and fixed-order contribution. The shoulder-compatible dijet profile is used in the calculation. The uncertainty bands are the envelope of individual variations of dijet and shoulder profile parameters. We observe good convergence in both processes.}
    \label{fig:full_profile_matched}
\end{figure}

Throughout this section we take $Q=m_H=125\,{\rm GeV}$ and normalize the spectrum to the Born total cross-section $\sigma_0$ in each channel. The perturbative uncertainty bands are obtained from the envelope of the dijet and shoulder profile variations discussed above, as well as the common hard-scale variation.
Fig.~\ref{fig:full_profile_matched} shows the matched distributions for $H\to gg$ and $H\to q\bar q$ at three successive orders.
In both channels the perturbative series converges once the dijet and shoulder resummations are combined. 
The gluon channel remains broader and peaks at larger $\rho$ than the quark channel, reflecting the larger color charge and potentially stronger non-perturbative corrections.
By contrast, the $H\to q\bar q$ distribution decreases monotonically over the same $\rho$ range, with a smaller magnitude of the distribution. 

\begin{figure}[!htbp]
    \centering
    \includegraphics[scale=0.38]{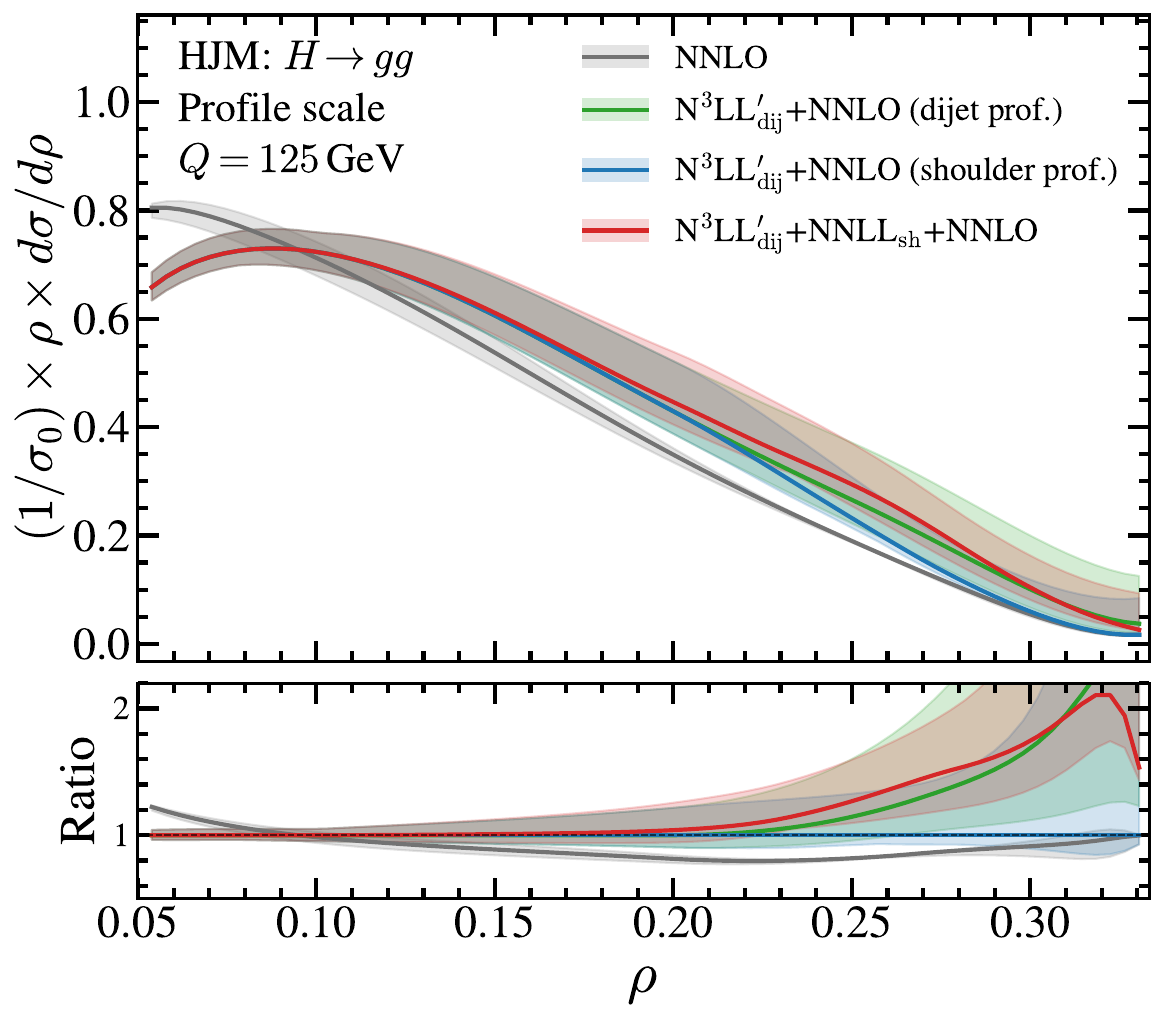}
    \includegraphics[scale=0.38]{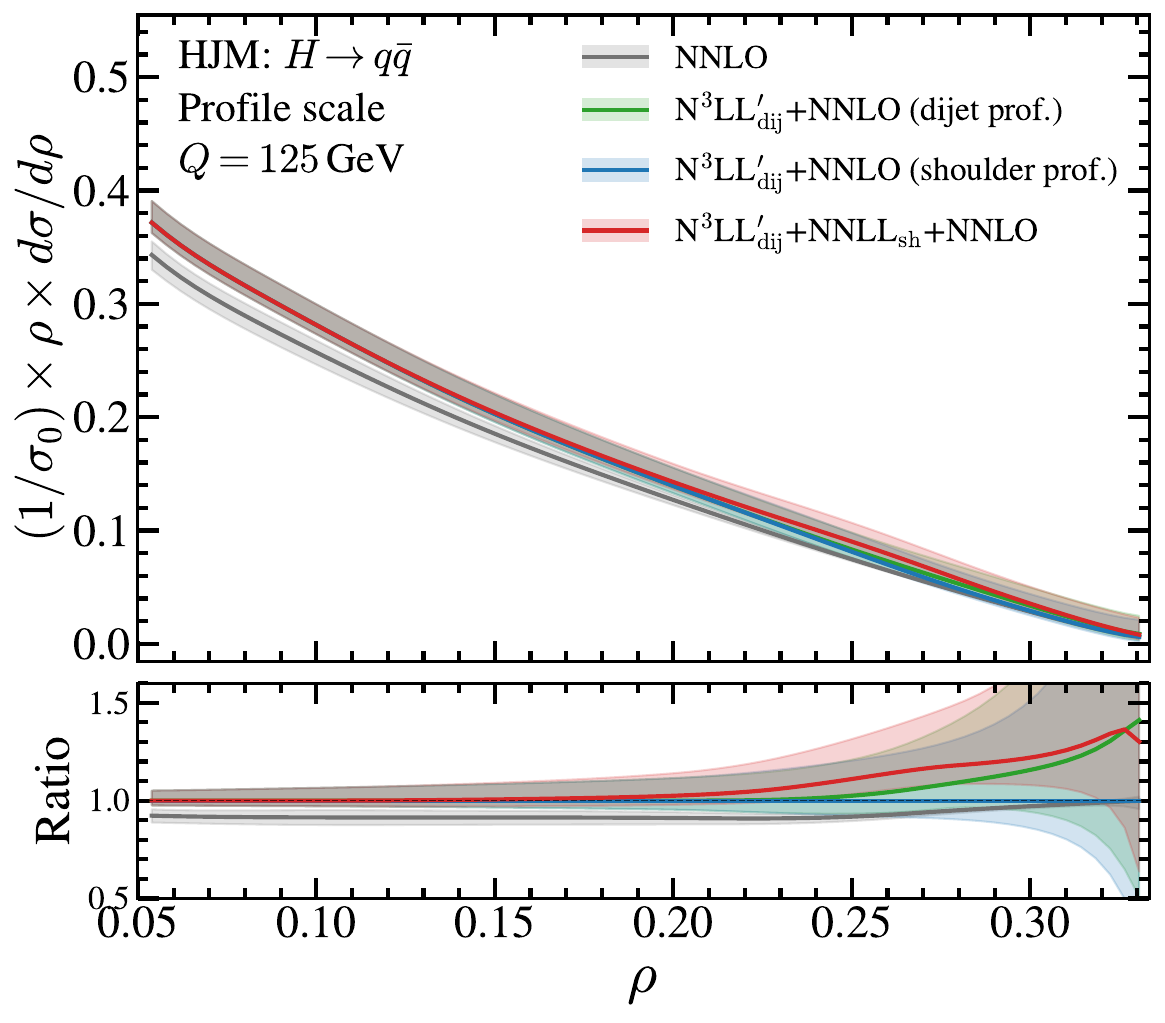}
    \caption{Impact of the shoulder resummation at the highest perturbative order. The gray band is NNLO fixed order, the green band is N$^{3}$LL${}^{\prime}_{\rm dij}+$NNLO with the standalone dijet profile, the blue band is the same dijet resummed prediction with the shoulder-compatible dijet profile, and the red band is the full N$^{3}$LL${}^{\prime}_{\rm dij}+$NNLL${}_{\rm sh}+$NNLO result. The lower panels show ratios to the central blue curve.}
    \label{fig:full_profile_ratio}
\end{figure}

To isolate the numerical impact of the shoulder resummation, Fig.~\ref{fig:full_profile_ratio} compares the highest-order prediction with several intermediate settings. 
The gray band is the NNLO fixed-order result. The green curve shows the N$^{3}$LL${}^{\prime}_{\rm dij}+$NNLO prediction with the standalone dijet profile used in Sec.~\ref{sec:dijet_profile}, while the blue curve uses the modified dijet profile of Eq.~\eqref{eq:sh_dijet_profile}, which gradually turns off the dijet resummation before the shoulder region.
The red curve is the full result, in which the missing trijet logarithms are restored by the NNLL shoulder resummation.
The ratios in the lower panels are normalized to the central value of the blue curve, so that the difference between the red and blue curves directly measures the impact of the shoulder resummed contribution with the same dijet profile.
Away from the shoulder, the red and blue curves are nearly identical as expected. In this region $g_{\rm sh}(\rho)$ turns off the shoulder scales and the matched result reduces to the pure dijet resummation.
Starting around the left-shoulder transition region, the shoulder contribution becomes visible and lifts the spectrum. In the $H\to q\bar q$ channel, the effect is moderate over most of the range, but becomes a sizable correction close to $\rho=1/3$. In the $H\to gg$ channel the correction is substantially larger, reaching tens of percent already before the endpoint and becoming order-one very close to the shoulder.
In the meantime, by comparing the green and red curves in the ratio figures, the dijet resummation with original dijet profile also lifts the distribution, which partially accounts for some trijet contribution. However, the difference between them is still sizable in most of the shoulder region and crosses zero around $\rho\simeq 0.32$.
In short, these comparisons demonstrate that the shoulder resummation is an important ingredient for getting reliable predictions in the trijet region.

\section{Conclusion}
\label{sec:conclusion}

In this work we present the resummed predictions for HJM in hadronic Higgs decays.  For both $H\to gg$ and $H\to q\bar q$ processes, we report the dijet factorization theorem in SCET and perform the resummation to N$^{3}$LL${}^{\prime}_\text{dij}$ accuracy, matched to the NNLO fixed-order distribution. In addition, we analyze the Sudakov shoulder in the trijet region, where the incomplete cancellation of soft and collinear divergences has given rise to a kink at $\rho=1/3$. Expanding both Higgs decay squared amplitudes and the four-particle phase space in the symmetric trijet limit, we are able to compute the leading power of thrust and HJM at NLO analytically. Then we extend the Sudakov shoulder factorization theorem to Higgs decays and verify it against both analytic NLO results and numerical NLO data from \textsc{Eerad3}.
To achieve the NNLL resummation, we match all two-loop $H\to 3$ partons amplitudes in the literature, including $H\to ggg$ and $H\to gq\bar{q}$ via HEFT vertices and $H\to q\bar{q}g$ via the Yukawa vertex, to SCET operators to extract the two-loop trijet hard functions.
With all ingredients, we also present the shoulder resummed distribution to NNLL${}_\text{sh}$ accuracy.

To properly match the resummation with fixed-order results, profile scales are used in the implementation. In the standalone dijet resummation, we use a profile function to turn off the dijet resummation in the far tail region, where singular dijet logarithms no longer dominate. For joint resummation, we use another profile function for shoulder scales to turn off shoulder resummation in the dijet region, and rescale the dijet profile such that the dijet resummation is mostly off before $\rho=1/3$. The resulting N$^{3}$LL${}^{\prime}_{\rm dij}+$NNLL${}_{\rm sh}+$NNLO curve provides the HJM distribution in hadronic Higgs decay with state-of-the-art perturbative accuracy. This is needed for phenomenological studies at future $e^+e^-$ colliders like FCC-ee. In particular, the accurate HJM distribution could benefit the extractions of strong coupling constant and Yukawa coupling in precision Higgs physics, contributing to both the lower $\alpha_s$-value issue and Yukawa coupling hierarchy.

Looking forward, there are several natural extensions of this work. The first is to incorporate non-perturbative corrections systematically. In the dijet region, the leading power correction is described by a shape function, whose first moment induces the $\mathcal{O}(\Lambda_{\rm QCD}/Q)$ shift of event shapes, namely the $\Omega_{1\kappa}$ contribution, with $\kappa=q,g$. For precision applications, this shape function should be implemented in a renormalon-free gap scheme, following Refs.~\cite{Hoang:2007vb,Abbate:2010xh,Benitez:2024nav,Benitez:2025vsp}, so that the perturbative soft function and the leading non-perturbative parameter are separately well defined. While $\Omega_{1q}$ has been extracted from many event shapes with LEP data, there is no available value for $\Omega_{1g}$. One of the main applications for hadronic Higgs decays at future colliders is the simultaneous measurement of strong coupling constant and  $\Omega_{1g}$, and testing how far it is from Casimir-scaling approximation.

A second direction is the treatment of hadron-mass effects. In a realistic experimental analysis, the event reconstruction is performed with some assumptions of hadron masses. For example, charged hadrons measured on the track have accurate three momenta, and their energy is determined by the on-shell condition, assuming the hadrons are all pions in most cases. This introduces ambiguities to event shape observables. In Ref.~\cite{Salam:2001bd}, the authors introduce the so-called $p$-scheme and $E$-scheme of observable definition, and study the hadron mass effects on $\alpha_s$ extraction. In Ref.~\cite{Mateu:2012nk}, the universality among different observables is explored. 
HJM is a perfect candidate for studying these effects since it is directly sensitive to hemisphere invariant masses. It will be interesting to investigate them also in hadronic Higgs decays. 

Lastly, the non-perturbative corrections in the trijet region lack a first-principle treatment from field theories, while there has been some progress on the renormalon-based calculations~\cite{Luisoni:2020efy,Caola:2021kzt,Caola:2022vea,Nason:2023asn,Nason:2025qbx}. It will be interesting to come up with a similar treatment as dijet power correction but involving three Wilson lines.
Overall, combining these additional ingredients with flavor-tagging at colliders would turn the perturbative predictions developed in this work into a phenomenological tool for precision Higgs studies.

\acknowledgments

We thank Arindam Bhattacharya for collaborating on the early stages of this project.
We also thank Anjie Gao, Christian T. Preuss and Tong-Zhi Yang for useful discussions.
LC was supported by NSF Graduate Fellowship Program.
XYZ was supported by the MIT Pappalardo Fellowship.  The computations in this paper
were run on the FASRC Cannon cluster at Harvard University.
This work was performed in part at Aspen Center for Physics, which is supported by NSF grant PHY-2210452.

%\clearpage
\appendix

\section{QCD running coupling}
\label{app:betaRG}

In this appendix, we list the expressions for running coupling. The beta function is given as,
\begin{equation}
	\label{eq:beta}
	\frac{\mathrm{d}\alpha_s(\mu)}{\mathrm{d}\ln \mu}= \beta (\alpha_s(\mu)),\quad 
	\beta (\alpha)=- 2 \alpha\, \left[ \left( \frac{\alpha}{4 \pi} \right) \beta_0 + \left(
	\frac{\alpha}{4 \pi} \right)^2 \beta_1 + \left( \frac{\alpha}{4 \pi}
	\right)^3 \beta_2 + \cdots \right]\,,
\end{equation}
where the coefficients up to five loops are given by~\cite{Tarasov:1980au,Larin:1993tp,vanRitbergen:1997va,Czakon:2004bu,Herzog:2017ohr}
\begin{align}
	\beta_0 &=\frac{11}{3} C_A - \frac{4}{3} T_F n_f 
	\,, \\
	\beta_1 &= \frac{34}{3} C_A^2 - \frac{20}{3} C_A T_F n_f - 4 C_F T_F n_f 
	\nn \,,\\
	\beta_2 &= n_f^2 T_F^2 \left(\frac{158 }{27}C_A+\frac{44
	}{9}C_F\right)+n_f T_F \left(2
	C_F^2-\frac{205 }{9}C_FC_A-\frac{1415 }{27}C_A^2\right)+\frac{2857 }{54}C_A^3
	\,,\nn\\
	% \beta_3 &= \frac{1093}{729} n_f^3
	% +\left(\frac{50065}{162} + \frac{6472}{81}\zeta_3\right) n_f^2
	% +\left(-\frac{1078361}{162} - \frac{6508}{27} \zeta_3 \right) n_f + 3564 \zeta_3 + \frac{149753}{6} \nn\\
 %    \beta_4 &= \left(\frac{1205}{2916}-\frac{152 \zeta_3}{81}\right) n_f^4+\left(-\frac{48722 \zeta
 %   (3)}{243}+\frac{460 \zeta_5}{9}-\frac{630559}{5832}+\frac{809 \pi ^4}{1215}\right)
 %   n_f^3\nn\\
 %   &+\left(\frac{698531 \zeta_3}{81}-\frac{381760 \zeta_5}{81}+\frac{25960913}{1944}-\frac{5263 \pi ^4}{405}\right)
 %   n_f^2\nn\\
 %   &+\left(-\frac{4811164 \zeta_3}{81}+\frac{1358995 \zeta_5}{27}-\frac{336460813}{1944}+\frac{6787 \pi ^4}{108}\right) n_f\nn\\
 %   &-288090 \zeta_5+\frac{621885 \zeta_3}{2}-\frac{9801 \pi ^4}{20}+\frac{8157455}{16}\,.
    \beta_{3}&=C_{A}^{4}\left(\frac{150653}{486}-\frac{44\zeta_{3}}{9}\right)+C_{A}^{3}T_{F}n_{f}\left(-\frac{39143}{81}+\frac{136\zeta_{3}}{3}\right)+C_{A}^{2}C_{F}T_{F}n_{f}\left(\frac{7073}{243}-\frac{656\zeta_{3}}{9}\right)\nn\\
    &+C_{A}C_{F}^{2}T_{F}n_{f}\left(-\frac{4204}{27}+\frac{352\zeta_{3}}{9}\right)+46C_{F}^{3}T_{F}n_{f}+C_{A}^{2}T_{F}^{2}n_{f}^{2}\left(\frac{7930}{81}+\frac{224\zeta_{3}}{9}\right)\nn\\
    &+C_{A}C_{F}T_{F}^{2}n_{f}^{2}\left(\frac{17152}{243}+\frac{448\zeta_{3}}{9}\right)+C_{F}^{2}T_{F}^{2}n_{f}^{2}\left(\frac{1352}{27}-\frac{704\zeta_{3}}{9}\right)+\frac{424C_{A}T_{F}^{3}n_{f}^{3}}{243}\nn\\
    &+\frac{1232C_{F}T_{F}^{3}n_{f}^{3}}{243}+\frac{d_{A}^{abcd}d_{A}^{abcd}}{N_{A}}\left(-\frac{80}{9}+\frac{704\zeta_{3}}{3}\right)+n_{f}\frac{d_{F}^{abcd}d_{A}^{abcd}}{N_{A}}\left(\frac{512}{9}-\frac{1664\zeta_{3}}{3}\right)\nn\\
    &+n_{f}^{2}\frac{d_{F}^{abcd}d_{F}^{abcd}}{N_{A}}\left(-\frac{704}{9}+\frac{512\zeta_{3}}{3}\right)\,,\nn\\
    %%%%%%%%%%%%%%%%%%%%%%%%
    \beta_{4}&=C_{A}^{5}\left(\frac{8296235}{3888}-\frac{1630\zeta_{3}}{81}-\frac{1045\zeta_{5}}{9}+\frac{121\pi^{4}}{540}\right)+C_{A}^{4}T_{F}n_{f}\left(-\frac{5048959}{972}+1230\zeta_{5}\right.\nn\\
    &\left.-\frac{583\pi^{4}}{270}+\frac{10505\zeta_{3}}{81}\right)+C_{A}^{3}C_{F}T_{F}n_{f}\left(\frac{8141995}{1944}+146\zeta_{3}-\frac{8720\zeta_{5}}{3}+\frac{451\pi^{4}}{135}\right)\nn\\
    &+C_{A}^{2}C_{F}^{2}T_{F}n_{f}\left(-\frac{548732}{81}-\frac{50581\zeta_{3}}{27}-\frac{242\pi^{4}}{135}+\frac{12820\zeta_{5}}{3}\right)+C_{A}C_{F}^{3}T_{F}n_{f}\left(3717\right.\nn\\
    &\left.-\frac{7480\zeta_{5}}{3}+\frac{5696\zeta_{3}}{3}\right)+C_{F}^{4}T_{F}n_{f}\left(-\frac{4157}{6}-128\zeta_{3}\right)+C_{A}^{3}T_{F}^{2}n_{f}^{2}\left(\frac{843067}{486}-\frac{2200\zeta_{5}}{3}\right.\nn\\
    &\left.-\frac{52\pi^{4}}{135}+\frac{18446\zeta_{3}}{27}\right)+C_{A}^{2}C_{F}T_{F}^{2}n_{f}^{2}\left(\frac{5701}{162}-\frac{472\pi^{4}}{135}+\frac{1600\zeta_{5}}{3}+\frac{26452\zeta_{3}}{27}\right)\nn\\
    &+C_{A}C_{F}^{2}T_{F}^{2}n_{f}^{2}\left(\frac{31583}{18}-\frac{28628\zeta_{3}}{27}-\frac{4400\zeta_{5}}{3}+\frac{572\pi^{4}}{135}\right)+C_{F}^{3}T_{F}^{2}n_{f}^{2}\left(-\frac{5018}{9}\right.\nn\\
    &\left.-\frac{2144\zeta_{3}}{3}+\frac{4640\zeta_{5}}{3}\right)+C_{A}^{2}T_{F}^{3}n_{f}^{3}\left(-\frac{2077}{27}-\frac{9736\zeta_{3}}{81}+\frac{56\pi^{4}}{135}+\frac{320\zeta_{5}}{9}\right)\nn\\
    &+C_{A}C_{F}T_{F}^{3}n_{f}^{3}\left(-\frac{736}{81}-\frac{5680\zeta_{3}}{27}+\frac{112\pi^{4}}{135}\right)+C_{F}^{2}T_{F}^{3}n_{f}^{3}\left(-\frac{9922}{81}-\frac{176\pi^{4}}{135}+\frac{7616\zeta_{3}}{27}\right)\nn\\
    &+C_{A}T_{F}^{4}n_{f}^{4}\left(\frac{916}{243}-\frac{640\zeta_{3}}{81}\right)+C_{F}T_{F}^{4}n_{f}^{4}\left(-\frac{856}{243}-\frac{128\zeta_{3}}{27}\right)+C_{A}\frac{d_{A}^{abcd}d_{A}^{abcd}}{N_{A}}\left(-\frac{514}{3}\right.\nn\\
    &\left.-\frac{15400\zeta_{5}}{3}-\frac{484\pi^{4}}{45}+\frac{18716\zeta_{3}}{3}\right)+C_{A}n_{f}\frac{d_{F}^{abcd}d_{A}^{abcd}}{N_{A}}\left(\frac{11312}{9}-\frac{127736\zeta_{3}}{9}+\frac{1144\pi^{4}}{45}\right.\nn\\
    &\left.+\frac{67520\zeta_{5}}{9}\right)+C_{F}n_{f}\frac{d_{F}^{abcd}d_{A}^{abcd}}{N_{A}}\left(-320+\frac{1280\zeta_{3}}{3}+\frac{6400\zeta_{5}}{3}\right)+T_{F}n_{f}\frac{d_{A}^{abcd}d_{A}^{abcd}}{N_{A}}\left(\frac{904}{9}\right.\nn\\
    &\left.-\frac{20752\zeta_{3}}{9}+\frac{176\pi^{4}}{45}+\frac{4000\zeta_{5}}{9}\right)+C_{A}n_{f}^{2}\frac{d_{F}^{abcd}d_{F}^{abcd}}{N_{A}}\left(-\frac{7184}{3}-\frac{352\pi^{4}}{45}+\frac{2240\zeta_{5}}{9}+\frac{40336\zeta_{3}}{9}\right)\nn\\
    &+C_{F}n_{f}^{2}\frac{d_{F}^{abcd}d_{F}^{abcd}}{N_{A}}\left(\frac{4160}{3}-\frac{12800\zeta_{5}}{3}+\frac{5120\zeta_{3}}{3}\right)+T_{F}n_{f}^{2}\frac{d_{F}^{abcd}d_{A}^{abcd}}{N_{A}}\left(-\frac{3680}{9}-\frac{1280\zeta_{5}}{9}\right.\nn\\
    &\left.-\frac{416\pi^{4}}{45}+\frac{40160\zeta_{3}}{9}\right)+T_{F}n_{f}^{3}\frac{d_{F}^{abcd}d_{F}^{abcd}}{N_{A}}\left(\frac{3520}{9}-\frac{2624\zeta_{3}}{3}+\frac{128\pi^{4}}{45}+\frac{1280\zeta_{5}}{3}\right)\,.\nn
\end{align}

At one-loop, Eq.~\eqref{eq:beta} can be solved exactly. At two loops and beyond, we use the iterative solution, where the order is controlled by $\beta_n/\beta_0$. At five loops, the result is
\begin{align}
    \alpha_s(\mu) &= \alpha_s(Q)\Bigg\{X+\frac{\alpha_s(Q)}{4\pi }\frac{\beta_1}{\beta_0}\ln X+\frac{\alpha_s^2(Q)}{16\pi^2}\left[\frac{\beta_2}{\beta_0}\left(1-\frac{1}{X}\right)+\frac{\beta_1^2}{\beta_0^2}\left(\frac{1}{X}-1+\frac{\ln X}{X}\right)\right]\nn\\
    &+\frac{\alpha_s^3(Q)}{64\pi^3}\left[\frac{\beta_3}{2\beta_0}\left(1-\frac{1}{X^2}\right)+\frac{\beta_1\beta_2}{\beta_0^2}\left(\frac{1-X}{X}+\frac{\ln X}{X^2}\right)+\frac{\beta_1^3}{2\beta_0^3}\left(1+\frac{1}{X^2}-\frac{2}{X}-\frac{\ln^2 X}{X^2}\right)\right]\nn\\
    &+\frac{\alpha_s^4(Q)}{256\pi^4}\Bigg[\frac{\beta_4}{3\beta_0}\left(1-\frac{1}{X^3}\right)+\frac{\beta _1 \beta _3}{\beta _0^2}\left(\frac{\ln X}{X^3}+\frac{1}{6X^3}+\frac{1}{2
   X}-\frac{2}{3}\right)\nn\\
   &+\frac{\beta _2^2}{\beta _0^2}\left(-\frac{2}{3 X^3}+\frac{1}{X^2}-\frac{1}{3}\right)+\frac{\beta _1^2 \beta _2}{\beta _0^3}\left(\frac{-\ln^2 X+\ln X+1}{X^3}-\frac{\ln X+1}{X^2}-\frac{1}{X}+1\right)\nn\\
   &+\frac{\beta _1^4}{\beta
   _0^4}\left(\frac{\ln^3 X}{3X^3}-\frac{\ln^2 X}{2X^3}-\frac{\ln X}{X^3}-\frac{1}{6X^3}+\frac{\ln X}{X^2}+\frac{1}{2 X}-\frac{1}{3}\right)\Bigg]\Bigg\}^{-1}\,,
\end{align}
where
\begin{equation}
    X\equiv 1+\frac{\alpha_s(Q)}{2\pi} \beta_0\ln\frac{\mu}{Q}\,.
\end{equation}
For the resummation presented in the main text, we choose $\alpha_s(m_Z)=0.118$. 

\section{RG kernels}

For both dijet and shoulder resummation, we introduce the RG kernels in Eq.~\eqref{eq:SAfunctions}. Here we quote the equations again,
\begin{align}
S_{i}(\nu,\mu)&=-\int_{\alpha_{s}(\nu)}^{\alpha_{s}(\mu)}d\alpha\,\frac{\Gamma_{i}^{\text{cusp}}(\alpha)}{\beta(\alpha)}\int_{\alpha_{s}(\nu)}^{\alpha}\frac{d\alpha'}{\beta(\alpha')}\,,\nn\\
A_{\Gamma,i}(\nu,\mu)&=-\int_{\alpha_{s}(\nu)}^{\alpha_{s}(\mu)}d\alpha\,\frac{\Gamma_{i}^{\text{cusp}}(\alpha)}{\beta(\alpha)}\,,\qquad
A_{i}(\nu,\mu)=-\int_{\alpha_{s}(\nu)}^{\alpha_{s}(\mu)}d\alpha\,\frac{\gamma_{i}(\alpha)}{\beta(\alpha)}\,,
\end{align}
The orders for these RG kernels are defined through the orders of $\beta$ function and anomalous dimensions. Below we list the results for N${}^3$LL and N${}^3$LL${}^\prime$ resummation, which are summarized in Refs.~\cite{Billis:2019evv,Almeida:2014uva}.

For $S_i(\nu,\mu)$, we have
\begin{equation}
	S_i(\nu,\mu)=S_i^{\text{LL}}(\nu,\mu)+S_i^{\text{NLL}}(\nu,\mu)+S_i^{\text{NNLL}}(\nu,\mu)+S_i^{\text{N}^3\text{LL}}(\nu,\mu)+\cdots\,,
\end{equation}
with 
\begin{align}
	S_i^{\text{LL}}(\nu,\mu)&=\frac{\Gamma_0\pi}{\beta_0^2 \alpha_s(\nu)}\left(1-\frac{1}{r}-\ln r\right)
    \,, \notag\\
	S_i^{\text{NLL}}(\nu,\mu)&=\frac{\Gamma_0}{4\beta_0^2}\left[\left(\frac{\Gamma _1}{\Gamma _0}-\frac{\beta
   _1}{\beta _0}\right) (1-r+\ln r)+\frac{\beta
   _1 }{2 \beta _0}\ln^2 r\right]
   \,, \notag \\
   S_i^{\text{NNLL}}(\nu,\mu)&=\frac{\Gamma_0 \alpha_s(\nu)}{32\pi\beta_0^2}\bigg[\left(\frac{\beta _1^2}{\beta _0^2}-\frac{\beta
   _2}{\beta _0}\right) \left(1-r^2+2 \ln r\right)+ \left(\frac{\beta _1 \Gamma
   _1}{\beta _0 \Gamma _0}-\frac{\Gamma _2}{\Gamma
   _0}\right)(1-r)^2
   \notag \\
   &+2 \left(\frac{\beta _1 \Gamma
   _1}{\beta _0 \Gamma _0}-\frac{\beta _1^2}{\beta
   _0^2}\right) (1-r+r \ln r)\bigg]\nn\\
   S_i^{\text{N}^3\text{LL}}(\nu,\mu)&=\frac{\Gamma_0 \alpha_s(\nu)^2}{64\pi^2\beta_0^2}\Bigg[\left(\frac{\beta
   _2}{\beta _0}-\frac{\beta _1^2}{\beta _0^2}\right)\left(\frac{\Gamma _1}{\Gamma _0}-\frac{\beta
   _1}{\beta _0}\right)\frac{(1-r)^2(2+r)}{3}+\left(\frac{\Gamma_3}{\Gamma_0}-\frac{\beta_3}{\beta_0}-\frac{\beta_1\Gamma_2}{\beta_0\Gamma_0}+\frac{\beta_1\beta_2}{\beta_0^2}\right)\nn\\
   &\left(\frac{1-r^3}{3}-\frac{1-r^2}{2}\right)+\frac{\beta_1}{\beta_0}\left(\frac{\Gamma_2}{\Gamma_0}-\frac{\beta_2}{\beta_0}-\frac{\beta_1\Gamma_1}{\beta_0\Gamma_0}+\frac{\beta_1^2}{\beta_0^2}\right)\left(\frac{1-r^2}{4}+\frac{r^2\ln r}{2}\right)\nn\\
   &+\left(-\frac{\beta_3}{\beta_0}+\frac{2\beta_1\beta_2}{\beta_0^2}-\frac{\beta_1^3}{\beta_0^3}\right)\left(\frac{1-r^2}{4}+\frac{\ln r}{2}\right)\Bigg]\,.
   \end{align}
where $\Gamma_{0,1,2,3,\cdots}$ are the cusp anomalous dimensions and we have omitted the $i=q,g$ subscript.
We also define $r\equiv \frac{\alpha_s(\mu)}{\alpha_s(\nu)}$ to save space. Similarly, we have
\begin{align}
	A_{\Gamma,i}(\nu,\mu)=A_{\Gamma,i}^{\text{LL}}(\nu,\mu)+A_{\Gamma,i}^{\text{NLL}}(\nu,\mu)+A_{\Gamma,i}^{\text{NNLL}}(\nu,\mu)+A_{\Gamma,i}^{\text{N}^3\text{LL}}(\nu,\mu)+\cdots
	\,, \notag \\
	A_{i}(\nu,\mu)=A_{i}^{\text{LL}}(\nu,\mu)+A_{i}^{\text{NLL}}(\nu,\mu)+A_{i}^{\text{NNLL}}(\nu,\mu)+A_{i}^{\text{N}^3\text{LL}}(\nu,\mu)+\cdots
\,.\end{align}
The corresponding term at each order is
\begin{align}
	A_{\Gamma,i}^{\text{LL}}(\nu,\mu)&=\frac{\Gamma_0}{2\beta_0}\ln r
	\,, \notag \\
	A_{\Gamma,i}^{\text{NLL}}(\nu,\mu)&=\frac{\Gamma_0\alpha_s(\nu)}{8\pi\beta_0}\left(\frac{\beta_1}{\beta_0}-\frac{\Gamma_1}{\Gamma_0}\right)(1-r)
	\,, \notag \\
	A_{\Gamma,i}^{\text{NNLL}}(\nu,\mu)&=\frac{\Gamma _0  \alpha (\nu )^2}{64 \pi ^2 \beta _0}\left(\frac{\beta _1^2}{\beta _0^2}-\frac{\beta
   _2}{\beta _0}+\frac{\Gamma _2}{\Gamma_0}-\frac{\beta _1 \Gamma _1}{\beta _0 \Gamma_0}\right)\left(r^2-1\right)\,,\nn\\
   A_{\Gamma,i}^{\text{N}^3\text{LL}}(\nu,\mu)&=\frac{\Gamma _0  \alpha (\nu )^3}{384 \pi ^3 \beta _0}\left[\frac{\Gamma_3}{\Gamma_0}-\frac{\Gamma_2\beta_1}{\Gamma_0\beta_0}+\frac{\Gamma_1}{\Gamma_0}\left(\frac{\beta_1^2}{\beta_0^2}-\frac{\beta_2}{\beta_0}\right)-\frac{\beta_1^3}{\beta_0^3}+\frac{2\beta_1\beta_2}{\beta_0^2}-\frac{\beta_3}{\beta_0}\right](r^3-1)\,.
\end{align}

For $A_{i}(\nu,\mu)$, the expressions are similar with the replacement of anomalous dimensions.

\section{Anomalous dimensions}
In this appendix, we summarize both cusp and regular anomalous dimensions for dijet resummation and shoulder resummation. The dijet anomalous dimensions agree with Ref.~\cite{Ju:2023dfa}.
The trijet hemisphere soft functions are the same as Ref.~\cite{Bhattacharya:2023qet} and have been presented in Sec.~\ref{sec:trijet_soft_func}, so we will not repeat here.

\subsection*{Cusp anomalous dimensions}
The cusp anomalous dimension has been computed up to four loops in Refs.~\cite{Korchemsky:1987wg, Moch:2004pa,Moch:2017uml, Moch:2018wjh, Davies:2016jie,Henn:2019swt,vonManteuffel:2020vjv}. If we write $\Gamma^{\text{cusp}}_i(\alpha_s)=C_i \Gamma^{\text{cusp}}(\alpha_s)$ with $C_i=C_F$ or $C_A$, then
\begin{align} 
\Gamma^{\text{cusp}}(\alpha_s)&=\left(\frac{\alpha_s}{4\pi}\right)\Gamma_0+\left(\frac{\alpha_s}{4\pi}\right)^2 \Gamma_1+\left(\frac{\alpha_s}{4\pi}\right)^3 \Gamma_2+\left(\frac{\alpha_s}{4\pi}\right)^4 \Gamma_{3,i}+\cdots\notag 
 \,,
\end{align}
where
\begin{align}
    \Gamma_0&=4 
      \,, \notag\\
	\Gamma_1&=4\left[C_A\left(\frac{67}{9}-\frac{\pi^2}{3}\right)-\frac{20}{9}T_F n_f\right]
      \,, \notag\\
	\Gamma_2&=4 \bigg[\left(-\frac{56 \zeta_3}{3}-\frac{418}{27}+\frac{40 \pi ^2}{27}\right)
   C_A n_f T_F+\left(\frac{22 \zeta_3}{3}+\frac{245}{6}-\frac{134 \pi ^2}{27}+\frac{11
   \pi ^4}{45}\right) C_A^2
      \notag\\
   &\qquad+\left(16 \zeta_3-\frac{55}{3}\right) C_F n_f T_F-\frac{16}{27}
   n_f^2 T_F^2\bigg]\,,\nn\\
   \Gamma_{3,q} &=    15526.5-3878.93 n_f +146.683  n_f^2+ 2.454 n_f^3\,,\nn\\
   \Gamma_{3,g}  &= 13626.7-3904.67 n_f+146.683  n_f^2+ 2.454  n_f^3\,.
\end{align}
For four-loop cusp $\Gamma_3$, we only provide the numerical values and the full analytic result can be found in Refs.~\cite{Henn:2019swt,vonManteuffel:2020vjv}. Note that the Casimir scaling between quark and gluon cusp anomalous dimensions breaks at four-loop.
\subsection*{Dijet hard anomalous dimensions}
For $Hgg$ dijet hard function, the anomalous dimensions are
\begin{align}
    \gamma^\text{dij}_{H,g}=\left(\frac{\alpha_s}{4\pi}\right) \gamma^{\text{dij},(0)}_{H,g}+\left(\frac{\alpha_s}{4\pi}\right) ^2\gamma^{\text{dij},(1)}_{H,g}+\left(\frac{\alpha_s}{4\pi}\right) ^3\gamma^{\text{dij},(2)}_{H,g}+\cdots\,,
\end{align}
with
\begin{align}
    \gamma^{\text{dij},(0)}_{H,g}&=0,\nn\\
    \gamma^{\text{dij},(1)}_{H,g}&=-\frac{1240}{9} + \frac{23\pi^2}{3} + 36\zeta_3\,, \nn\\
    \gamma^{\text{dij},(2)}_{H,g}&=-\frac{617585}{243} + \frac{4159\pi^2}{9} - \frac{1801\pi^4}{45} + 2396\zeta_3 - 120\pi^2\zeta_3 - 864\zeta_5\,.
\end{align}
We have plugged in the color $C_F=4/3$, $C_A=3$, $T_F=1/2$ and $n_f=5$ to save space.
Similarly, for $Hq\bar{q}$ dijet hard function, we have
\begin{align}
    \gamma^{\text{dij},(0)}_{H,q}&=0,\nn\\
    \gamma^{\text{dij},(1)}_{H,q}&=-\frac{256}{81} - \frac{28\pi^2}{9} + \frac{368\zeta_3}{3}\,,\nn\\
    \gamma^{\text{dij},(2)}_{H,q}&=\frac{1510468}{2187} + \frac{9268\pi^2}{243} - \frac{11716\pi^4}{1215} + \frac{124304\zeta_3}{27} - \frac{9760}{81}\pi^2\zeta_3 - \frac{30656\zeta_5}{9}\,.
\end{align}

\subsection*{Inclusive jet anomalous dimensions}

Both dijet resummation and shoulder resummation use the same jet function, and thus the same jet anomalous dimensions. From Refs.~\cite{Lunghi:2002ju,Bosch:2004th,Becher:2006qw,Becher:2009th,Becher:2010pd}, we have
\begin{align}
\gamma_{jq}^{(0)}=&-3C_F\,, \nn\\
\gamma_{jq}^{(1)}=&C_F^2 \left(-\frac{3}{2} + 2\pi^2 - 24\zeta_3\right) + C_F C_A \left(-\frac{1769}{54} - \frac{11\pi^2}{9} + 40\zeta_3\right) + C_F T_F n_f \left(\frac{242}{27} + \frac{4\pi^2}{9}\right) \,, \nn\\
\gamma_{jq}^{(2)}=&C_F^3 \left(-\frac{29}{2} - 3\pi^2 - \frac{8\pi^4}{5} - 68\zeta_3 + \frac{16\pi^2}{3}\zeta_3 + 240\zeta_5\right) \nn\\
&\quad + C_F^2 C_A \left(-\frac{151}{4} + \frac{205\pi^2}{9} + \frac{247\pi^4}{135} - \frac{844}{3}\zeta_3 - \frac{8\pi^2}{3}\zeta_3 - 120\zeta_5\right) \nn\\
&\quad + C_F C_A^2 \left(-\frac{412907}{2916} - \frac{419\pi^2}{243} - \frac{19\pi^4}{10} + \frac{5500}{9}\zeta_3 - \frac{88\pi^2}{9}\zeta_3 - 232\zeta_5\right) \nn\\
&\quad + C_F^2 T_F n_f \left(\frac{4664}{27} - \frac{32\pi^2}{9} - \frac{164\pi^4}{135} + \frac{208}{9}\zeta_3\right) \nn\\
&\quad + C_F C_A T_F n_f \left(-\frac{5476}{729} + \frac{1180\pi^2}{243} + \frac{46\pi^4}{45} - \frac{2656}{27}\zeta_3\right) \nn\\
&\quad + C_F T_F^2 n_f^2 \left(\frac{13828}{729} - \frac{80\pi^2}{81} - \frac{256}{27}\zeta_3\right)\,,  \nn\\
\gamma_{jg}^{(0)}=&-\beta_0 \,, \nn\\
\gamma_{jg}^{(1)}=&C_A^2 \left(-\frac{1096}{27} + \frac{11\pi^2}{9} + 16\zeta_3\right) + C_A T_F n_f \left(\frac{368}{27} - \frac{4\pi^2}{9}\right) + 4 C_F T_F n_f\,,  \nn\\
\gamma_{jg}^{(2)}=&C_A^3 \left(-\frac{331153}{1458} + \frac{6217\pi^2}{243} + 260\zeta_3 - \frac{583\pi^4}{270} - \frac{64\pi^2}{9}\zeta_3 - 112\zeta_5\right) \nn\\
&\quad + C_A^2 T_F n_f \left(\frac{42557}{729} - \frac{2612\pi^2}{243} - \frac{16}{27}\zeta_3 + \frac{154\pi^4}{135}\right) \nn\\
&\quad + C_A T_F^2 n_f^2 \left(\frac{3622}{729} + \frac{80\pi^2}{81} - \frac{448}{27}\zeta_3\right) \nn\\
&\quad + C_A C_F T_F n_f \left(\frac{4145}{27} - \frac{4\pi^2}{3} - \frac{608}{9}\zeta_3 - \frac{16\pi^4}{45}\right) \nn\\
&\quad - 2 C_F^2 T_F n_f - \frac{44}{9} C_F T_F^2 n_f^2\,.
\end{align}
\subsection*{Dijet soft anomalous dimensions}
For dijet soft functions, we also have both quark and gluon channels. The quark channel anomalous dimensions up to three loops are
\begin{align}
\gamma^{\text{dij},(0)}_{s,q}=&0 \,, \nn\\
\gamma^{\text{dij},(1)}_{s,q}=&C_F n_f T_F \left(-\frac{224}{27} + \frac{4\pi^2}{9}\right) + C_A C_F \left(\frac{808}{27} - \frac{11\pi^2}{9} - 28\zeta_3\right)\,, \nn\\
\gamma^{\text{dij},(2)}_{s,q}=&C_F n_f^2 T_F^2 \left(-\frac{8320}{729} - \frac{80\pi^2}{81} + \frac{448}{27}\zeta_3\right) + C_F^2 n_f T_F \left(-\frac{3422}{27} + \frac{4\pi^2}{3} + \frac{16\pi^4}{45} + \frac{608}{9}\zeta_3\right)\nn\\
+& C_A C_F n_f T_F \left(-\frac{23684}{729} + \frac{2828\pi^2}{243} - \frac{16\pi^4}{15} + \frac{1456}{27}\zeta_3\right) \nn\\
+& C_A^2 C_F \left(\frac{136781}{729} - \frac{6325\pi^2}{243} + \frac{88\pi^4}{45} - \frac{1316}{3}\zeta_3 + \frac{88}{9}\pi^2\zeta_3 + 192\zeta_5\right)\,,
\end{align}
and the gluon anomalous dimensions can be obtained by Casimir scaling.

\subsection*{Trijet hard anomalous dimensions}
In Eq.~\eqref{eq:sh_hard_rge}, we express the trijet hard functions in terms of quark and gluon anomalous dimensions. To three loops in terms of $\alpha_s/(4\pi)$ expansion, they are~\cite{Moch:2005id,Moch:2005tm,Idilbi:2005ni,Idilbi:2006dg,Becher:2006mr}
\begin{align}
  \gamma^{q}_0 = & \, -3 C_F \,,
\nn\\ 
  \gamma^{q}_1 = & \, C_A C_F \left(-11 \zeta_2+26
    \zeta_3-\frac{961}{54}\right)
+C_F^2 \left(12 \zeta_2-24 \zeta_3-\frac{3}{2}\right)+C_F n_f \left(2 \zeta_2+\frac{65}{27}\right) \,,
\nn\\
  \gamma_2^{q} = & \,   
     -\frac{4880 \pi ^2 \zeta _3}{81}+\frac{82072 \zeta _3}{27}-\frac{15328 \zeta
   _5}{9}-\frac{2066 \pi ^4}{405}-\frac{5062 \pi
   ^2}{81}-\frac{196621}{243}
   \nn\\ &
   +\left(-\frac{7472 \zeta _3}{81}+\frac{68 \pi ^4}{1215}+\frac{4564 \pi
   ^2}{243}+\frac{36236}{729}\right) n_f+\left(-\frac{32 \zeta _3}{81}-\frac{40 \pi ^2}{81}+\frac{9668}{2187}\right)
   n_f^2 \,,
\nn\\
  \gamma_0^{g} = &\, - \beta_0 \,,
\nn\\
  \gamma_1^{g} = &\, \,C_A^2 \left(\frac{11 \zeta_2}{3}+2 \zeta_3-\frac{692}{27}\right)+C_A n_f \left(\frac{128}{27}-\frac{2 \zeta_2}{3}\right)+2 C_F n_f \,,
  \nn\\
  \gamma_2^{g} = &\,  -60 \pi ^2 \zeta _3+1098 \zeta _3-432 \zeta
   _5-\frac{319 \pi ^4}{10}+\frac{6109 \pi ^2}{18}-\frac{97186}{27}
   \nn\\ &
  +\left(\frac{460 \zeta _3}{9}+\frac{107 \pi ^4}{45}-\frac{635 \pi
   ^2}{27}+\frac{59635}{162}\right) n_f+\left(-\frac{56 \zeta _3}{9}+\frac{10 \pi ^2}{27}-\frac{1061}{486}\right)
   n_f^2 \,.
\end{align}

\subsection*{Other couplings}
For Yukawa coupling, we have the following dimensions~\cite{Vermaseren:1997fq,Chetyrkin:1997dh},
\begin{align}
\gamma_y^{(0)} &= -6C_F\,,  \nn\\
\gamma_y^{(1)} &= -\frac{97}{3}C_A C_F - 3C_F^2 + \frac{10}{3}C_F n_f\,,  \nn\\
\gamma_y^{(2)} &= -\frac{11413}{54}C_A^2 C_F + \frac{129}{2}C_A C_F^2 - 129 C_F^3 + \frac{70}{27}C_F n_f^2 \nn\\
&\qquad + C_F^2 n_f \left(46 - 48\zeta_3\right) + C_A C_F n_f \left(\frac{556}{27} + 48\zeta_3\right)\,.
\end{align}
For the effective $Hgg$ coupling, we can determine the anomalous dimensions from RG invariance of the $Hgg$ dijet factorization formula. In the end, we use
\begin{align}
\gamma_\lambda^{(n)} &= -2(n+1)\beta_{n}
\end{align}

\section{Hard function boundaries}

In this section, we summarize the fixed-order expansion of hard functions, including dijet and shoulder hard functions. Since the logarithmic term at each order can be predicted directly from the hard RGE, i.e. Eq.~\eqref{eq:hard_rge} and Eq.~\eqref{eq:sh_hard_rge}, we will only present the constants.

For $Hgg$ dijet hard function, we have up to three loops, with the values of colors and $n_f$ plugged in,
\begin{align}
    H_g^{\text{dij}}(Q^2,Q^2)&=1 + \left(\frac{\alpha_s(Q)}{4\pi}\right)7\pi^2 + \left(\frac{\alpha_s(Q)}{4\pi}\right)^2 \left(\frac{2135}{27} + \frac{755\pi^2}{6} + \frac{37\pi^4}{2} - \frac{998\zeta_3}{3}\right) \nn\\
    &+ \left(\frac{\alpha_s(Q)}{4\pi}\right)^3 \left(\frac{28761827}{4374} + \frac{1767937\pi^2}{486} + \frac{585928\pi^4}{1215} + \frac{47437\pi^6}{1890} - \frac{2038448\zeta_3}{81}\right.\nn\\
    &\left.\hspace{2cm} - \frac{28456}{9}\pi^2\zeta_3 - 624\zeta_3^2 + \frac{157228\zeta_5}{9}\right)\,,
\end{align}
and for $Hq\bar{q}$, we get
\begin{align}
    H_q^{\text{dij}}(Q^2,Q^2)&=1 + \left(\frac{\alpha_s(Q)}{4\pi}\right)\left(-\frac{16}{3} + \frac{28\pi^2}{9}\right) + \left(\frac{\alpha_s(Q)}{4\pi}\right)^2 \left(\frac{3704}{243} + \frac{2530\pi^2}{81} + \frac{88\pi^4}{27} + \frac{824\zeta_3}{27}\right) \nn\\
    &+ \left(\frac{\alpha_s(Q)}{4\pi}\right)^3 \left(\frac{82632322}{19683} + \frac{1479814\pi^2}{2187} + \frac{513826\pi^4}{10935} + \frac{329408\pi^6}{229635} + \frac{2833504\zeta_3}{729}\right.\nn\\
    &\left.\hspace{2cm}- \frac{75976}{81}\pi^2\zeta_3 - \frac{7808\zeta_3^2}{3} - \frac{46064\zeta_5}{81}\right)\,.
\end{align}

Shoulder hard functions are the generic trijet hard functions evaluated at the symmetric configuration. For $Hgg$ process, we have both $H{ggg}$ and $H{gq\bar{q}}$. The first one, $H{ggg}$, reads~\cite{Gehrmann:2023etk,Chen:2025utl}
\begin{align}
    H^{\text{sh}}_{Hggg} &= 1+\left(\frac{\alpha_s(Q)}{4\pi}\right)\left[-\frac{5}{21}n_f + C_A \left(\frac{5}{21} + \frac{5\pi^2}{2} - 6\ln^2 2 - 3\ln^2 3 - 12\,\text{Li}_2\!\left(-\frac{1}{2}\right)\right)\right]\nn\\
    &+\left(\frac{\alpha_s(Q)}{4\pi}\right)^2\left[516.915041\, C_A^2 - 95.622247\, C_A n_f - 1.591840\, C_F n_f + 0.454806\, n_f^2\right]\,.
\end{align}
There are several comments on the hard function expressions. First of all, we always normalize the trijet hard function (or shoulder hard function) to start with 1, and the normalization (which is proportional to $\mathcal{O}(\alpha_s)$) will be absorbed into $\sigma_{\text{LO}}$. Secondly, at $\mathcal{O}(\alpha_s^2)$, there exists an analytic result in terms of GPLs up to weight-4; however, we list the numbers for convenience. Lastly, for NNLL shoulder resummation, we only need up to $\mathcal{O}(\alpha_s)$ hard function boundary; $\mathcal{O}(\alpha_s^2)$ is only needed for NNLL${}^\prime$ or N${}^3$LL.

For $H{gq\bar{q}}$, we get~\cite{Gehrmann:2023etk}
\begin{align}
    H^{\text{sh}}_{Hgq\bar{q}} &= 1+\left(\frac{\alpha_s(Q)}{4\pi}\right)\Bigg[n_f \left(-\frac{20}{9} - \frac{4\ln 3}{3}\right) \nn\\
    &\qquad + C_F \left(-14 + \frac{5\pi^2}{3} - 4\ln^2 2 - 6\ln 3 - 2\ln^2 3 - 8\,\text{Li}_2\!\left(-\frac{1}{2}\right)\right) \nn\\
    &\qquad + C_A \left(\frac{134}{9} + \frac{5\pi^2}{6} - 2\ln^2 2 + \frac{22\ln 3}{3} - \ln^2 3 - 4\,\text{Li}_2\!\left(-\frac{1}{2}\right)\right)\Bigg]\nn\\
    &+\left(\frac{\alpha_s(Q)}{4\pi}\right)^2\left[17.308615 + 643.702454\, C_A^2 - 53.664810\, C_A C_F - 12.527560\, C_F^2 \right.\nn\\
    &\left. \hspace{2cm}- 197.782285\, C_A n_f + 21.911979\, C_F n_f + 5.809199\, n_f^2\right]\,,
\end{align}
and for $Hq\bar{q}g$ in the $Hq\bar{q}$ process, the hard function reads~\cite{Ahmed:2014pka,Mondini:2019vub},
\begin{align}
    H^{\text{sh}}_{Hq\bar{q}g} &= 1+\left(\frac{\alpha_s(Q)}{4\pi}\right)\Bigg[-\frac{2}{3} n_f \ln 3 + C_F \left(-\frac{26}{5} - \frac{4\pi^2}{3} - 8\ln^2 2 + 6\ln 3 - 16\,\text{Li}_2\!\left(-\frac{1}{2}\right)\right) \nn\\
    &\qquad + C_A \left(\frac{6}{5} + \frac{\pi^2}{3} + 2\ln^2 2 + \frac{11\ln 3}{3} + 4\,\text{Li}_2\!\left(-\frac{1}{2}\right)\right)\Bigg]\nn\\
    &+\left(\frac{\alpha_s(Q)}{4\pi}\right)^2\left[49.458563\, C_A^2 - 131.281641\, C_A C_F - 58.862150\, C_F^2 - 17.328458\, C_A n_f\right. \nn\\
    &\left.\hspace{2cm}+ 51.650870\, C_F n_f - 0.660266\, n_f^2\right]\,.
\end{align}

\bibliographystyle{jhep}
\bibliography{biblio}

\end{document}